\documentclass[english,preprint]{revtex4-1}
\usepackage[T1]{fontenc}
\usepackage[latin9]{inputenc}
\usepackage{mathtools}
\usepackage{amsmath}
\usepackage{esint}
\usepackage{babel}
\begin{document}
\title{Non-Markovian Quantum State Diffusion in a Fermionic Bath}
\author{Evgeny A. Polyakov$^{1}$, Alexey N. Rubtsov$^{1,2}$}
\address{$^{1}$Russian Quantum Center, 100 Nonaya St., Skolkovo, Moscow 143025,
Russia}
\address{$^{2}$Department of Physics, Lomonosov Moskow State University, Leninskie
gory 1, 119991 Moscow, Russia}
\begin{abstract}
We present a stochastic approach for the description of the quantum
dynamics of open system in a fermionic environment (bath). The full
quantum evolution as provided by the Schrodinger equation is reformulated
\textit{exactly} as a probabilistic average over the so-called dressed
quantum trajectories. The latter are defined as follows. The fermionic
environment can be represented as a fermi sea whose ``surface''
is covered by the ripples of quantum fluctuations. If we consider
these fluctuations in the basis of the particle-hole coherent states,
then these fluctuations produce a \textit{classical particle-hole
noise. }The probability distribution of this noise is provided by
the generalized particle-hole Husimi function of the vacuum. Then
we define the dressed quantum trajectory as the evolution of the open
system and the bath which is conditioned on a particular particle-hole
noise sample. The resulting description resembles the non-Markovian
quantum state diffusion for the bosonic bath. Therefore, we expect
that our fermionic approach will share its favourable propeties like
the possibility to carry out Monte-Carlo simulations of non-Markovian
quantum dynamics on long times.
\end{abstract}
\maketitle

\section{INTRODUCTION}

During the last decades, the stochastic description of the dynamics
of open quantum systems (OQS) in a bosonic non-Markovian bath \citep{Diosi1997,Wang2019}
receives increasing attention. There are several reasons for this.
First of all, such stochastic methods as the non-Markovian stochastic
Schrodinger equations (NMSSE) \citep{Shao2004,Yan2004,Shao2008,Yan2016},
the hierarchy of the pure states methods (HOPS) \citep{Suess2014},
and the recently proposed approach of the dressed quantum trajectories
\citep{Polyakov2019}, showed themselves as very promising approaches
for the computations on large times and in the cases of strong couplings
\citep{Hartmann2017}. Second, these stochastic descriptions are being
considered as a possible approach to the continous measurement problem
\citep{Shabani2014} and to the interpretation of the quantum mechanics
\citep{Gambetta2002,Gambetta2003}. 

At the same time for OQS in a fermionic environment the research on
stochastic description is in its infancy: there only a recent work
\citep{Han2019} in which a way of approximate mapping of the grassman
numbers to $c$-numbers is proposed. There are two reasons for such
a dramatic difference between the cases of the bosonic and fermionic
bath. The first reason is that the bosonic environment in a Gaussian
state has the same Wick theorem as a classical Gaussian process. This
allows one to represent the quantum bosonic fluctuations as a classical
noise. This fact is heavily exploited in the derivation of the NMSSE
and HOPS. At the same time, the Wick theom for a fermionic environment
is not equivalent to any classical stochastic process. This hinders
the derivation of the fermionic analogs of NMSSE and HOPS. 

The second reason is that in the conventional theory of open systems,
OQS is considered to be a local part of the bath, whose internal degress
of freedom commute with the bath's degrees of freedom \citep{Breuer2011}.
Physical motivation is that a local change of the internal state of
OQS is not expected to directly affect the amplidutes of the distant
bath regions. Otherwise it will be conceptually and computationally
difficult to draw a border between OQS and the environment. However,
in the case of a fermionic OQS in a fermionic bath (e.g. the Anderson
imputiry model (AIM)) exactly this difficulty happens: whatever far
we go into the bath, its degrees of freedom anticommute with the OQS
internal ones.

In this work we present a solution to the aforementioned two problems
of the fermionic bath case, and derive a fermionic variant of the
quantum state diffusion, namely of its variant called the dressed
quantum trajectories \citep{Polyakov2019}. 

The first problem, how to represent the fermionic fluctuations as
a classical noise, is solved by considering the particle-hole coherent
states and their Husimi function \citep{Zhang1990,RosalesZarate2013,RosalesZarate2015,Drummond2019}.
It turns out that the Husimi function of the fermi sea (the vacuum)
leads to a classical probability distribution for the particle-hole
noise. 

The second problem, that internal DOF of fermionic OQS anticommute
with the DOF of the bath, is solved by noting that this problem is
an artifact of this approximate OQS model: on a fundamental level,
there is no fermionic nonlinearities in Nature: fermionic interactions
are mediated by bosonic force fields. As a consequence, there is no
such problem on a fundamental level, if we consider all the fermions
as a part of the integral and undivisible bath, and identify the OQS
with the states of these mediating bosonic fields. Operationally this
means that we need to bosonize the fermionic nonlinearity term in
the OQS.

While in this work we present the derivation of the formalism, we
expect it to yield novel efficient Monte Carlo simulation techniques
as it happens in the bosonic case. Let us briefly recapitulate the
reasoning benind this \citep{Polyakov2019}. According to the method
of dressed quantum trajectories, we represent the bath evolution as
the master equation for its Husimi function. This master equation
is solved by first sampling stochastically from the initial Husimi
function of the vacuum (i.e. by sampling the classical particle-hole
noise of the fermionic vacuum fluctuations). For each noise sample,
we obtain a joint OQS-bath state which is conditioned on this sample.
This pair of the noise sample and the condtional OQS-bath state is
called the quantum trajectory. Then, the quantum trajectory is self-consistently
propagated in time. The average over all the ensemble of quantum trajectories
yields an \textit{exact} evolution as provided by the solution of
the full Schrodinger equation for the joint OQS-bath wavefunction.
Each individual quantum trajectory is expected to have much less complexity
then the full joint OQS-bath wavefunction. This is because the conditional
state is the state which is projected on a partical-hole coherent
state. With respect to this projection operation, the quantum field
of the bath is divided into the two parts: the observable field of
irreversibly emitted quanta, which contribute to the projection, and
the virtual field of quanta which were emitted by the open system,
but later will be reabsorbed by the open system, so that they do not
contirubute to the projection. The idea is that the observable field
may be highly populated and correlated, but it has classical statistics,
which follows from the Husimi function master equation. Therefore,
its dynamics may be simulated by Monte Carlo methods. On the contrary,
the virtual field, which is inherently quantum object, is expected
to have bounded population with time, since physically it is a retarded
polarization of the bath around the open system. Therefore, the dynamics
of virtual field can be calculated in a truncated Fock space, which
is expected to has uniform convergence on large time scales. 

This work is structured as follows. In sec \ref{sec:FERMIONIC-VACUUM-AND}
we discuss the vacuum of the fermionic bath (the fermi sea) and the
coherent states as its displacements. We introduce the classical particle-hole
field which is carried by these states. In sec. \ref{sec:CLASSICAL-STATISTICS-OF}
we introduce the particle-hole Husimi function as the probability
measure for the random samples of the partile-hole field. The vacuum
fluctuations of this field are derived. The master equation for the
particle-hole Husimi function is presented, and it is shown that non-zero
expected values of the OQS coupling operators lead to a deterministic
drift of the particle-hole field. In sec. \ref{sec:DRESSED-QUANTUM-TRAJECTORIES:}
we present the equations of motion for the dressed quantum trajectories.
A numerical Monte Carlo simulation algorithm is presented in sec.
\ref{sec:NUMERICAL-SIMULATION-FOR}. In sec. \ref{sec:DISCRETIZATION-OF-A}
a discretization of the continuous (infinite) bath is discussed. In
sec. \ref{sec:ANDERSON-IMPURITY-PROBLEM} we discuss the bosonization
of the Anderson impurity problem. We conclude in sec. \ref{sec:CONCLUSION}.
In appendix \ref{sec:Derivation-of-the} we present the derivation
of the volume element in the phase space of particle-hole fields.
In appendix \ref{sec:Differential-correspondences-for} we derive
differential correspondences for the particle-hole coherent states.
In appendix \ref{sec:CALCULATION-OF-THE} we derive the dressed Hamiltonian
which guides the temporal evolution of the quantum trajectory. 

\section{\label{sec:FERMIONIC-VACUUM-AND}FERMIONIC VACUUM AND PARTICLE-HOLE
COHERENT STATES}

\subsection{Reference (vacuum) state of the environment}

The quantum states of the environment are specified with respect to
a certain reference (vacuum) state. In the simplest case (considered
here) it is also assumed to be the initial state (at time moment $t=0$)
of the environment. We define this reference state as follows. Suppose
that there are in total $M$ modes of the environment (different spin
direction are counted as different modes: $M$ is even). We will present
equations for a finite $M$, and later discuss the limit $M\to\infty$.
We assume that all the states below $\varepsilon_{F}$ (the number
of which $M_{\textrm{h}}$) are occupied. All the states above $\varepsilon_{F}$
(the number of which is $M_{\textrm{p}}$) are free. In total we have:
$M=M_{\textrm{h}}+M_{\textrm{p}}$. We denote this state as $\left|\textrm{vac}\right\rangle $.
Then, we have the standard creation/annihilation operators $\widehat{c}_{k\sigma}^{\dagger}$
and $\widehat{c}_{k\sigma}$ in the mode labeled $k$ with the spin
direction $\sigma$ with the canonical anticommutation relations
\begin{equation}
\left\{ \widehat{c}_{k\sigma},\widehat{c}_{k^{\prime}\sigma^{\prime}}^{\dagger}\right\} =\delta_{\sigma\sigma^{\prime}}\delta_{kk^{\prime}}.
\end{equation}
We introduce the hole and particle operators: 
\begin{equation}
\widehat{c}_{k\sigma}=\begin{cases}
\widehat{c}_{k\sigma}^{\textrm{p}} & \textrm{for}\,\,\,\varepsilon_{k}\geq\varepsilon_{F},\\
\widehat{c}_{k\sigma}^{\textrm{h}\dagger} & \textrm{for}\,\,\,\varepsilon_{k}<\varepsilon_{F}.
\end{cases}
\end{equation}
Below we will also employ the greek multiindices to denote the pair
of the mode and spin index e.g. $\gamma=\left(k\sigma\right)$ . For
particles there are $M_{\textrm{p}}$ different values of multiindex
$\gamma$, and for holes there are $M_{\textrm{h}}$ different values
of multiindex $\gamma$. We denote by $k_{F}$ the mode which corresponds
to $\varepsilon_{F}$: $\varepsilon\left(k_{F}\right)=\varepsilon_{F}$.
Sometimes in order to make the notation compact, we will employ the
notation $\varepsilon_{\gamma}\equiv\varepsilon_{k\sigma}\equiv\varepsilon_{k}$.

\subsection{Model of Open System in a Fermionic Environment}

\subsubsection{Hamiltonian of the Kondo model}

The symmetry transformations of the reference state follow from the
Hamiltonian of the joint open system + bath. For example, the Kondo
model has the Hamiltonian 
\begin{equation}
\widehat{H}=\sum_{k}\varepsilon_{k}\widehat{c}_{k\sigma}^{\dagger}\widehat{c}_{k\sigma}-J\sum_{a}\sum_{kk^{\prime}\sigma\sigma^{\prime}}\widehat{c}_{k\sigma}^{\dagger}\sigma_{\sigma\sigma^{\prime}}^{a}\widehat{c}_{k^{\prime}\sigma^{\prime}}\widehat{\sigma}^{a},
\end{equation}
where $a=x,y,z$; $\sigma=\downarrow,\uparrow$; $\sigma_{\sigma\sigma^{\prime}}^{a}$
are the Pauli matrices with matrix elements $\sigma$ and $\sigma^{\prime}$.
$\widehat{\sigma}^{a}$ are the operators in the impurity (open system)
Hilbert space. $J$ is a scalar coupling coefficient. We rewrite this
Hamiltonian with respect to the chosen reference state:
\begin{multline}
\widehat{H}=\sum_{k\geq k_{F}}\varepsilon_{k}\widehat{c}_{k\sigma}^{\textrm{p}\dagger}\widehat{c}_{k\sigma}^{\textrm{p}}-\sum_{k<k_{F}}\varepsilon_{k}\widehat{c}_{k\sigma}^{\textrm{h}\dagger}\widehat{c}_{k\sigma}^{\textrm{h}}+2\sum_{\varepsilon_{k}<\varepsilon_{F}}\varepsilon_{k}\\
-J\sum_{a\sigma\sigma^{\prime}}\sum_{\begin{array}{c}
k\geq k_{F}\\
k^{\prime}\geq k_{F}
\end{array}}\widehat{c}_{k\sigma}^{\textrm{p}\dagger}\sigma_{\sigma\sigma^{\prime}}^{a}\widehat{c}_{k^{\prime}\sigma^{\prime}}^{\textrm{p}}\widehat{\sigma}^{a}+J\sum_{a\sigma\sigma^{\prime}}\sum_{\begin{array}{c}
k<k_{F}\\
k^{\prime}<k_{F}
\end{array}}\widehat{c}_{k\sigma}^{\textrm{h}\dagger}\sigma_{\sigma\sigma^{\prime}}^{a*}\widehat{c}_{k^{\prime}\sigma^{\prime}}^{\textrm{h}}\widehat{\sigma}^{a}\\
-J\sum_{a\sigma\sigma^{\prime}}\sum_{\begin{array}{c}
k\geq k_{F}\\
k^{\prime}<k_{F}
\end{array}}\widehat{c}_{k\sigma}^{\textrm{p}\dagger}\sigma_{\sigma\sigma^{\prime}}^{a}\widehat{c}_{k^{\prime}\sigma^{\prime}}^{\textrm{h}\dagger}\widehat{\sigma}^{a}-J\sum_{a\sigma\sigma^{\prime}}\sum_{\begin{array}{c}
k<k_{F}\\
k^{\prime}\geq k_{F}
\end{array}}\widehat{c}_{k\sigma}^{\textrm{h}}\sigma_{\sigma\sigma^{\prime}}^{a}\widehat{c}_{k^{\prime}\sigma^{\prime}}^{\textrm{p}}\widehat{\sigma}^{a}.
\end{multline}
Discarding the constant $2\sum_{\varepsilon_{k}<\varepsilon_{F}}\varepsilon_{k}$,
the Kondo model Hamiltonian has the general form
\begin{multline}
\widehat{H}=\widehat{H}_{s}+\widehat{H}_{\textrm{b}}+\sum_{\gamma\gamma^{\prime}}\widehat{h}_{\gamma\gamma^{\prime}}^{\textrm{pp}}\widehat{c}_{\gamma}^{\textrm{p}\dagger}\widehat{c}_{\gamma^{\prime}}^{\textrm{p}}-\sum_{\gamma\gamma^{\prime}}\widehat{h}_{\gamma\gamma^{\prime}}^{\textrm{hh}}\widehat{c}_{\gamma}^{\textrm{h}\dagger}\widehat{c}_{\gamma^{\prime}}^{\textrm{h}}\\
+\sum_{\gamma\gamma^{\prime}}\widehat{h}_{\gamma\gamma^{\prime}}^{\textrm{ph}}\widehat{c}_{\gamma}^{\textrm{p}\dagger}\widehat{c}_{\gamma^{\prime}}^{\textrm{h}\dagger}+\sum_{\gamma\gamma^{\prime}}\widehat{h}_{\gamma\gamma^{\prime}}^{\textrm{ph}\dagger}\widehat{c}_{\gamma}^{\textrm{h}}\widehat{c}_{\gamma^{\prime}}^{\textrm{p}},\label{eq:general_OQS_in_fermionic_bath}
\end{multline}
where $\widehat{H}_{\textrm{s}}$ is the Hamiltonian of the impurity
(open system); $\widehat{H}_{\textrm{b}}$ is a free bath Hamiltonian
\begin{equation}
\widehat{H}_{\textrm{b}}=\sum_{k\sigma}\varepsilon_{k}^{\textrm{p}}\widehat{c}_{k\sigma}^{\textrm{p}\dagger}\widehat{c}_{k\sigma}^{\textrm{p}}-\sum_{k\sigma}\varepsilon_{k}^{\textrm{h}}\widehat{c}_{k\sigma}^{\textrm{h}\dagger}\widehat{c}_{k\sigma}^{\textrm{h}}.
\end{equation}
For simplicity of notation, we adopt the convention of the impicitly
assumed ranges (e.g. in the $\gamma=\left(k\sigma\right)$ for a hole
creation operator it is assumed that $k<k_{F}$) . The coupling operators
$\widehat{h}_{\gamma\gamma^{\prime}}^{\textrm{pp}}$, $\widehat{h}_{\gamma\gamma^{\prime}}^{\textrm{hh}}$,
and $\widehat{h}_{\gamma\gamma^{\prime}}^{\textrm{ph}}$ act in the
impurity Hilbert space. Additionally, the operators $\widehat{h}_{\gamma\gamma^{\prime}}^{\textrm{pp}}$
and $\widehat{h}_{\gamma\gamma^{\prime}}^{\textrm{hh}}$ are supposed
to satisfy the ``generalized Hermiticity'' property
\begin{equation}
\widehat{h}_{\gamma\gamma^{\prime}}^{\textrm{pp}}=\widehat{h}_{\gamma^{\prime}\gamma}^{\textrm{pp}\dagger}\,\,\,\textrm{and}\,\,\,\widehat{h}_{\gamma\gamma^{\prime}}^{\textrm{hh}}=\widehat{h}_{\gamma^{\prime}\gamma}^{\textrm{hh}\dagger}.\label{eq:generalized_Hermite}
\end{equation}

\subsubsection{Interaction picture with respect to the bath}

For the following, it will be convenient to switch to the interaction
picture with respect to the free bath $\widehat{H}_{\textrm{b}}$:
\begin{equation}
\widehat{c}_{k\sigma}^{\textrm{p}}\to\widehat{c}_{k\sigma}^{\textrm{p}}\left(t\right)=\widehat{c}_{k\sigma}^{\textrm{p}}\exp\left(-i\varepsilon_{k}^{\textrm{p}}t\right),\,\,\,\widehat{c}_{k\sigma}^{\textrm{h}}\to\widehat{c}_{k\sigma}^{\textrm{h}}\left(t\right)=\widehat{c}_{k\sigma}^{\textrm{h}}\exp\left(+i\varepsilon_{k}^{\textrm{h}}t\right).
\end{equation}
It will be convenient to change the notation so that the time dependent
exponential factors are transfered from bath operators to open system
coupling operators: in the interaction picture the Hamiltonian (\ref{eq:general_OQS_in_fermionic_bath})
assumes the form
\begin{equation}
\widehat{H}\left(t\right)=\widehat{H}_{\textrm{s}}+\widehat{H}_{\textrm{int}}\left(t\right),
\end{equation}
where the coupling term
\begin{multline}
\widehat{H}_{\textrm{int}}\left(t\right)=\widehat{h}_{k\sigma k^{\prime}\sigma^{\prime}}^{\textrm{pp}}\left(t\right)\widehat{c}_{k\sigma}^{\textrm{p}\dagger}\widehat{c}_{k^{\prime}\sigma^{\prime}}^{\textrm{p}}-\widehat{h}_{k\sigma k^{\prime}\sigma^{\prime}}^{\textrm{hh}}\left(t\right)\widehat{c}_{k\sigma}^{\textrm{h}\dagger}\widehat{c}_{k^{\prime}\sigma^{\prime}}^{\textrm{h}}\\
+\widehat{h}_{k\sigma k^{\prime}\sigma^{\prime}}^{\textrm{ph}}\left(t\right)\widehat{c}_{k\sigma}^{\textrm{p}\dagger}\widehat{c}_{k^{\prime}\sigma^{\prime}}^{\textrm{h}\dagger}+\widehat{h}_{k\sigma k^{\prime}\sigma^{\prime}}^{\textrm{ph}\dagger}\left(t\right)\widehat{c}_{k\sigma}^{\textrm{h}}\widehat{c}_{k^{\prime}\sigma^{\prime}}^{\textrm{p}},
\end{multline}
and 
\begin{equation}
\widehat{h}_{k\sigma k^{\prime}\sigma^{\prime}}^{\textrm{pp}}\left(t\right)=\widehat{h}_{k\sigma k^{\prime}\sigma^{\prime}}^{\textrm{pp}}e^{it\left(\varepsilon_{k}^{\textrm{p}}-\varepsilon_{k^{\prime}}^{\textrm{p}}\right)},\,\,\,\widehat{h}_{k\sigma k^{\prime}\sigma^{\prime}}^{\textrm{hh}}\left(t\right)=\widehat{h}_{k\sigma k^{\prime}\sigma^{\prime}}^{\textrm{hh}}e^{it\left(-\varepsilon_{k}^{\textrm{h}}+\varepsilon_{k^{\prime}}^{\textrm{h}}\right)},
\end{equation}
\begin{equation}
\widehat{h}_{k\sigma k^{\prime}\sigma^{\prime}}^{\textrm{ph}}\left(t\right)=\widehat{h}_{k\sigma k^{\prime}\sigma^{\prime}}^{\textrm{ph}}e^{it\left(\varepsilon_{k}^{\textrm{p}}-\varepsilon_{k^{\prime}}^{\textrm{h}}\right)}.
\end{equation}
It is seen that the ``generalized Hermicitity'' property (\ref{eq:generalized_Hermite})
is still satisfied by $\widehat{h}_{k\sigma k^{\prime}\sigma^{\prime}}^{\textrm{pp}}\left(t\right)$
and $\widehat{h}_{k\sigma k^{\prime}\sigma^{\prime}}^{\textrm{hh}}\left(t\right)$.

Hereinafter we always assume the interaction picture with respect
to the bath.

\subsection{The Quantum Phase Space of The Fermionic Bath }

The Hamiltonian is composed of the $M_{\textrm{p}}^{2}$ elements
$\widehat{c}_{\alpha}^{\textrm{p}\dagger}\widehat{c}_{\alpha^{\prime}}^{\textrm{p}}$,
of the $M_{\textrm{h}}^{2}$ elements $\widehat{c}_{\alpha}^{\textrm{h}\dagger}\widehat{c}_{\alpha^{\prime}}^{\textrm{h}}$,
and of the ``anomaleous'' terms $\widehat{c}_{\alpha}^{\textrm{p}\dagger}\widehat{c}_{\alpha^{\prime}}^{\textrm{h}\dagger}$
and $\widehat{c}_{\alpha}^{\textrm{h}}\widehat{c}_{\alpha^{\prime}}^{\textrm{p}}$.
If we consider an arbitrary Hamiltonian $\widehat{g}$ which is composed
of these terms, we obtain a group $G$ of unitary evolutions $\exp\left(-i\widehat{g}\right)$.
This group is known to be isomorphic to the unitary matrix group $U\left(M\right)$
in the space of dimension $M$\citep{Zhang1990}. If the bath is initially
in the reference state $\left|\textrm{vac}\right\rangle $, the action
of the group $G$ will in general tranform $\left|\textrm{vac}\right\rangle $
into some other reference state $\left|\textrm{vac}^{\prime}\right\rangle $:
\begin{equation}
\left|\textrm{vac}^{\prime}\right\rangle =\exp\left(-i\widehat{g}\right)\left|\textrm{vac}\right\rangle .\label{eq:group_coherent_state}
\end{equation}
Such a \textit{displaced} reference states are called the particle-hole
coherent states. The appearance of the unitary group $U\left(M\right)$
can be understood that the evolution under arbitrary Hamiltonian $\widehat{g}$
leads to a certain unitary rotation of all the particle and hole orbitals. 

Let us recall the case of the bosonic bath: the dispaced vacuum state
$\left|z\right\rangle $ corresponds to a classical signal $z$, which
is the desired classical noise of the bosonic quantum fluctuations.
Then we conclude that in our case of the fermionic bath, the displaced
state (\ref{eq:group_coherent_state}) also contains some classical
signal. The procedure of how to extract such a signal is known \citep{Zhang1990}.
First we study the symmetry group of $\left|\textrm{vac}\right\rangle $
i.e. we describe the subgoup $H$ of all such quantum evolutions in
$G$ which leave the vacuum invariant up to a phase factor: 
\begin{equation}
\exp\left(-i\widehat{h}\right)\left|\textrm{vac}\right\rangle =\exp\left(-i\phi\right)\left|\textrm{vac}\right\rangle \,\,\,\textrm{for all}\,\,\,\exp\left(-i\widehat{h}\right)\,\,\textrm{in }H.
\end{equation}
Evidently, such displacements do not lead to any classical signal.
It can be seen that such a group $H$ of ``zero classical signal''
is generated by all the Hamiltonians composed of the terms $\widehat{c}_{\alpha}^{\textrm{p}\dagger}\widehat{c}_{\alpha^{\prime}}^{\textrm{p}}$
and $\widehat{c}_{\alpha}^{\textrm{h}\dagger}\widehat{c}_{\alpha^{\prime}}^{\textrm{h}}$.
It is known that such a group is isomorphic to arbitrary independent
separate unitary rotations of the hole and of the particle orbitals
$U\left(M_{\textrm{p}}\right)\otimes U\left(M_{\textrm{h}}\right)$
\citep{Zhang1990}. The dispacements outside $H$ lead to non-zero
classical signal. However, they are in a many-to-one correspondence:
if $\exp\left(-i\widehat{k}\right)\left|\textrm{vac}\right\rangle $
has non-zero classical signal $\boldsymbol{\xi}$, then any other
state of the form $\exp\left(-i\widehat{k}\right)\exp\left(-i\widehat{h}\right)\left|\textrm{vac}\right\rangle $
corresponds to the same signal $\boldsymbol{\xi}$, provided $\exp\left(-i\widehat{h}\right)$
belongs to ``zero-signal'' subgroup $H$. We obtain an equivalence
relation in the group $G$: displacements are equivalent if they lead
to the same classical signal. In other words, the displacements are
equivalent if they differ by a factor $\exp\left(-i\widehat{h}\right)$
on the right. Such a construction is called the coset space and denoted
as
\begin{equation}
\Gamma=U\left(M_{\textrm{p}}+M_{\textrm{h}}\right)/U\left(M_{\textrm{p}}\right)\otimes U\left(M_{\textrm{h}}\right).
\end{equation}
It is known that $\Gamma$ can be interpreted as a curved space whose
points are labeled by the signals $\boldsymbol{\xi}$, which actually
play the role of a coordinate system on $\Gamma$. Expcicitly, each
point $\boldsymbol{\xi}$ in $\Gamma$ corresponds to a unique transformation
of the vacuum 
\begin{equation}
\left|\boldsymbol{\xi}\right\rangle =\mathcal{N}^{-1/2}\left(\boldsymbol{\xi},\boldsymbol{\xi}^{\dagger}\right)\exp\left(\widehat{\boldsymbol{c}}^{\textrm{h}\dagger}\boldsymbol{\xi}^{\dagger}\widehat{\boldsymbol{c}}^{\textrm{p}\dagger}\right)\left|\textrm{vac}\right\rangle ,
\end{equation}
where the classial signal $\boldsymbol{\xi}^{\dagger}$ is found to
be a $M_{\textrm{h}}\times M_{\textrm{p}}$complex matrix; $\mathcal{N}\left(\boldsymbol{\xi},\boldsymbol{\xi}^{\dagger}\right)$
is the normalization factor of the state
\begin{equation}
\left\Vert \boldsymbol{\xi}\right\rangle =\exp\left(\widehat{\boldsymbol{c}}^{\textrm{h}\dagger}\boldsymbol{\xi}^{\dagger}\widehat{\boldsymbol{c}}^{\textrm{p}\dagger}\right)\left|\textrm{vac}\right\rangle ,
\end{equation}
namely 
\begin{equation}
\mathcal{N}\left(\boldsymbol{\xi},\boldsymbol{\xi}^{\dagger}\right)=\left\langle \boldsymbol{\xi}\left\Vert \boldsymbol{\xi}\right.\right\rangle =\det\left(\boldsymbol{I}_{\textrm{h}}+\boldsymbol{\xi}^{\dagger}\boldsymbol{\xi}\right)=\det\left(\boldsymbol{I}_{\textrm{p}}+\boldsymbol{\xi}\boldsymbol{\xi}^{\dagger}\right),
\end{equation}
where $\boldsymbol{I}_{\textrm{h}}$ is the $M_{\textrm{h}}\times M_{\textrm{h}}$
identity matrix, and $\boldsymbol{I}_{\textrm{p}}$ is the $M_{\textrm{p}}\times M_{\textrm{p}}$
identity matrix. 

Surprisingly enough for anyone who is not familiar with the theory
of general coherent states, the logarithm of this normalization factor,
\begin{equation}
K\left(\boldsymbol{\xi},\boldsymbol{\xi}^{\dagger}\right)=\ln\mathcal{N}\left(\boldsymbol{\xi},\boldsymbol{\xi}^{\dagger}\right)=\ln\det\left(\boldsymbol{I}_{\textrm{h}}+\boldsymbol{\xi}^{\dagger}\boldsymbol{\xi}\right)=\ln\det\left(\boldsymbol{I}_{\textrm{p}}+\boldsymbol{\xi}\boldsymbol{\xi}^{\dagger}\right),
\end{equation}
also called the K�hler potential, completely determines the geometry
of the space $\Gamma$. In particular, the complex metric tensor of
the space $\Gamma$ is 
\begin{equation}
g_{\alpha\alpha^{\prime},\overline{\beta\beta^{\prime}}}=\frac{\partial^{2}}{\partial\boldsymbol{\xi}_{\alpha\alpha^{\prime}}\partial\boldsymbol{\xi}_{\beta\beta^{\prime}}^{*}}K\left(\boldsymbol{\xi},\boldsymbol{\xi}^{\dagger}\right)=\left(\boldsymbol{I}_{\textrm{p}}+\boldsymbol{\xi}^{*}\boldsymbol{\xi}^{T}\right)_{\alpha\beta}^{-1}\left(\boldsymbol{I}_{\textrm{h}}+\boldsymbol{\xi}^{\dagger}\boldsymbol{\xi}\right)_{\alpha^{\prime}\beta^{\prime}}^{-1},\label{eq:metric_from_Kahler}
\end{equation}
which is important because it defines the volume element for the probability
distributions on $\Gamma$:
\begin{equation}
d\mu\left(\boldsymbol{\xi},\boldsymbol{\xi}^{\dagger}\right)=\det\left[g_{\alpha\alpha^{\prime},\overline{\beta\beta^{\prime}}}\right]\prod_{\gamma\gamma^{\prime}}d\xi_{\gamma\gamma^{\prime}}d\xi_{\gamma\gamma^{\prime}}^{*},
\end{equation}
where 
\begin{equation}
\det\left[g_{\alpha\alpha^{\prime},\overline{\beta\beta^{\prime}}}\right]=\textrm{det}\left(\boldsymbol{I}_{\textrm{h}}+\boldsymbol{\xi}^{\dagger}\boldsymbol{\xi}\right)^{-M}=\det\left(\boldsymbol{I}_{\textrm{p}}+\boldsymbol{\xi}\boldsymbol{\xi}^{\dagger}\right)^{-M}.
\end{equation}
See Appendix \ref{sec:Derivation-of-the} for the calculations. 

Analogously to the case of the bosonic coherent states, the fermionic
particle-hole states also possess the resolution of identity property:
\begin{equation}
V^{-1}\int\left|\boldsymbol{\xi}\right\rangle \left\langle \boldsymbol{\xi}\right|d\mu\left(\boldsymbol{\xi}\right)=\widehat{1}_{\textrm{bath}},
\end{equation}

where $\widehat{1}_{\textrm{bath}}$ is the identity operator in the
number-conserving (i.e. with equal number of particles and holes)
Hilbert subspace of the discretized fermionic bath; $V$ is irrelevant
normalization constant.

\section{\label{sec:CLASSICAL-STATISTICS-OF}CLASSICAL STATISTICS OF OBSERVABLE
QUANTUM FLUCTUATIONS: THE HUSIMI FUNCTION}

\subsection{Husimi function}

Now we know what are the states $\left|\boldsymbol{\xi}\right\rangle $
of the fermi bath which bear a classical signal $\boldsymbol{\xi}$.
Our next step is to answer the question: given a joint OQS-bath wavefunction
$\left|\Psi\left(t\right)\right\rangle $ from the number-conserving
Hilbert space, what is the probability of observing a particular instance
of the signal $\boldsymbol{\xi}$? This is provided by the \textit{particle-hole}
Husimi function 
\begin{equation}
Q\left(\boldsymbol{\xi},\boldsymbol{\xi}^{\dagger}\right)=\det\left[g_{\alpha\alpha^{\prime},\overline{\beta\beta^{\prime}}}\right]\textrm{Tr}_{\textrm{s}}\left\{ \left|\boldsymbol{\xi}\right\rangle \left\langle \boldsymbol{\xi}\right|\otimes\widehat{1}_{s}\times\left|\Psi\left(t\right)\right\rangle \left\langle \Psi\left(t\right)\right|\right\} ,\label{eq:noninvariant_Husimi_function}
\end{equation}
so that the probability to observe a signal instance in an infinitesimal
volume $dV=\prod_{\gamma\gamma^{\prime}}d\xi_{\gamma\gamma^{\prime}}d\xi_{\gamma\gamma^{\prime}}^{*}$
centered at $\boldsymbol{\xi}$ is $Q\left(\boldsymbol{\xi},\boldsymbol{\xi}^{\dagger}\right)dV$. 

\subsection{Master equation for Husimi function}

In order to simulate the time evolution of the observable quantum
field statistics, we need to find the master equation for the Husimi
function (\ref{eq:noninvariant_Husimi_function}). We do this by differentiating
$Q\left(\boldsymbol{\xi},\boldsymbol{\xi}^{\dagger}\right)$ with
respect to time:
\begin{multline}
\partial_{t}Q\left(\boldsymbol{\xi},\boldsymbol{\xi}^{\dagger};t\right)\\
=\det g_{\alpha,\overline{\beta}}\textrm{Tr}\left\{ \left|\boldsymbol{\xi}\right\rangle \left\langle \boldsymbol{\xi}\right|\otimes\widehat{1}_{s}\times\left(-i\left[\widehat{H}_{\textrm{s}}+\widehat{H}_{\textrm{int}}\left(t\right),\left|\Psi\left(t\right)\right\rangle \left\langle \Psi\left(t\right)\right|\right]\right)\right\} \\
=\textrm{Tr}\left\{ +i\left[\widehat{H}_{\textrm{int}}\left(t\right),\det g_{\alpha,\overline{\beta}}\left|\boldsymbol{\xi}\right\rangle \left\langle \boldsymbol{\xi}\right|\otimes\widehat{1}_{s}\right]\times\left|\Psi\left(t\right)\right\rangle \left\langle \Psi\left(t\right)\right|\right\} ,\label{eq:Husimi_derivative}
\end{multline}
where we have employed the cyclic trace property
\begin{equation}
\textrm{Tr}\left\{ A\left[B,C\right]\right\} =\textrm{Tr}\left\{ \left[A,B\right]C\right\} .
\end{equation}
Before we continue the calcualtions in (\ref{eq:Husimi_derivative}),
let us discuss what we want to achieve. We need to obtain a causal
master equation which can be interpreted as a one-time probability
distribution of a certain stochastic process. Otherwise we can neither
interpret the non-Markovian quantum dissipative evolution classically
nor to perform Monte Carlo calculations of observables. We have tree
types of probabilistic master equations: (i) convection (drift), (ii)
diffusion, and (iii) jumps. The terms of types (ii) and (iii) are
not compatible with the time-reversal symmetry of the Schrodinger
equation. Therefore, the desired form is
\begin{equation}
\partial_{t}Q\left(\boldsymbol{\xi},\boldsymbol{\xi}^{\dagger};t\right)=\partial_{\xi_{\alpha\beta}}\left\{ i\mathcal{A}_{\alpha\beta}\left(\boldsymbol{\xi},\boldsymbol{\xi}^{\dagger};t\right)Q\left(\boldsymbol{\xi},\boldsymbol{\xi}^{\dagger};t\right)\right\} +\partial_{\xi_{\alpha\beta}^{*}}\left\{ \left(-i\right)\mathcal{A}_{\alpha\beta}^{*}\left(\boldsymbol{\xi},\boldsymbol{\xi}^{\dagger};t\right)Q\left(\boldsymbol{\xi},\boldsymbol{\xi}^{\dagger};t\right)\right\} ,\label{eq:Husimi_master_equation}
\end{equation}
where $\mathcal{A}_{\alpha\beta}\left(\boldsymbol{\xi},\boldsymbol{\xi}^{\dagger};t\right)$
is a convection velocity field on $\Gamma$. In order to arrive at
this convection form, we employ the fact that the commutators of the
Hamiltonian terms with the projections

\begin{equation}
\widehat{p}\left(\boldsymbol{\xi},\boldsymbol{\xi}^{\dagger}\right)=\det g_{\alpha,\overline{\beta}}\left|\boldsymbol{\xi}\right\rangle \left\langle \boldsymbol{\xi}\right|
\end{equation}
can be represented as a differential operators. In Appendix \ref{sec:Differential-correspondences-for}
it is shown that
\begin{equation}
i\left[\widehat{H}_{\textrm{int}}\left(t\right),\widehat{p}\left(\boldsymbol{\xi},\boldsymbol{\xi}^{\dagger}\right)\right]=i\partial_{\xi_{\alpha\beta}}\left\{ \widehat{\mathcal{A}}_{\alpha\beta}\left(\boldsymbol{\xi};t\right)\widehat{p}\left(\boldsymbol{\xi},\boldsymbol{\xi}^{\dagger}\right)\right\} -i\partial_{\xi_{\alpha\beta}^{*}}\left\{ \mathcal{\widehat{A}}_{\alpha\beta}^{\dagger}\left(\boldsymbol{\xi};t\right)\widehat{p}\left(\boldsymbol{\xi},\boldsymbol{\xi}^{\dagger}\right)\right\} ,\label{eq:differential_correspondence}
\end{equation}
where $\widehat{\mathcal{A}}_{\alpha\beta}$ is an operator in an
impurity Hilbert space, whose explicit form is 
\begin{equation}
\widehat{\mathcal{A}}_{\alpha\beta}\left(\boldsymbol{\xi};t\right)=-\widehat{h}_{\alpha\beta}^{\textrm{ph}}\left(t\right)+\xi_{\gamma\beta}\widehat{h}_{\gamma\alpha}^{\textrm{pp}}\left(t\right)-\xi_{\alpha\gamma}\widehat{h}_{\gamma\beta}^{\textrm{hh}}\left(t\right)+\xi_{\gamma\beta}\xi_{\alpha\delta}\widehat{h}_{\gamma\delta}^{\textrm{ph}}\left(t\right).
\end{equation}
We substitute (\ref{eq:differential_correspondence}) into (\ref{eq:Husimi_derivative})
and find:
\begin{equation}
\partial_{t}Q\left(\boldsymbol{\xi},\boldsymbol{\xi}^{\dagger};t\right)=i\partial_{\xi_{\alpha\beta}}\left\langle \mathcal{\widehat{A}}_{\alpha\beta}\right\rangle _{Q}^{\textrm{u}}\left(\boldsymbol{\xi},\boldsymbol{\xi}^{\dagger};t\right)-i\partial_{\xi_{\alpha\beta}^{*}}\left\langle \mathcal{\widehat{A}}_{\alpha\beta}\right\rangle _{Q}^{\textrm{u}*}\left(\boldsymbol{\xi},\boldsymbol{\xi}^{\dagger};t\right),
\end{equation}
where the unnormalized Husimi average $\left\langle \mathcal{\widehat{A}}_{\alpha\beta}\right\rangle _{Q}^{\textrm{u}}$
of operator $\widehat{\mathcal{A}}_{\alpha\beta}$ is
\begin{equation}
\mathcal{A}_{\alpha\beta}^{\textrm{u}}\left(\boldsymbol{\xi},\boldsymbol{\xi}^{\dagger};t\right)=\textrm{Tr}\left\{ \widehat{p}\left(\boldsymbol{\xi},\boldsymbol{\xi}^{\dagger}\right)\widehat{\mathcal{A}}_{\alpha\beta}\left(\boldsymbol{\xi};t\right)\left|\Psi\left(t\right)\right\rangle \left\langle \Psi\left(t\right)\right|\right\} .
\end{equation}
By introducing the normalized Husimi averages
\begin{equation}
\left\langle \widehat{o}\right\rangle _{Q}\left(\boldsymbol{\xi},\boldsymbol{\xi}^{\dagger};t\right)=\frac{\left\langle \textrm{Tr}\left\{ \widehat{p}\left(\boldsymbol{\xi},\boldsymbol{\xi}^{\dagger}\right)\widehat{o}\left(t\right)\left|\Psi\left(t\right)\right\rangle \left\langle \Psi\left(t\right)\right|\right\} \right\rangle }{\textrm{Tr}\left\{ \widehat{p}\left(\boldsymbol{\xi},\boldsymbol{\xi}^{\dagger}\right)\left|\Psi\left(t\right)\right\rangle \left\langle \Psi\left(t\right)\right|\right\} },
\end{equation}
we obtain the desired result (\ref{eq:Husimi_master_equation}), where
the convection is
\begin{multline}
\mathcal{A}_{\alpha\beta}\left(\boldsymbol{\xi},\boldsymbol{\xi}^{\dagger};t\right)=-\left\langle \widehat{h}_{\alpha\beta}^{\textrm{ph}}\right\rangle _{Q}\left(\boldsymbol{\xi},\boldsymbol{\xi}^{\dagger};t\right)+\xi_{\gamma\beta}\left\langle \widehat{h}_{\gamma\alpha}^{\textrm{pp}}\right\rangle _{Q}\left(\boldsymbol{\xi},\boldsymbol{\xi}^{\dagger};t\right)\\
-\xi_{\alpha\gamma}\left\langle \widehat{h}_{\gamma\beta}^{\textrm{hh}}\right\rangle _{Q}\left(\boldsymbol{\xi},\boldsymbol{\xi}^{\dagger};t\right)+\xi_{\gamma\beta}\xi_{\alpha\delta}\left\langle \widehat{h}_{\gamma\delta}^{\textrm{ph}}\right\rangle _{Q}\left(\boldsymbol{\xi},\boldsymbol{\xi}^{\dagger};t\right).\label{eq:resulting_convection}
\end{multline}

\subsection{Stochastic interpretation of the Husimi master equation}

The master equation (\ref{eq:Husimi_master_equation}) has the form
of the convection equation in the space $\boldsymbol{\xi}$ with velocity
field $\mathcal{A}_{\alpha\beta}\left(\boldsymbol{\xi},\boldsymbol{\xi}^{\dagger};t\right)$.
That means we can simulate the evolution of the observable field by
sampling stochastically the initial conditions $\boldsymbol{\xi}\left(0\right)$
from the probability distribution
\begin{equation}
P\left(\boldsymbol{\xi}\left(0\right)\right)\propto\det g_{\alpha,\overline{\beta}}\mathcal{N}^{-1}\left(\boldsymbol{\xi}\left(0\right),\boldsymbol{\xi}^{\dagger}\left(0\right)\right)=\textrm{det}\left(\boldsymbol{I}_{\textrm{h}}+\boldsymbol{\xi}^{\dagger}\left(0\right)\boldsymbol{\xi}\left(0\right)\right)^{-M-1}=\textrm{det}\left(\boldsymbol{I}_{\textrm{p}}+\boldsymbol{\xi}\left(0\right)\boldsymbol{\xi}^{\dagger}\left(0\right)\right)^{-M-1}.\label{eq:probability_distribution_for_vacuum_fluctuations}
\end{equation}
Then, each $\boldsymbol{\xi}$ is propagated in time according to
\begin{equation}
\dot{\xi}_{\alpha\beta}\left(t\right)=i\mathcal{A}_{\alpha\beta}\left(\boldsymbol{\xi},\boldsymbol{\xi}^{*};t\right).\label{eq:Husimi_drift}
\end{equation}
The average over the Husimi function of an arbitrary function $O\left(\boldsymbol{\xi},\boldsymbol{\xi}^{\dagger}\right)$
at a time moment $t$ is provided by the stochastic sampling:
\begin{equation}
\overline{O}=\int\prod_{\gamma\gamma^{\prime}}d\xi_{\gamma\gamma^{\prime}}d\xi_{\gamma\gamma^{\prime}}^{*}Q\left(\boldsymbol{\xi},\boldsymbol{\xi}^{\dagger};t\right)O\left(\boldsymbol{\xi},\boldsymbol{\xi}^{\dagger}\right)=\overline{\left\{ O\left(\boldsymbol{\xi}\left(t\right),\boldsymbol{\xi}^{\dagger}\left(t\right)\right)\right\} }_{\boldsymbol{\xi}\left(0\right)},
\end{equation}
where $\boldsymbol{\xi}\left(t\right)$ is a solution of (\ref{eq:Husimi_drift}). 

Physical interpretation of this result is that there are vacuum fluctuations
of the environment which manifest themselves as a fluctuating classical
partile-hole noise $\xi_{\alpha\beta}\left(0\right)$. Each noise
instance is propagated in time classically due to interaction with
the QOS through its averaged observables (\ref{eq:resulting_convection}).

\section{\label{sec:DRESSED-QUANTUM-TRAJECTORIES:}DRESSED QUANTUM TRAJECTORIES:
VIRTUAL FIELD}

\subsection{Dressed quantum state}

In order to solve the drift equation (\ref{eq:Husimi_drift}), we
need to know the projections
\begin{equation}
\left|\Psi\left(\boldsymbol{\xi};t\right)\right\rangle =\left\langle \boldsymbol{\xi}\left\Vert \Psi\left(t\right)\right.\right\rangle .
\end{equation}
Therefore, the Husimi master equation is not a closed set of equations.
Some additional (\textit{virtual}) degrees of freedom are coupled
to the observable field statistics. Using the definition of coherent
states, we write 
\begin{equation}
\left\langle \boldsymbol{\xi}\left\Vert \Psi\left(t\right)\right.\right\rangle =\left\langle \textrm{vac}\left|\Psi_{\textrm{dress}}\left(\boldsymbol{\xi};t\right)\right.\right\rangle ,\label{eq:vacuum_projection}
\end{equation}
where the dressed state in introduced as
\begin{equation}
\left|\Psi_{\textrm{dress}}\left(\boldsymbol{\xi};t\right)\right\rangle =\exp\left(\widehat{\boldsymbol{c}}^{\textrm{p}}\boldsymbol{\xi}\widehat{\boldsymbol{c}}^{\textrm{h}}\right)\left|\Psi\left(t\right)\right\rangle .
\end{equation}
Meaning of the last two equations is the following. To reach the time
moment $t$, we evolve the joint wavefunction up to this time moment.
Then, sitting in the phase space $\Gamma$ at the measurement outcome
(classical signal) $\boldsymbol{\xi}$, we ``center'' the wavefunction
at this outcome by applying the operation $\exp\left(\widehat{\boldsymbol{c}}^{\textrm{p}}\boldsymbol{\xi}\widehat{\boldsymbol{c}}^{\textrm{h}}\right)$.
All the quanta which ``stick out'' from this center (i.e. incompatible
with the outcome $\boldsymbol{\xi}$) are discarded by the vacuum
projection in Eq. (\ref{eq:vacuum_projection}). Such quanta do not
contribute to the measurement: if they were emitted, the only way
for them to contribute to the dynamics is to be reabsorbed by the
open system. These quanta are called ``virtual quanta'', and the
wavefunction centered at the phase space outcome $\boldsymbol{\xi}$
is called the dressed quantum state corresponding to outcome $\boldsymbol{\xi}$
(an open system state dressed by unobservable quanta (incompatible
with the outcome $\boldsymbol{\xi}$)). 

So, the Husimi master equations are coupled to the quantum dynamics
of virtual quanta. The Husimi master equations must be supplemented
by the equations for $\left|\Psi_{\textrm{dress}}\left(\boldsymbol{\xi};t\right)\right\rangle $.
Let us find the latter:
\begin{multline}
\partial_{t}\left|\Psi_{\textrm{dress}}\left(\boldsymbol{\xi};t\right)\right\rangle =\exp\left(\widehat{\boldsymbol{c}}^{\textrm{p}}\boldsymbol{\xi}\widehat{\boldsymbol{c}}^{\textrm{h}}\right)\widehat{H}\left(t\right)\left|\Psi\left(t\right)\right\rangle \\
=\exp\left(\widehat{\boldsymbol{c}}^{\textrm{p}}\boldsymbol{\xi}\widehat{\boldsymbol{c}}^{\textrm{h}}\right)\widehat{H}\left(t\right)\exp\left(-\widehat{\boldsymbol{c}}^{\textrm{p}}\boldsymbol{\xi}\widehat{\boldsymbol{c}}^{\textrm{h}}\right)\exp\left(\widehat{\boldsymbol{c}}^{\textrm{p}}\boldsymbol{\xi}\widehat{\boldsymbol{c}}^{\textrm{h}}\right)\left|\Psi\left(t\right)\right\rangle .
\end{multline}
Denoting the \textit{dressed Hamiltonian} as
\begin{equation}
\widehat{H}_{\textrm{dress}}\left(\boldsymbol{\xi};t\right)=\exp\left(\widehat{\boldsymbol{c}}^{\textrm{p}}\boldsymbol{\xi}\widehat{\boldsymbol{c}}^{\textrm{h}}\right)\widehat{H}\left(t\right)\exp\left(-\widehat{\boldsymbol{c}}^{\textrm{p}}\boldsymbol{\xi}\widehat{\boldsymbol{c}}^{\textrm{h}}\right),
\end{equation}
we find 
\begin{equation}
\partial_{t}\left|\Psi_{\textrm{dress}}\left(\boldsymbol{\xi};t\right)\right\rangle =-i\widehat{H}_{\textrm{dress}}\left(\boldsymbol{\xi};t\right)\left|\Psi_{\textrm{dress}}\left(\boldsymbol{\xi};t\right)\right\rangle .
\end{equation}
In appendix \ref{sec:CALCULATION-OF-THE} it is shown that
\begin{multline}
\widehat{H}_{\textrm{dress}}\left(\boldsymbol{\xi};t\right)=\left\{ \widehat{h}_{k\sigma k^{\prime}\sigma^{\prime}}^{\textrm{pp}}+\widehat{h}_{k\sigma\gamma}^{\textrm{ph}}\xi_{k^{\prime}\sigma^{\prime},\gamma}e^{it\left(\varepsilon_{k^{\prime}}^{\textrm{p}}-\varepsilon_{\gamma}^{\textrm{h}}\right)}\right\} \widehat{c}_{k\sigma}^{\textrm{p}\dagger}\left(t\right)\widehat{c}_{k^{\prime}\sigma^{\prime}}^{\textrm{p}}\left(t\right)\\
+\left\{ \widehat{h}_{\gamma k\sigma}^{\textrm{ph}}\xi_{\gamma,k^{\prime}\sigma^{\prime}}e^{it\left(\varepsilon_{\gamma}^{\textrm{p}}-\varepsilon_{k^{\prime}}^{\textrm{h}}\right)}-\widehat{h}_{k\sigma k^{\prime}\sigma^{\prime}}^{\textrm{hh}}\right\} \widehat{c}_{k\sigma}^{\textrm{h}\dagger}\left(t\right)\widehat{c}_{k^{\prime}\sigma^{\prime}}^{\textrm{h}}\left(t\right)\\
+\left\{ \widehat{h}_{k\sigma k^{\prime}\sigma^{\prime}}^{\textrm{ph}\dagger}-\widehat{h}_{\gamma k^{\prime}\sigma^{\prime}}^{\textrm{pp}}\xi_{\gamma,k\sigma}e^{it\left(\varepsilon_{\gamma}^{\textrm{p}}-\varepsilon_{k}^{\textrm{h}}\right)}+\widehat{h}_{\gamma k\sigma}^{\textrm{hh}}\xi_{k^{\prime}\sigma^{\prime},\gamma}+\widehat{h}_{\gamma\delta}^{\textrm{ph}}\xi_{\gamma,k\sigma}e^{it\left(\varepsilon_{\gamma}^{\textrm{p}}-\varepsilon_{k}^{\textrm{h}}\right)}\xi_{k^{\prime}\sigma^{\prime},\delta}e^{it\left(\varepsilon_{k^{\prime}}^{\textrm{p}}-\varepsilon_{\delta}^{\textrm{h}}\right)}\right\} \widehat{c}_{k\sigma}^{\textrm{h}}\left(t\right)\widehat{c}_{k^{\prime}\sigma^{\prime}}^{\textrm{p}}\left(t\right)\\
+\widehat{h}_{k\sigma k^{\prime}\sigma^{\prime}}^{\textrm{ph}}\left\{ \widehat{c}_{k\sigma}^{\textrm{p}\dagger}\left(t\right)\widehat{c}_{k^{\prime}\sigma^{\prime}}^{\textrm{h}\dagger}\left(t\right)-\xi_{k\sigma,k^{\prime}\sigma^{\prime}}e^{i\tau\left(\varepsilon_{k}^{\textrm{p}}-\varepsilon_{k^{\prime}}^{\textrm{h}}\right)}\right\} .
\end{multline}

\subsection{Dressed quantum trajectory}

Since $\boldsymbol{\xi}\left(t\right)$ is evolving according to the
drift equation (\ref{eq:Husimi_drift}), the total time derivative
of the dressed wavefunction is 
\begin{multline}
\partial_{t}\left|\Psi_{\textrm{dress}}\left(\boldsymbol{\xi}\left(t\right);t\right)\right\rangle =\\
-i\widehat{H}_{\textrm{dress}}\left(\boldsymbol{\xi}\left(t\right);t\right)\left|\Psi_{\textrm{dress}}\left(\boldsymbol{\xi}\left(t\right);t\right)\right\rangle +\dot{\xi}_{\alpha\beta}\left(t\right)\frac{\partial}{\partial\xi_{\alpha\beta}\left(t\right)}\left|\Psi_{\textrm{dress}}\left(\boldsymbol{\xi}\left(t\right);t\right)\right\rangle \\
=-i\left\{ \widehat{H}_{\textrm{dress}}\left(\boldsymbol{\xi}\left(t\right);t\right)-\mathcal{A}_{\alpha\beta}\left(\boldsymbol{\xi},\boldsymbol{\xi}^{*};t\right)\frac{\partial}{\partial\xi_{\alpha\beta}\left(t\right)}\right\} \left|\Psi_{\textrm{dress}}\left(\boldsymbol{\xi}\left(t\right);t\right)\right\rangle \label{eq:self-consistent_dressed_Schrodinger}
\end{multline}
Evaluating the derivative term in the last line 
\begin{equation}
\frac{\partial}{\partial\xi_{\alpha\beta}\left(t\right)}\left|\Psi_{\textrm{dress}}\left(\boldsymbol{\xi}\left(t\right);t\right)\right\rangle =\frac{\partial}{\partial\xi_{\alpha\beta}}\exp\left(\widehat{\boldsymbol{c}}^{\textrm{p}}\boldsymbol{\xi}\widehat{\boldsymbol{c}}^{\textrm{h}}\right)\left|\Psi\left(t\right)\right\rangle =\widehat{c}_{\alpha}^{\textrm{p}}\widehat{c}_{\beta}^{\textrm{h}}\left|\Psi_{\textrm{dress}}\left(\boldsymbol{\xi};t\right)\right\rangle ,
\end{equation}
we present the equation (\ref{eq:self-consistent_dressed_Schrodinger})
in the form 
\begin{equation}
\partial_{t}\left|\Psi_{\textrm{dress}}\left(\boldsymbol{\xi}\left(t\right);t\right)\right\rangle =-i\widehat{H}_{\textrm{dress}}^{\prime}\left(\boldsymbol{\xi}\left(t\right);t\right)\left|\Psi_{\textrm{dress}}\left(\boldsymbol{\xi}\left(t\right);t\right)\right\rangle ,
\end{equation}
where the self-consistent dressed Hamiltonian $\widehat{H}_{\textrm{dress}}^{\prime}\left(\boldsymbol{\xi}\left(t\right);t\right)$
is
\begin{multline}
\widehat{H}_{\textrm{dress}}^{\prime}\left(\boldsymbol{\xi}\left(t\right);t\right)=\left\{ \widehat{h}_{k\sigma k^{\prime}\sigma^{\prime}}^{\textrm{pp}}+\widehat{h}_{k\sigma\gamma}^{\textrm{ph}}\xi_{k^{\prime}\sigma^{\prime},\gamma}\left(t\right)e^{it\left(\varepsilon_{k^{\prime}}^{\textrm{p}}-\varepsilon_{\gamma}^{\textrm{h}}\right)}\right\} \widehat{c}_{k\sigma}^{\textrm{p}\dagger}\left(t\right)\widehat{c}_{k^{\prime}\sigma^{\prime}}^{\textrm{p}}\left(t\right)\\
+\left\{ \widehat{h}_{\gamma k\sigma}^{\textrm{ph}}\xi_{\gamma,k^{\prime}\sigma^{\prime}}\left(t\right)e^{it\left(\varepsilon_{\gamma}^{\textrm{p}}-\varepsilon_{k^{\prime}}^{\textrm{h}}\right)}-\widehat{h}_{k\sigma k^{\prime}\sigma^{\prime}}^{\textrm{hh}}\right\} \widehat{c}_{k\sigma}^{\textrm{h}\dagger}\left(t\right)\widehat{c}_{k^{\prime}\sigma^{\prime}}^{\textrm{h}}\left(t\right)\\
+\left\{ \widehat{h}_{k\sigma k^{\prime}\sigma^{\prime}}^{\textrm{ph}\dagger}-\widehat{h}_{\gamma k^{\prime}\sigma^{\prime}}^{\textrm{pp}}\xi_{\gamma,k\sigma}\left(t\right)e^{it\left(\varepsilon_{\gamma}^{\textrm{p}}-\varepsilon_{k}^{\textrm{h}}\right)}+\widehat{h}_{\gamma k\sigma}^{\textrm{hh}}\xi_{k^{\prime}\sigma^{\prime},\gamma}\right.\\
\left.+\widehat{h}_{\gamma\delta}^{\textrm{ph}}\xi_{\gamma,k\sigma}\left(t\right)e^{it\left(\varepsilon_{\gamma}^{\textrm{p}}-\varepsilon_{k}^{\textrm{h}}\right)}\xi_{k^{\prime}\sigma^{\prime},\delta}\left(t\right)e^{it\left(\varepsilon_{k^{\prime}}^{\textrm{p}}-\varepsilon_{\delta}^{\textrm{h}}\right)}+\mathcal{A}_{k^{\prime}\sigma^{\prime}k\sigma}\left(\boldsymbol{\xi},\boldsymbol{\xi}^{\dagger};t\right)\right\} \widehat{c}_{k\sigma}^{\textrm{h}}\left(t\right)\widehat{c}_{k^{\prime}\sigma^{\prime}}^{\textrm{p}}\left(t\right)\\
+\widehat{h}_{k\sigma k^{\prime}\sigma^{\prime}}^{\textrm{ph}}\left\{ \widehat{c}_{k\sigma}^{\textrm{p}\dagger}\left(t\right)\widehat{c}_{k^{\prime}\sigma^{\prime}}^{\textrm{h}\dagger}\left(t\right)-\xi_{k\sigma,k^{\prime}\sigma^{\prime}}\left(t\right)e^{it\left(\varepsilon_{k}^{\textrm{p}}-\varepsilon_{k^{\prime}}^{\textrm{h}}\right)}\right\} .\label{eq:self-consistent_dressed_Hamiltonian}
\end{multline}
Here we use the notation: if $\delta=\left(k\sigma\right)$, then
$\varepsilon_{\delta}^{\textrm{h}}\coloneqq\varepsilon_{k}^{\textrm{h}}$.
We call the self-consistent evolution of the pair$\left|\Psi_{\textrm{dress}}\left(\boldsymbol{\xi}\left(t\right);t\right)\right\rangle $
and $\boldsymbol{\xi}\left(t\right)$ the \textit{Husimi dressed quantum
trajectory}. 

\section{\label{sec:NUMERICAL-SIMULATION-FOR}NUMERICAL SIMULATION FOR A FINITE
BATH DISCRETIZATION}

We generate random $\boldsymbol{\xi}\left(0\right)$ from the distribution
(\ref{eq:probability_distribution_for_vacuum_fluctuations}), which
we repreat here:
\begin{equation}
P\left(\xi_{\alpha\beta}\left(0\right)\right)\propto\textrm{det}\left(\boldsymbol{I}_{\textrm{h}}+\boldsymbol{\xi}^{\dagger}\left(0\right)\boldsymbol{\xi}\left(0\right)\right)^{-M-1}=\textrm{det}\left(\boldsymbol{I}_{\textrm{p}}+\boldsymbol{\xi}\left(0\right)\boldsymbol{\xi}^{\dagger}\left(0\right)\right)^{-M-1}.\label{eq:probability_measure}
\end{equation}
Then, starting from the initial condition 
\begin{equation}
\left|\Psi_{\textrm{dress}}\left(\boldsymbol{\xi}\left(0\right);0\right)\right\rangle =\left|\phi\right\rangle _{\textrm{sys}}\otimes\left|\textrm{vac}\right\rangle ,
\end{equation}
we solve the Schrodinger equation
\begin{equation}
\partial_{t}\left|\Psi_{\textrm{dress}}\left(\boldsymbol{\xi}\left(t\right);t\right)\right\rangle =-i\widehat{H}_{\textrm{dress}}^{\prime}\left(\boldsymbol{\xi}\left(t\right);t\right)\left|\Psi_{\textrm{dress}}\left(\boldsymbol{\xi}\left(t\right);t\right)\right\rangle ,
\end{equation}
where the Hamiltonian $\widehat{H}_{\textrm{dress}}^{\prime}\left(\boldsymbol{\xi}\left(t\right);t\right)$
is eq. (\ref{eq:self-consistent_dressed_Hamiltonian}), which we repeat
here:
\begin{multline}
\widehat{H}_{\textrm{dress}}^{\prime}\left(\boldsymbol{\xi}\left(t\right);t\right)=\left\{ \widehat{h}_{k\sigma k^{\prime}\sigma^{\prime}}^{\textrm{pp}}+\widehat{h}_{k\sigma\gamma}^{\textrm{ph}}\xi_{k^{\prime}\sigma^{\prime},\gamma}e^{it\left(\varepsilon_{k^{\prime}}^{\textrm{p}}-\varepsilon_{\gamma}^{\textrm{h}}\right)}\right\} \widehat{c}_{k\sigma}^{\textrm{p}\dagger}\left(t\right)\widehat{c}_{k^{\prime}\sigma^{\prime}}^{\textrm{p}}\left(t\right)\\
+\left\{ \widehat{h}_{\gamma k\sigma}^{\textrm{ph}}\xi_{\gamma,k^{\prime}\sigma^{\prime}}e^{it\left(\varepsilon_{\gamma}^{\textrm{p}}-\varepsilon_{k^{\prime}}^{\textrm{h}}\right)}-\widehat{h}_{k\sigma k^{\prime}\sigma^{\prime}}^{\textrm{hh}}\right\} \widehat{c}_{k\sigma}^{\textrm{h}\dagger}\left(t\right)\widehat{c}_{k^{\prime}\sigma^{\prime}}^{\textrm{h}}\left(t\right)\\
+\left\{ \widehat{h}_{k\sigma k^{\prime}\sigma^{\prime}}^{\textrm{ph}\dagger}-\widehat{h}_{\gamma k^{\prime}\sigma^{\prime}}^{\textrm{pp}}\xi_{\gamma,k\sigma}e^{it\left(\varepsilon_{\gamma}^{\textrm{p}}-\varepsilon_{k}^{\textrm{h}}\right)}+\widehat{h}_{\gamma k\sigma}^{\textrm{hh}}\xi_{k^{\prime}\sigma^{\prime},\gamma}\right.\\
\left.+\widehat{h}_{\gamma\delta}^{\textrm{ph}}\xi_{\gamma,k\sigma}e^{it\left(\varepsilon_{\gamma}^{\textrm{p}}-\varepsilon_{k}^{\textrm{h}}\right)}\xi_{k^{\prime}\sigma^{\prime},\delta}e^{it\left(\varepsilon_{k^{\prime}}^{\textrm{p}}-\varepsilon_{\delta}^{\textrm{h}}\right)}+\mathcal{A}_{k^{\prime}\sigma^{\prime}k\sigma}\left(\boldsymbol{\xi},\boldsymbol{\xi}^{\dagger};t\right)\right\} \widehat{c}_{k\sigma}^{\textrm{h}}\left(t\right)\widehat{c}_{k^{\prime}\sigma^{\prime}}^{\textrm{p}}\left(t\right)\\
+\widehat{h}_{k\sigma k^{\prime}\sigma^{\prime}}^{\textrm{ph}}\left\{ \widehat{c}_{k\sigma}^{\textrm{p}\dagger}\left(t\right)\widehat{c}_{k^{\prime}\sigma^{\prime}}^{\textrm{h}\dagger}\left(t\right)-\xi_{k\sigma,k^{\prime}\sigma^{\prime}}e^{it\left(\varepsilon_{k}^{\textrm{p}}-\varepsilon_{k^{\prime}}^{\textrm{h}}\right)}\right\} .
\end{multline}
The classical signal $\boldsymbol{\xi}\left(t\right)$ is evolved
in time according to the drift equation (\ref{eq:Husimi_drift}):
\begin{multline}
-i\dot{\xi}_{\alpha\beta}\left(t\right)=-\left\langle \widehat{h}_{\alpha\beta}^{\textrm{ph}}\right\rangle _{Q}\left(\boldsymbol{\xi},\boldsymbol{\xi}^{\dagger};t\right)+\xi_{\gamma\beta}\left\langle \widehat{h}_{\gamma\alpha}^{\textrm{pp}}\right\rangle _{Q}\left(\boldsymbol{\xi},\boldsymbol{\xi}^{\dagger};t\right)\\
-\xi_{\alpha\gamma}\left\langle \widehat{h}_{\gamma\beta}^{\textrm{hh}}\right\rangle _{Q}\left(\boldsymbol{\xi},\boldsymbol{\xi}^{\dagger};t\right)+\xi_{\gamma\beta}\xi_{\alpha\delta}\left\langle \widehat{h}_{\gamma\delta}^{\textrm{ph}}\right\rangle _{Q}\left(\boldsymbol{\xi},\boldsymbol{\xi}^{\dagger};t\right)
\end{multline}
where the normalized averages of the system coupling operators are
defined as
\begin{equation}
\left\langle \widehat{o}\right\rangle _{Q}\left(\boldsymbol{\xi},\boldsymbol{\xi}^{\dagger};t\right)=\frac{\left\langle \left.\Psi_{\textrm{dress}}\left(\boldsymbol{\xi}\left(t\right);t\right)\right|\textrm{vac}\right\rangle \widehat{o}\left\langle \textrm{vac}\left|\Psi_{\textrm{dress}}\left(\boldsymbol{\xi}\left(t\right);t\right)\right.\right\rangle }{\left\Vert \left\langle \textrm{vac}\left|\Psi_{\textrm{dress}}\left(\boldsymbol{\xi}\left(t\right);t\right)\right.\right\rangle \right\Vert ^{2}}.
\end{equation}
The average values $s\left(t\right)$ for the impurity (OQS) observables
$\widehat{s}$ are computed by averaging over many instances of the
signal initial values $\boldsymbol{\xi}_{1}\left(0\right)\ldots\boldsymbol{\xi}_{M}\left(0\right)$:
\begin{equation}
s\left(t\right)=\frac{1}{N}\sum_{i=1}^{N}\left.\left\langle \widehat{s}\right\rangle _{Q}\left(\boldsymbol{\xi}\left(t\right),\boldsymbol{\xi}^{\dagger}\left(t\right);t\right)\right|_{\boldsymbol{\xi}\left(0\right)=\boldsymbol{\xi}_{i}\left(0\right)}.
\end{equation}

\section{\label{sec:DISCRETIZATION-OF-A}DISCRETIZATION OF A CONTINUOUS BATH}

\subsection{Discretization of Hamiltonian}

In this section we remember that our $M$-mode Hamiltonian (\ref{eq:general_OQS_in_fermionic_bath})
is a result of the discretization of an infinite bath with a continuous
specral density. In the simplest case, given a certain mode range
$\left[k_{\textrm{min}},k_{\textrm{max}}\right]$, we discretize it
with a grid of $M$ values $k_{i}$ with a step $\Delta k$. Then,
a continuous-mode creation/annihlation operators $\widehat{c}_{\sigma}\left(k\right)$
are substituted with the discrete ones $\widehat{c}_{k\sigma}$ via
the rule 
\begin{equation}
\sqrt{\Delta k}\widehat{c}_{\sigma}\left(k_{i}\right)\to\widehat{c}_{k_{i}\sigma},
\end{equation}
so that the free bath part is discretized as 
\begin{equation}
\intop_{k\geq k_{F}}dk\varepsilon_{k}\widehat{c}_{\sigma}^{\textrm{p}\dagger}\left(k\right)\widehat{c}_{\sigma}^{\textrm{p}}\left(k\right)\to\sum_{k\geq k_{F}}\varepsilon_{k}\widehat{c}_{k\sigma}^{\textrm{p}\dagger}\widehat{c}_{k\sigma}^{\textrm{p}},
\end{equation}
and the coupling terms are discretized as 
\begin{equation}
\intop_{k,k^{\prime}\geq k_{F}}dkdk^{\prime}\widehat{h}_{\sigma\sigma^{\prime}}^{\textrm{pp}}\left(k,k^{\prime}\right)\widehat{c}_{\sigma}^{\textrm{p}\dagger}\left(k\right)\widehat{c}_{\sigma}^{\textrm{p}}\left(k\right)\to\Delta k\sum_{k,k^{\prime}\geq k_{F}}\widehat{h}_{\sigma\sigma^{\prime}}^{\textrm{pp}}\left(k,k^{\prime}\right)\widehat{c}_{k\sigma}^{\textrm{p}\dagger}\widehat{c}_{k^{\prime}\sigma^{\prime}}^{\textrm{p}}.
\end{equation}
That means that the terms $\widehat{h}_{\gamma\gamma^{\prime}}^{\textrm{pp}}$,
$\widehat{h}_{\gamma\gamma^{\prime}}^{\textrm{hh}}$, $\widehat{h}_{\gamma\gamma^{\prime}}^{\textrm{ph}}$,
and $\widehat{h}_{\gamma\gamma^{\prime}}^{\textrm{ph}}$ in the Hamiltonian
(\ref{eq:general_OQS_in_fermionic_bath}) have the scale of $\Delta k$:
\begin{equation}
\widehat{h}_{\gamma\gamma^{\prime}}^{\textrm{pp}},\ldots\,\,\propto\Delta k\propto M^{-1}k_{\textrm{max}}.
\end{equation}

\subsection{Discretization of the vacuum noise probability distribution}

The continuous probability distribution for the vacuum noise fluctuations
$P\left(\boldsymbol{\xi}\left(0\right)\right)$, eq. (\ref{eq:probability_distribution_for_vacuum_fluctuations}),
can be written as
\begin{multline}
P\left(\boldsymbol{\xi}\left(0\right)\right)\propto\lim_{M\to\infty}\textrm{det}\left(\boldsymbol{I}_{\textrm{h}}+\boldsymbol{\xi}^{\dagger}\left(0\right)\boldsymbol{\xi}\left(0\right)\right)^{-M-1}=\lim_{M\to\infty}e^{-\left(M+1\right)\textrm{Tr}\ln\left(\boldsymbol{I}_{\textrm{h}}+\boldsymbol{\xi}^{\dagger}\left(0\right)\boldsymbol{\xi}\left(0\right)\right)}.\label{eq:as_a_logarithm}
\end{multline}
In order to have a non-vanishing limit, the Frobenius norm of the
matrix $\boldsymbol{\xi}^{\dagger}\left(0\right)\boldsymbol{\xi}\left(0\right)$
should scale with $M$ as 
\begin{equation}
\textrm{Tr}\boldsymbol{\xi}^{\dagger}\left(0\right)\boldsymbol{\xi}\left(0\right)\propto M^{-1},
\end{equation}
which follows from the Taylor expansion of the logarithm in (\ref{eq:as_a_logarithm}):
\begin{equation}
\textrm{Tr}\ln\left(\boldsymbol{I}_{\textrm{h}}+\boldsymbol{\xi}^{\dagger}\left(0\right)\boldsymbol{\xi}\left(0\right)\right)\approx\textrm{Tr}\boldsymbol{\xi}^{\dagger}\left(0\right)\boldsymbol{\xi}\left(0\right)+O\left(M^{-2}\right),
\end{equation}
where $O\left(M^{-2}\right)$ is a consequence of the inequality 
\begin{equation}
\textrm{Tr}\left[\left(\boldsymbol{\xi}^{\dagger}\left(0\right)\boldsymbol{\xi}\left(0\right)\right)^{q}\right]\leq\left(\textrm{Tr}\boldsymbol{\xi}^{\dagger}\left(0\right)\boldsymbol{\xi}\left(0\right)\right)^{q}
\end{equation}
for any integer $q>0$. From this we conjecture that in the continuum
limit, the noise of the fermionic fluctuations becomes Gaussian, with
a probability distribution
\begin{equation}
P\left(\boldsymbol{\xi}\left(0\right)\right)\propto\lim_{M\to\infty}\exp\left(M\textrm{Tr}\boldsymbol{\xi}^{\dagger}\left(0\right)\boldsymbol{\xi}\left(0\right)\right).
\end{equation}
In a numerical Monte Carlo simulation, this amounts to generating
$M_{\textrm{h}}\times M_{\textrm{p}}$ complex independent random
numbers $\xi_{\gamma\gamma^{\prime}}\left(0\right)$ which have a
complex normal distribution with a variance
\begin{equation}
\overline{\xi_{\gamma\gamma^{\prime}}\left(0\right)\xi_{\gamma\gamma^{\prime}}^{*}\left(0\right)}=M^{-1}
\end{equation}

\section{\label{sec:ANDERSON-IMPURITY-PROBLEM}ANDERSON IMPURITY PROBLEM}

\subsection{Model of the Fermionic Open System}

In this section we deal with the second problem mention in the introduction,
namely the description of fermionic OQS in a fermionic bath. As a
typical model we consider the Anderson impurity model (AIM), with
the following Hamiltonian
\begin{equation}
\widehat{H}=\varepsilon_{d}\widehat{d}_{\sigma}^{\dagger}\widehat{d}_{\sigma}+U\widehat{d}_{\uparrow}^{\dagger}\widehat{d}_{\uparrow}\widehat{d}_{\downarrow}^{\dagger}\widehat{d}_{\downarrow}+\varepsilon_{k}\widehat{c}_{k\sigma}^{\dagger}\widehat{c}_{k\sigma}+V_{k}\widehat{d}_{\sigma}^{\dagger}\widehat{c}_{k\sigma}+V_{k}^{*}\widehat{c}_{k\sigma}^{\dagger}\widehat{d}_{\sigma}.
\end{equation}
The index $k$ runs over the modes of the conduction band. The label
$d$ denotes the impurity degrees of freedom. As a free ``bath''
we consider the term
\begin{equation}
\widehat{H}_{\textrm{b}}=\varepsilon_{k}\widehat{c}_{k\sigma}^{\dagger}\widehat{c}_{k\sigma}.\label{eq:AIM_bath}
\end{equation}
As the open system we consider the Hamiltonian 
\begin{equation}
\widehat{H}_{\textrm{s}}=\varepsilon_{d}\widehat{d}_{\sigma}^{\dagger}\widehat{d}_{\sigma}+U\widehat{d}_{\uparrow}^{\dagger}\widehat{d}_{\uparrow}\widehat{d}_{\downarrow}^{\dagger}\widehat{d}_{\downarrow}.\label{eq:AIM_oqs}
\end{equation}
The coupling term is
\begin{equation}
\widehat{H}_{\textrm{int}}=V_{k}\widehat{d}_{\sigma}^{\dagger}\widehat{c}_{k\sigma}+V_{k}^{*}\widehat{c}_{k\sigma}^{\dagger}\widehat{d}_{\sigma}.\label{eq:AIM_int}
\end{equation}

\subsection{The difficulty of the Fermionic Open System}

The Anderson impurity is a more difficult case. The reason is the
non-locality of the fermion vacuum. Indeed, in the bosonic case, when
there is a local degree of freedom $\widehat{b}$, and ``distant''
degrees of freedom $\widehat{f}_{k}$, then if we create locally some
excitation, distant part does not feel it. This is reflected in the
commutation relations: 
\begin{equation}
\left[\widehat{b}^{\dagger},\widehat{f}_{k}\right]=0,\,\,\,\left[\widehat{b}^{\dagger},\widehat{f}_{k}^{\dagger}\right]=0.
\end{equation}
In the same way, a local bosonic degree of freedom is also independent
of the fermionic distant modes. However, the local fermionic mode
$\widehat{c}$ is always entangled to fermionic modes $\widehat{f}$
whatever far they are. This is reflected in the non-zero commutation
relations 
\begin{equation}
\left[\widehat{c}^{\dagger},\widehat{f}_{k}\right]=2\widehat{c}^{\dagger}\widehat{f}_{k},\,\,\,\left[\widehat{c}^{\dagger},\widehat{f}_{k}^{\dagger}\right]=2\widehat{c}^{\dagger}\widehat{f}_{k}^{\dagger}.
\end{equation}
Physically, this means that whereas the bosonic vacuum is dynamically
``divisible'', i.e. we can factorize it as a tensor product of ``subvacuums''
(with the ensuing factorization of their evolution generators), the
fermionic vacuum is an integral entity, we cannot divide it into parts. 

On a formal level, this leads to the impossibility of deriving a closed
probabilistic Husimi master equation for such kind of systems.

\subsection{The Proposed Approach}

Of course, at a first glance this is a bad news for the open system
theory: we cannot break off a piece of fermi vacuum which carries
the open system, so as to devise a reduced description (i.e. to forget
about the remaining degrees of freedom). However, we can learn a lesson
from the previous section: everything dynamically local in physics
is bosonic; everything dynamically global is fermionic. If we want
to obtain a local dynamical description of the open system, we should
do it in the bosonic terms. Therefore, our strategy is not to break
the fermi vacuum. Instead, while keeping its integrity, we factor
it out: let us represent the open system dynamics as a bosonic one
on top of the free fermionic vacuum. Then, the dynamics of the fermion
vacuum can be simulated stochastically by means of the previous section.
And the local open system bosonic evolution will be described by a
truncated basis dressed Schrodinger equation.

After all, there are no fermionic non-linearities in the Nature: it
is enough to look at the Lagrangian of the standard model. Every interaction
between the fermions is mediated by a bosonic force field. Fermionic
nonlinearities may be considered as the artifacts of the way we are
formulating our approximations. In this respect, the bosonization
of the non-linear fermionic open systems may be considered as the
``repair'' of a certain important physical structure which is broken
by low-energy approximate models like AIM. 

Let us realize this proposal. There are numeruous ways {[}{]} to bosonize
the interaction term in AIM: some of them may be more or less suitable
for the stochastic simulation. Here we present one way of doing it
as a prove-of-concept. We reformulate the Anderson impurity Hamiltonian
by introducing the ``global background vacuum'' fermionic degrees
of freedom $\widehat{c}_{d\sigma}$, ($d$ is a label, not index)
and the local bosonic ones $\widehat{b}_{\sigma}$. Explicitly, we
decompose
\begin{equation}
\widehat{d}_{\sigma}=\widehat{c}_{d\sigma}\widehat{b}_{\sigma}.
\end{equation}
Then, we obtain that in the original Anderson impurity model (\ref{eq:AIM_bath})-(\ref{eq:AIM_int}),
the interaction term becomes
\begin{equation}
\widehat{H}_{\textrm{int}}=V_{k}\widehat{b}_{\sigma}^{\dagger}\widehat{c}_{d\sigma}^{\dagger}\widehat{c}_{k\sigma}+V_{k}^{*}\widehat{c}_{k\sigma}^{\dagger}\widehat{c}_{d\sigma}\widehat{b}_{\sigma}
\end{equation}
and the OQS term becomes
\begin{equation}
\widehat{H}_{\textrm{s}}=\varepsilon_{d}\widehat{c}_{d\sigma}^{\dagger}\widehat{c}_{d\sigma}\widehat{b}_{\sigma}^{\dagger}\widehat{b}_{\sigma}+U\widehat{c}_{d\uparrow}^{\dagger}\widehat{c}_{d\uparrow}\widehat{c}_{d\downarrow}^{\dagger}\widehat{c}_{d\downarrow}\widehat{b}_{\uparrow}^{\dagger}\widehat{b}_{\uparrow}\widehat{b}_{\downarrow}^{\dagger}\widehat{b}_{\downarrow}.
\end{equation}
Such a decomposition is valid in the invariant ``physical'' subspace
$L_{\textrm{s}}$ spanned by the states 
\begin{equation}
\left|0\right\rangle _{\downarrow-}\otimes\left|0\right\rangle _{\downarrow+},\,\,\,\left|1\right\rangle _{\downarrow-}\otimes\left|1\right\rangle _{\downarrow+},\,\,\,\left|0\right\rangle _{\uparrow-}\otimes\left|0\right\rangle _{\uparrow+},\,\,\,\left|1\right\rangle _{\uparrow-}\otimes\left|1\right\rangle _{\uparrow+}.
\end{equation}
Here ``-'' means the fermionic species (created by $\widehat{c}_{d\sigma}$),
and ``+'' the bosonic ones (created by $\widehat{b}_{\sigma}$).
In other words, if the number of fermionic species $\sigma$ was equal
to the number of bosonic species $\sigma$ at a certain time, then
it will be so at any later time. Moreover, in this subspace we have
identities: 
\begin{equation}
\widehat{c}_{d\sigma}^{\dagger}\widehat{c}_{d\sigma}\widehat{b}_{\sigma}^{\dagger}\widehat{b}_{\sigma}=\widehat{c}_{d\sigma}^{\dagger}\widehat{c}_{d\sigma}=\widehat{b}_{\sigma}^{\dagger}\widehat{b}_{\sigma},
\end{equation}
and 
\begin{equation}
\widehat{c}_{d\uparrow}^{\dagger}\widehat{c}_{d\uparrow}\widehat{c}_{d\downarrow}^{\dagger}\widehat{c}_{d\downarrow}\widehat{b}_{\uparrow}^{\dagger}\widehat{b}_{\uparrow}\widehat{b}_{\downarrow}^{\dagger}\widehat{b}_{\downarrow}=\widehat{c}_{d\uparrow}^{\dagger}\widehat{c}_{d\uparrow}\widehat{c}_{d\downarrow}^{\dagger}\widehat{c}_{d\downarrow}=\widehat{b}_{\uparrow}^{\dagger}\widehat{b}_{\uparrow}\widehat{b}_{\downarrow}^{\dagger}\widehat{b}_{\downarrow}=\widehat{c}_{d\downarrow}^{\dagger}\widehat{c}_{d\downarrow}\widehat{b}_{\uparrow}^{\dagger}\widehat{b}_{\uparrow}=\widehat{c}_{d\uparrow}^{\dagger}\widehat{c}_{d\uparrow}\widehat{b}_{\downarrow}^{\dagger}\widehat{b}_{\downarrow}.
\end{equation}
Therefore, we can choose the following form of Hamiltonian terms:
\begin{equation}
\widehat{H}_{\textrm{s}}\to\widehat{H}_{\textrm{s}}^{\prime}=\varepsilon_{d}\widehat{b}_{\sigma}^{\dagger}\widehat{b}_{\sigma}+U\widehat{b}_{\uparrow}^{\dagger}\widehat{b}_{\uparrow}\widehat{b}_{\downarrow}^{\dagger}\widehat{b}_{\downarrow}.
\end{equation}
or: 
\begin{equation}
\widehat{H}_{\textrm{s}}\to\widehat{H}_{\textrm{s}}^{\prime}=\varepsilon_{d}\widehat{c}_{d\sigma}^{\dagger}\widehat{c}_{d\sigma}+U\widehat{b}_{\uparrow}^{\dagger}\widehat{b}_{\uparrow}\widehat{b}_{\downarrow}^{\dagger}\widehat{b}_{\downarrow}.
\end{equation}
The isomorphism between the original system and its bosonised form
is established by assigning the states
\begin{equation}
\left|0\right\rangle _{\sigma}\to\left|0\right\rangle _{\sigma-}\otimes\left|0\right\rangle _{\sigma+},\,\,\,\left|1\right\rangle _{\sigma}\to\left|1\right\rangle _{\sigma-}\otimes\left|1\right\rangle _{\sigma+},
\end{equation}
and by changing the canonical operators
\begin{equation}
\widehat{d}_{\sigma}\to\widehat{c}_{d\sigma}\widehat{b}_{\sigma}.
\end{equation}
Then, the matrix elements (in the global space) are the same under
such a mapping: 
\begin{multline}
\left\langle \textrm{vac}_{-}\right|\widehat{c}_{k_{1}\sigma_{1}}\ldots\widehat{c}_{k_{p}\sigma_{p}}\widehat{d}_{\varsigma_{1}}\ldots\widehat{d}_{\varsigma_{q}}\left\{ \widehat{d}_{\sigma}\right\} \widehat{d}_{\varsigma_{r}^{\prime}}^{\dagger}\ldots\widehat{d}_{\varsigma_{1}^{\prime}}^{\dagger}\widehat{c}_{k_{s}^{\prime}\sigma_{s}^{\prime}}\ldots\widehat{c}_{k_{1}^{\prime}\sigma_{1}^{\prime}}\left|\textrm{vac}_{-}\right\rangle \\
=\left\{ 0\,\,\textrm{if set }\varsigma\cup\sigma\,\,\textrm{is not equal to the set }\varsigma^{\prime}\right\} \\
\times\left\langle \textrm{vac}_{-}\right|\widehat{c}_{k_{1}\sigma_{1}}\ldots\widehat{c}_{k_{p}\sigma_{p}}\widehat{d}_{\varsigma_{1}}\ldots\widehat{d}_{\varsigma_{q}}\left\{ \widehat{d}_{\sigma}\right\} \widehat{d}_{\varsigma_{r}^{\prime}}^{\dagger}\ldots\widehat{d}_{\varsigma_{1}^{\prime}}^{\dagger}\widehat{c}_{k_{s}^{\prime}\sigma_{s}^{\prime}}^{\dagger}\ldots\widehat{c}_{k_{1}^{\prime}\sigma_{1}^{\prime}}^{\dagger}\left|\textrm{vac}_{-}\right\rangle \\
=\left\langle \textrm{vac}_{+}\left|\widehat{b}_{\varsigma_{1}}\ldots\widehat{b}_{\varsigma_{q}}\widehat{b}_{\sigma}\widehat{b}_{\varsigma_{r}^{\prime}}^{\dagger}\ldots\widehat{b}_{\varsigma_{1}^{\prime}}^{\dagger}\right|\textrm{vac}_{+}\right\rangle \\
\times\left\langle \textrm{vac}_{-}\right|\widehat{c}_{k_{1}\sigma_{1}}\ldots\widehat{c}_{k_{p}\sigma_{p}}\widehat{d}_{\varsigma_{1}}\ldots\widehat{d}_{\varsigma_{q}}\left\{ \widehat{d}_{\sigma}\right\} \widehat{d}_{\varsigma_{r}^{\prime}}^{\dagger}\ldots\widehat{d}_{\varsigma_{1}^{\prime}}^{\dagger}\widehat{c}_{k_{s}^{\prime}\sigma_{s}^{\prime}}^{\dagger}\ldots\widehat{c}_{k_{1}^{\prime}\sigma_{1}^{\prime}}^{\dagger}\left|\textrm{vac}_{-}\right\rangle \\
=\left\langle \textrm{vac}_{+}\right|\left\langle \textrm{vac}_{-}\right|\widehat{c}_{k_{1}\sigma_{1}}\ldots\widehat{c}_{k_{p}\sigma_{p}}\widehat{b}_{\varsigma_{1}}\widehat{d}_{\varsigma_{1}}\ldots\widehat{b}_{\varsigma_{q}}\widehat{d}_{\varsigma_{q}}\left\{ \widehat{b}_{\sigma}\widehat{d}_{\sigma}\right\} \widehat{b}_{\varsigma_{r}^{\prime}}^{\dagger}\widehat{d}_{\varsigma_{r}^{\prime}}^{\dagger}\ldots\widehat{b}_{\varsigma_{1}^{\prime}}^{\dagger}\widehat{d}_{\varsigma_{1}^{\prime}}^{\dagger}\widehat{c}_{k_{s}^{\prime}\sigma_{s}^{\prime}}^{\dagger}\ldots\widehat{c}_{k_{1}^{\prime}\sigma_{1}^{\prime}}^{\dagger}\left|\textrm{vac}_{+}\right\rangle \left|\textrm{vac}_{-}\right\rangle .
\end{multline}
Every number-conserving impurity Hamiltonian can be cast into such
a bosonized form.

\subsection{Computing averages}

Since the physical space $L_{\textrm{s}}$ is invariant, then, given
that an inital bosonized state of the impurity and the bath is physical,
it will be so at any later time, e.g. it will be orthogonal to the
unphysical components like
\begin{equation}
\left|0\right\rangle _{\downarrow-}\otimes\left|1\right\rangle _{\downarrow+}\,\,\,\textrm{or}\,\,\,\left|1\right\rangle _{\downarrow-}\otimes\left|0\right\rangle _{\downarrow+}.
\end{equation}
This means that when computing the averages, we can take the trace
over the all Hilbert space: unphysical components will not contribute:
\begin{multline}
\left\langle \widehat{d}_{\sigma}^{\dagger}\widehat{d}_{\sigma^{\prime}}\left(t\right)\right\rangle =\textrm{Tr}_{+}\textrm{Tr}_{-}\left\{ \widehat{c}_{d\sigma}^{\dagger}\widehat{c}_{d\sigma^{\prime}}\left|\Psi\left(t\right)\right\rangle \left\langle \Psi\left(t\right)\right|\right\} \\
=\textrm{Tr}_{+}\textrm{Tr}_{-}\left\{ \widehat{b}_{\sigma}^{\dagger}\widehat{b}_{\sigma^{\prime}}\left|\Psi\left(t\right)\right\rangle \left\langle \Psi\left(t\right)\right|\right\} \\
=\textrm{Tr}_{+}\textrm{Tr}_{-}\left\{ \widehat{b}_{\sigma}^{\dagger}\widehat{b}_{\sigma^{\prime}}\widehat{c}_{d\sigma}^{\dagger}\widehat{c}_{d\sigma^{\prime}}\left|\Psi\left(t\right)\right\rangle \left\langle \Psi\left(t\right)\right|\right\} ,\label{eq:bosonized_averages}
\end{multline}
where 
\begin{equation}
\left|\Psi\left(t\right)\right\rangle =\exp\left(-it\widehat{H}_{\textrm{s}}^{\prime}\right)\left|\Psi\left(0\right)\right\rangle ,\,\,\,\textrm{with}\,\,\,\left|\Psi\left(0\right)\right\rangle \in L_{\textrm{s}}.
\end{equation}
all the three averages are equivalent in Eq. (\ref{eq:bosonized_averages})
when the calculations are exact. However, when approximations are
done, their performance may differ.

\subsection{The resulting Hamiltonian is Kondo-like}

Here we again state the resulting bosonized Hamiltonian
\begin{equation}
\widehat{H}=U\widehat{b}_{\uparrow}^{\dagger}\widehat{b}_{\uparrow}\widehat{b}_{\downarrow}^{\dagger}\widehat{b}_{\downarrow}+V_{k}\widehat{b}_{\sigma}^{\dagger}\widehat{c}_{d\sigma}^{\dagger}\widehat{c}_{k\sigma}+V_{k}^{*}\widehat{c}_{k\sigma}^{\dagger}\widehat{c}_{d\sigma}\widehat{b}_{\sigma}+\varepsilon_{d}\widehat{c}_{d\sigma}^{\dagger}\widehat{c}_{d\sigma}+\varepsilon_{k}\widehat{c}_{k\sigma}^{\dagger}\widehat{c}_{k\sigma}.
\end{equation}
We see this Hamiltonian has the Kondo-like form. We consider $\widehat{b}_{\sigma}$
as the open system degrees of freedom, and $\widehat{c}_{k\sigma},\widehat{c}_{d\sigma}$
as the bath. Then, we can simulate this dynamics of this model by
applying the formalism of section \ref{sec:NUMERICAL-SIMULATION-FOR}.

\section{\label{sec:CONCLUSION}CONCLUSION}

In this work we propose a solution to the long standing problem of
a stochastic description of non-Markovian dissipative processes in
a fermionic bath. This is the problem of creating a formalism of stochastic
non-Markovian quantum state diffusion in a fermionic bath. At the
heart of such a description is the indentification of quantum states
of environment which carry classical signals. In the bosonic case
such an indentification is easy: the coherent states. In this work
we propose to consider the particle-hole coherent states as carrying
the signal of a particle-hole field. Then, the Husimi function leads
to a representation of the bath quantum state as a classical probability
ensemble of partile-hole field samples. The master equation for Husimi
turns out to be a classical convection equation, which we believe
allows for Monte Carlo simulation methods for real-time non-Markovian
open system quantum dynamics.

The probability measure of the vacuum fluctuations of the particle-hole
field is highly non-Gaussian for a finite number of the bath modes.
However it is conjectured to become Gaussian in the infinite-bath
limit. 

Here we present a mathematical derivation of the approach. Numerical
evaluation is an ongoing work and will be published elsewhere.

\appendix

\section{\label{sec:Derivation-of-the}Derivation of the metric tensor and
volume element}

Let us evaluate (\ref{eq:metric_from_Kahler}):
\begin{multline}
g_{\alpha,\overline{\beta}}=\frac{\partial}{\partial\boldsymbol{\xi}_{\alpha}}\left(\det\left(\boldsymbol{I}_{\textrm{h}}+\boldsymbol{\xi}^{\dagger}\boldsymbol{\xi}\right)^{-1}\frac{\partial}{\partial\boldsymbol{\xi}_{\beta}^{*}}\det\left(\boldsymbol{I}_{\textrm{h}}+\boldsymbol{\xi}^{\dagger}\boldsymbol{\xi}\right)\right)\\
=\frac{\partial}{\partial\boldsymbol{\xi}_{\alpha}}\left(\det\left(\boldsymbol{I}_{\textrm{h}}+\boldsymbol{\xi}^{\dagger}\boldsymbol{\xi}\right)^{-1}\det\left(\boldsymbol{I}_{\textrm{h}}+\boldsymbol{\xi}^{\dagger}\boldsymbol{\xi}\right)\textrm{Tr}\left(\left(\boldsymbol{I}_{\textrm{h}}+\boldsymbol{\xi}^{\dagger}\boldsymbol{\xi}\right)^{-1}\left|l^{\prime}\varsigma^{\prime}\right\rangle \left\langle l\varsigma\right|\boldsymbol{\xi}\right)\right)\\
=\frac{\partial}{\partial\boldsymbol{\xi}_{\alpha}}\left\langle l\varsigma\right|\boldsymbol{\xi}\left(\boldsymbol{I}_{\textrm{h}}+\boldsymbol{\xi}^{\dagger}\boldsymbol{\xi}\right)^{-1}\left|l^{\prime}\varsigma^{\prime}\right\rangle \\
=\delta_{kl}\delta_{\sigma\varsigma}\left\langle k^{\prime}\sigma^{\prime}\right|\left(\boldsymbol{I}_{\textrm{h}}+\boldsymbol{\xi}^{\dagger}\boldsymbol{\xi}\right)^{-1}\left|l^{\prime}\varsigma^{\prime}\right\rangle -\left\langle l\varsigma\right|\boldsymbol{\xi}\left(\boldsymbol{I}_{\textrm{h}}+\boldsymbol{\xi}^{\dagger}\boldsymbol{\xi}\right)^{-1}\boldsymbol{\xi}^{\dagger}\left|k\sigma\right\rangle \left\langle k^{\prime}\sigma^{\prime}\right|\left(\boldsymbol{I}_{\textrm{h}}+\boldsymbol{\xi}^{\dagger}\boldsymbol{\xi}\right)^{-1}\left|l^{\prime}\varsigma^{\prime}\right\rangle \\
=\left(\boldsymbol{I}_{\textrm{p}}-\boldsymbol{\xi}\left(\boldsymbol{I}_{\textrm{h}}+\boldsymbol{\xi}^{\dagger}\boldsymbol{\xi}\right)^{-1}\boldsymbol{\xi}^{\dagger}\right)_{l\varsigma,k\sigma}\left(\boldsymbol{I}_{\textrm{h}}+\boldsymbol{\xi}^{\dagger}\boldsymbol{\xi}\right)_{k^{\prime}\sigma^{\prime},l^{\prime}\varsigma^{\prime}}^{-1}\\
=\left(\boldsymbol{I}_{\textrm{p}}-\boldsymbol{\xi}\left(\boldsymbol{I}-\boldsymbol{\xi}^{\dagger}\boldsymbol{\xi}+\left(\boldsymbol{\xi}^{\dagger}\boldsymbol{\xi}\right)^{2}+\ldots+\left(-\boldsymbol{\xi}^{\dagger}\boldsymbol{\xi}\right)^{k}\right)\boldsymbol{\xi}^{\dagger}\right)_{l\varsigma,k\sigma}\left(\boldsymbol{I}_{\textrm{h}}+\boldsymbol{\xi}^{\dagger}\boldsymbol{\xi}\right)_{k^{\prime}\sigma^{\prime},l^{\prime}\varsigma^{\prime}}^{-1}\\
=\left(\boldsymbol{I}_{\textrm{p}}-\boldsymbol{\xi}\boldsymbol{\xi}^{\dagger}+\boldsymbol{\xi}\boldsymbol{\xi}^{\dagger}\boldsymbol{\xi}\boldsymbol{\xi}^{\dagger}+\ldots+\left(-\boldsymbol{\xi}\boldsymbol{\xi}^{\dagger}\right)^{k+1}+\ldots\right)_{l\varsigma,k\sigma}\left(\boldsymbol{I}_{\textrm{h}}+\boldsymbol{\xi}^{\dagger}\boldsymbol{\xi}\right)_{k^{\prime}\sigma^{\prime},l^{\prime}\varsigma^{\prime}}^{-1}\\
=\left(\boldsymbol{I}_{\textrm{p}}+\boldsymbol{\xi}\boldsymbol{\xi}^{\dagger}\right)_{l\varsigma,k\sigma}^{-1}\left(\boldsymbol{I}_{\textrm{h}}+\boldsymbol{\xi}^{\dagger}\boldsymbol{\xi}\right)_{k^{\prime}\sigma^{\prime},l^{\prime}\varsigma^{\prime}}^{-1}.
\end{multline}
The metric is 
\begin{multline}
ds^{2}=\textrm{Tr}\left\{ d\boldsymbol{\xi}^{\dagger}\left(\boldsymbol{I}_{\textrm{p}}+\boldsymbol{\xi}\boldsymbol{\xi}^{\dagger}\right)^{-1}d\boldsymbol{\xi}\left(\boldsymbol{I}_{\textrm{h}}+\boldsymbol{\xi}^{\dagger}\boldsymbol{\xi}\right)^{-1}\right\} ,
\end{multline}
so that 
\begin{multline}
g_{\left(k\sigma,k^{\prime}\sigma^{\prime}\right),\overline{\left(l\varsigma,l^{\prime}\varsigma^{\prime}\right)}}=\left(\boldsymbol{I}_{\textrm{p}}+\boldsymbol{\xi}\boldsymbol{\xi}^{\dagger}\right)_{l\varsigma,k\sigma}^{-1}\left(\boldsymbol{I}_{\textrm{h}}+\boldsymbol{\xi}^{\dagger}\boldsymbol{\xi}\right)_{k^{\prime}\sigma^{\prime},l^{\prime}\varsigma^{\prime}}^{-1}\\
=\left(\boldsymbol{I}_{\textrm{p}}+\boldsymbol{\xi}^{*}\boldsymbol{\xi}^{T}\right)_{k\sigma,l\varsigma}^{-1}\left(\boldsymbol{I}_{\textrm{h}}+\boldsymbol{\xi}^{\dagger}\boldsymbol{\xi}\right)_{k^{\prime}\sigma^{\prime},l^{\prime}\varsigma^{\prime}}^{-1}\\
=\left\{ \left(\boldsymbol{I}_{\textrm{p}}+\boldsymbol{\xi}^{*}\boldsymbol{\xi}^{T}\right)^{-1}\otimes\left(\boldsymbol{I}_{\textrm{h}}+\boldsymbol{\xi}^{\dagger}\boldsymbol{\xi}\right)^{-1}\right\} _{\left(k\sigma,l\varsigma\right),\left(k^{\prime}\sigma^{\prime},l^{\prime}\varsigma^{\prime}\right)},
\end{multline}
where in the last line the operation $\otimes$ is a kronecker product. 

The volume element is provided by 
\begin{multline*}
\sqrt{g}=\det g_{\alpha,\overline{\beta}}\\
=\det\left\{ \left(\boldsymbol{I}_{\textrm{p}}+\boldsymbol{\xi}^{*}\boldsymbol{\xi}^{T}\right)^{-1}\otimes\left(\boldsymbol{I}_{\textrm{h}}+\boldsymbol{\xi}^{\dagger}\boldsymbol{\xi}\right)^{-1}\right\} \\
=\det\left\{ \left(\boldsymbol{I}_{\textrm{p}}+\boldsymbol{\xi}^{*}\boldsymbol{\xi}^{T}\right)^{-1}\right\} ^{M_{\textrm{h}}}\det\left\{ \left(\boldsymbol{I}_{\textrm{h}}+\boldsymbol{\xi}^{\dagger}\boldsymbol{\xi}\right)^{-1}\right\} ^{M_{\textrm{p}}}(\textrm{determinant of kronecker product})\\
=\det\left\{ \left(\boldsymbol{I}_{\textrm{p}}+\boldsymbol{\xi}\boldsymbol{\xi}^{\dagger}\right)^{-T}\right\} ^{M_{\textrm{h}}}\det\left\{ \left(\boldsymbol{I}_{\textrm{h}}+\boldsymbol{\xi}^{\dagger}\boldsymbol{\xi}\right)^{-1}\right\} ^{M_{\textrm{p}}}\\
=\det\left\{ \left(\boldsymbol{I}_{\textrm{p}}+\boldsymbol{\xi}\boldsymbol{\xi}^{\dagger}\right)^{-1}\right\} ^{M_{\textrm{h}}}\det\left\{ \left(\boldsymbol{I}_{\textrm{h}}+\boldsymbol{\xi}^{\dagger}\boldsymbol{\xi}\right)^{-1}\right\} ^{M_{\textrm{p}}}\\
=\det\left\{ \left(\boldsymbol{I}_{\textrm{h}}+\boldsymbol{\xi}^{\dagger}\boldsymbol{\xi}\right)^{-1}\right\} ^{M_{\textrm{h}}}\det\left\{ \left(\boldsymbol{I}_{\textrm{h}}+\boldsymbol{\xi}^{\dagger}\boldsymbol{\xi}\right)^{-1}\right\} ^{M_{\textrm{p}}}(\textrm{by Sylvester's theorem}\textrm{)}\\
=\det\left\{ \left(\boldsymbol{I}_{\textrm{h}}+\boldsymbol{\xi}^{\dagger}\boldsymbol{\xi}\right)^{-1}\right\} ^{M}.
\end{multline*}

\section{\label{sec:Differential-correspondences-for}Differential correspondences
for the commutator}

Here we provice the detailed calculations of the differential correspondences
for the particle-hole coherent state projections. 

\subsection{$D$-algebra of unnormalized coherent states}

We have:
\begin{multline}
\left\langle \boldsymbol{\xi}\right\Vert \widehat{c}_{k\sigma}^{\textrm{p}\dagger}\widehat{c}_{k^{\prime}\sigma^{\prime}}^{\textrm{p}}=\left\langle \textrm{vac}\right|\exp\left(\widehat{\boldsymbol{c}}^{\textrm{p}}\boldsymbol{\xi}\widehat{\boldsymbol{c}}^{\textrm{h}}\right)\widehat{c}_{k\sigma}^{\textrm{p}\dagger}\widehat{c}_{k^{\prime}\sigma^{\prime}}^{\textrm{p}}\\
=\left\langle \textrm{vac}\right|\left\{ \left(\widehat{c}_{k\sigma}^{\textrm{p}\dagger}-\sum_{\beta}\xi_{k\sigma,\beta}\widehat{c}_{\beta}^{\textrm{h}}\right)\widehat{c}_{k^{\prime}\sigma^{\prime}}^{\textrm{p}}\right\} \exp\left(\widehat{\boldsymbol{c}}^{\textrm{p}}\boldsymbol{\xi}\widehat{\boldsymbol{c}}^{\textrm{h}}\right)\\
=\sum_{\beta}\xi_{k\sigma,\beta}\left\langle \textrm{vac}\right|\left\{ \widehat{c}_{k^{\prime}\sigma^{\prime}}^{\textrm{p}}\widehat{c}_{\beta}^{\textrm{h}}\right\} \exp\left(\widehat{\boldsymbol{c}}^{\textrm{p}}\boldsymbol{\xi}\widehat{\boldsymbol{c}}^{\textrm{h}}\right)\\
=\sum_{\beta}\xi_{k\sigma,\beta}\frac{\partial}{\partial\xi_{k^{\prime}\sigma^{\prime}\beta}}\left\langle \boldsymbol{\xi}\right\Vert ,
\end{multline}
\begin{multline}
\left\langle \boldsymbol{\xi}\right\Vert \widehat{c}_{k\sigma}^{\textrm{h}\dagger}\widehat{c}_{k^{\prime}\sigma^{\prime}}^{\textrm{h}}=\left\langle \textrm{vac}\right|\exp\left(\widehat{\boldsymbol{c}}^{\textrm{p}}\boldsymbol{\xi}\widehat{\boldsymbol{c}}^{\textrm{h}}\right)\widehat{c}_{k\sigma}^{\textrm{h}\dagger}\widehat{c}_{k^{\prime}\sigma^{\prime}}^{\textrm{h}}\\
=\left\langle \textrm{vac}\right|\left\{ \left(\widehat{c}_{k\sigma}^{\textrm{h}\dagger}+\sum_{\alpha}\widehat{c}_{\alpha}^{\textrm{p}}\xi_{\alpha,k\sigma}\right)\widehat{c}_{k^{\prime}\sigma^{\prime}}^{\textrm{h}}\right\} \exp\left(\widehat{\boldsymbol{c}}^{\textrm{p}}\boldsymbol{\xi}\widehat{\boldsymbol{c}}^{\textrm{h}}\right)\\
=\sum_{\alpha}\xi_{\alpha,k\sigma}\left\langle \textrm{vac}\right|\left\{ \widehat{c}_{\alpha}^{\textrm{p}}\widehat{c}_{k^{\prime}\sigma^{\prime}}^{\textrm{h}}\right\} \exp\left(\widehat{\boldsymbol{c}}^{\textrm{p}}\boldsymbol{\xi}\widehat{\boldsymbol{c}}^{\textrm{h}}\right)\\
=\sum_{\alpha}\xi_{\alpha,k\sigma}\frac{\partial}{\partial\xi_{\alpha k^{\prime}\sigma^{\prime}}}\left\langle \boldsymbol{\xi}\right\Vert ,
\end{multline}
\begin{multline}
\left\langle \boldsymbol{\xi}\right\Vert \widehat{c}_{k\sigma}^{\textrm{p}\dagger}\widehat{c}_{k^{\prime}\sigma^{\prime}}^{\textrm{h}\dagger}=\left\langle \textrm{vac}\right|\exp\left(\widehat{\boldsymbol{c}}^{\textrm{p}}\boldsymbol{\xi}\widehat{\boldsymbol{c}}^{\textrm{h}}\right)\widehat{c}_{k\sigma}^{\textrm{p}\dagger}\widehat{c}_{k^{\prime}\sigma^{\prime}}^{\textrm{h}\dagger}\\
=\left\langle \textrm{vac}\right|\left\{ \left(\widehat{c}_{k\sigma}^{\textrm{p}\dagger}-\sum_{\beta}\xi_{k\sigma,\beta}\widehat{c}_{\beta}^{\textrm{h}}\right)\left(\widehat{c}_{k^{\prime}\sigma^{\prime}}^{\textrm{h}\dagger}+\sum_{\alpha}\widehat{c}_{\alpha}^{\textrm{p}}\xi_{\alpha,k^{\prime}\sigma^{\prime}}\right)\right\} \exp\left(\widehat{\boldsymbol{c}}^{\textrm{p}}\boldsymbol{\xi}\widehat{\boldsymbol{c}}^{\textrm{h}}\right)\\
=\sum_{\beta}\xi_{k\sigma,\beta}\sum_{\alpha}\xi_{\alpha,k^{\prime}\sigma^{\prime}}\left\langle \textrm{vac}\right|\left\{ \widehat{c}_{\alpha}^{\textrm{p}}\widehat{c}_{\beta}^{\textrm{h}}\right\} \exp\left(\widehat{\boldsymbol{c}}^{\textrm{p}}\boldsymbol{\xi}\widehat{\boldsymbol{c}}^{\textrm{h}}\right)-\xi_{k\sigma,k^{\prime}\sigma^{\prime}}\left\langle \boldsymbol{\xi}\right\Vert \\
=\left\{ \sum_{\alpha\beta}\xi_{k\sigma,\beta}\xi_{\alpha,k^{\prime}\sigma^{\prime}}\frac{\partial}{\partial\xi_{\alpha\beta}}-\xi_{k\sigma,k^{\prime}\sigma^{\prime}}\right\} \left\langle \boldsymbol{\xi}\right\Vert \\
=\left\{ \sum_{\alpha\neq k\sigma,\beta\neq k^{\prime}\sigma^{\prime}}\xi_{\alpha,k^{\prime}\sigma^{\prime}}\xi_{k\sigma,\beta}\frac{\partial}{\partial\xi_{\alpha\beta}}+\xi_{k\sigma,k^{\prime}\sigma^{\prime}}\left(\xi_{k\sigma,k^{\prime}\sigma^{\prime}}\frac{\partial}{\partial\xi_{\xi_{k\sigma,k^{\prime}\sigma^{\prime}}}}-1\right)\right\} \left\langle \boldsymbol{\xi}\right\Vert ,
\end{multline}
\begin{multline}
\left\langle \boldsymbol{\xi}\right\Vert \widehat{c}_{k\sigma}^{\textrm{h}}\widehat{c}_{k^{\prime}\sigma^{\prime}}^{\textrm{p}}=-\frac{\partial}{\partial\xi_{k^{\prime}\sigma^{\prime},k\sigma}}\left\langle \boldsymbol{\xi}\right\Vert .
\end{multline}

\subsection{Action of Hamiltonian on unnormalized coherent states}

Therefore, the general Hamiltonian
\begin{multline}
\widehat{H}=h_{k\sigma k^{\prime}\sigma^{\prime}}^{\textrm{pp}}\widehat{c}_{k\sigma}^{\textrm{p}\dagger}\widehat{c}_{k^{\prime}\sigma^{\prime}}^{\textrm{p}}-h_{k\sigma k^{\prime}\sigma^{\prime}}^{\textrm{hh}}\widehat{c}_{k\sigma}^{\textrm{h}\dagger}\widehat{c}_{k^{\prime}\sigma^{\prime}}^{\textrm{h}}\\
+h_{k\sigma k^{\prime}\sigma^{\prime}}^{\textrm{ph}}\widehat{c}_{k\sigma}^{\textrm{p}\dagger}\widehat{c}_{k^{\prime}\sigma^{\prime}}^{\textrm{h}\dagger}+h_{k\sigma k^{\prime}\sigma^{\prime}}^{\textrm{ph}\dagger}\widehat{c}_{k\sigma}^{\textrm{h}}\widehat{c}_{k^{\prime}\sigma^{\prime}}^{\textrm{p}}
\end{multline}
acts on the coherent state as
\begin{multline}
\left\langle \boldsymbol{\xi}\right\Vert \widehat{H}\\
=\left\{ h_{k\sigma k^{\prime}\sigma^{\prime}}^{\textrm{pp}}\sum_{\beta}\xi_{k\sigma,\beta}\frac{\partial}{\partial\xi_{k^{\prime}\sigma^{\prime}\beta}}-h_{k\sigma k^{\prime}\sigma^{\prime}}^{\textrm{hh}}\sum_{\alpha}\xi_{\alpha,k\sigma}\frac{\partial}{\partial\xi_{\alpha k^{\prime}\sigma^{\prime}}}\right.\\
\left.+h_{k\sigma k^{\prime}\sigma^{\prime}}^{\textrm{ph}}\left[\sum_{\alpha\beta}\xi_{k\sigma,\beta}\xi_{\alpha,k^{\prime}\sigma^{\prime}}\frac{\partial}{\partial\xi_{\alpha\beta}}-\xi_{k\sigma,k^{\prime}\sigma^{\prime}}\right]-h_{k\sigma k^{\prime}\sigma^{\prime}}^{\textrm{ph}\dagger}\frac{\partial}{\partial\xi_{k^{\prime}\sigma^{\prime},k\sigma}}\right\} \left\langle \boldsymbol{\xi}\right\Vert .
\end{multline}

\subsection{$D$-algebra for normalized coherent states}

We have the relation:
\begin{multline}
\mathcal{N}^{-1}\left(\boldsymbol{\xi}\right)\frac{\partial}{\partial\xi_{\alpha\beta}}=\frac{\partial}{\partial\xi_{\alpha\beta}}\mathcal{N}^{-1}\left(\boldsymbol{\xi}\right)-\left(\frac{\partial}{\partial\xi_{\alpha\beta}}\mathcal{N}^{-1}\left(\boldsymbol{\xi}\right)\right)\\
=\frac{\partial}{\partial\xi_{\alpha\beta}}\mathcal{N}^{-1}\left(\boldsymbol{\xi}\right)-\left(\frac{\partial}{\partial\xi_{\alpha\beta}}\textrm{det}\left(\left[\boldsymbol{I}_{\textrm{h}}+\boldsymbol{\xi}^{\dagger}\boldsymbol{\xi}\right]^{-1}\right)\right)\\
=\frac{\partial}{\partial\xi_{\alpha\beta}}\mathcal{N}^{-1}\left(\boldsymbol{\xi}\right)+\mathcal{N}^{-1}\left(\boldsymbol{\xi}\right)\left\langle \beta\right|\left[\boldsymbol{I}_{\textrm{h}}+\boldsymbol{\xi}^{\dagger}\boldsymbol{\xi}\right]^{-1}\boldsymbol{\xi}^{\dagger}\left|\alpha\right\rangle 
\end{multline}
and, alternatively, 
\begin{multline}
\mathcal{N}^{-1}\left(\boldsymbol{\xi}\right)\frac{\partial}{\partial\xi_{\alpha\beta}}\\
=\frac{\partial}{\partial\xi_{\alpha\beta}}\mathcal{N}^{-1}\left(\boldsymbol{\xi}\right)-\left(\frac{\partial}{\partial\xi_{\alpha\beta}}\textrm{det}\left(\left[\boldsymbol{I}_{\textrm{p}}+\boldsymbol{\xi}\boldsymbol{\xi}^{\dagger}\right]^{-1}\right)\right)\\
=\frac{\partial}{\partial\xi_{\alpha\beta}}\mathcal{N}^{-1}\left(\boldsymbol{\xi}\right)+\mathcal{N}^{-1}\left(\boldsymbol{\xi}\right)\left\langle \beta\right|\boldsymbol{\xi}^{\dagger}\left[\boldsymbol{I}_{\textrm{p}}+\boldsymbol{\xi}\boldsymbol{\xi}^{\dagger}\right]^{-1}\left|\alpha\right\rangle .
\end{multline}
Therefore, 
\begin{multline}
h_{k\sigma k^{\prime}\sigma^{\prime}}^{\textrm{pp}}\sum_{\beta}\xi_{k\sigma,\beta}\frac{\partial}{\partial\xi_{k^{\prime}\sigma^{\prime}\beta}}\\
\to h_{k\sigma k^{\prime}\sigma^{\prime}}^{\textrm{pp}}\sum_{\beta}\xi_{k\sigma,\beta}\frac{\partial}{\partial\xi_{k^{\prime}\sigma^{\prime}\beta}}+h_{k\sigma k^{\prime}\sigma^{\prime}}^{\textrm{pp}}\sum_{\beta}\xi_{k\sigma,\beta}\left\langle \beta\right|\boldsymbol{\xi}^{\dagger}\left[\boldsymbol{I}_{\textrm{p}}+\boldsymbol{\xi}\boldsymbol{\xi}^{\dagger}\right]^{-1}\left|k^{\prime}\sigma^{\prime}\right\rangle \\
=h_{k\sigma k^{\prime}\sigma^{\prime}}^{\textrm{pp}}\sum_{\beta}\xi_{k\sigma,\beta}\frac{\partial}{\partial\xi_{k^{\prime}\sigma^{\prime}\beta}}+h_{k\sigma k^{\prime}\sigma^{\prime}}^{\textrm{pp}}\left\langle k\sigma\right|\boldsymbol{\xi}\boldsymbol{\xi}^{\dagger}\left[\boldsymbol{I}_{\textrm{p}}+\boldsymbol{\xi}\boldsymbol{\xi}^{\dagger}\right]^{-1}\left|k^{\prime}\sigma^{\prime}\right\rangle \\
=h_{k\sigma k^{\prime}\sigma^{\prime}}^{\textrm{pp}}\sum_{\beta}\xi_{k\sigma,\beta}\frac{\partial}{\partial\xi_{k^{\prime}\sigma^{\prime}\beta}}+\textrm{Tr}\left\{ \boldsymbol{h}^{\textrm{pp}T}\boldsymbol{\xi}\boldsymbol{\xi}^{\dagger}\left[\boldsymbol{I}_{\textrm{p}}+\boldsymbol{\xi}\boldsymbol{\xi}^{\dagger}\right]^{-1}\right\} .
\end{multline}
Next,
\begin{multline}
h_{k\sigma k^{\prime}\sigma^{\prime}}^{\textrm{hh}}\sum_{\alpha}\xi_{\alpha,k\sigma}\frac{\partial}{\partial\xi_{\alpha k^{\prime}\sigma^{\prime}}}\\
\to h_{k\sigma k^{\prime}\sigma^{\prime}}^{\textrm{hh}}\sum_{\alpha}\xi_{\alpha,k\sigma}\frac{\partial}{\partial\xi_{\alpha k^{\prime}\sigma^{\prime}}}+h_{k\sigma k^{\prime}\sigma^{\prime}}^{\textrm{hh}}\sum_{\alpha}\xi_{\alpha,k\sigma}\left\langle k^{\prime}\sigma^{\prime}\right|\left[\boldsymbol{I}_{\textrm{h}}+\boldsymbol{\xi}^{\dagger}\boldsymbol{\xi}\right]^{-1}\boldsymbol{\xi}^{\dagger}\left|\alpha\right\rangle \\
=h_{k\sigma k^{\prime}\sigma^{\prime}}^{\textrm{hh}}\sum_{\alpha}\xi_{\alpha,k\sigma}\frac{\partial}{\partial\xi_{\alpha k^{\prime}\sigma^{\prime}}}+h_{k\sigma k^{\prime}\sigma^{\prime}}^{\textrm{hh}}\left\langle k^{\prime}\sigma^{\prime}\right|\left[\boldsymbol{I}_{\textrm{h}}+\boldsymbol{\xi}^{\dagger}\boldsymbol{\xi}\right]^{-1}\boldsymbol{\xi}^{\dagger}\boldsymbol{\xi}\left|k\sigma\right\rangle \\
=h_{k\sigma k^{\prime}\sigma^{\prime}}^{\textrm{hh}}\sum_{\alpha}\xi_{\alpha,k\sigma}\frac{\partial}{\partial\xi_{\alpha k^{\prime}\sigma^{\prime}}}+\textrm{Tr}\left\{ \boldsymbol{h}^{\textrm{hh}}\left[\boldsymbol{I}_{\textrm{h}}+\boldsymbol{\xi}^{\dagger}\boldsymbol{\xi}\right]^{-1}\boldsymbol{\xi}^{\dagger}\boldsymbol{\xi}\right\} .
\end{multline}
Next, 
\begin{multline}
h_{k\sigma k^{\prime}\sigma^{\prime}}^{\textrm{ph}\dagger}\frac{\partial}{\partial\xi_{k^{\prime}\sigma^{\prime},k\sigma}}\\
\to h_{k\sigma k^{\prime}\sigma^{\prime}}^{\textrm{ph}\dagger}\frac{\partial}{\partial\xi_{k^{\prime}\sigma^{\prime},k\sigma}}+h_{k\sigma k^{\prime}\sigma^{\prime}}^{\textrm{ph}\dagger}\left\langle k\sigma\right|\left[\boldsymbol{I}_{\textrm{h}}+\boldsymbol{\xi}^{\dagger}\boldsymbol{\xi}\right]^{-1}\boldsymbol{\xi}^{\dagger}\left|k^{\prime}\sigma^{\prime}\right\rangle =h_{k\sigma k^{\prime}\sigma^{\prime}}^{\textrm{ph}\dagger}\frac{\partial}{\partial\xi_{k^{\prime}\sigma^{\prime},k\sigma}}+\textrm{Tr}\left\{ \boldsymbol{h}^{\textrm{ph}*}\left[\boldsymbol{I}_{\textrm{h}}+\boldsymbol{\xi}^{\dagger}\boldsymbol{\xi}\right]^{-1}\boldsymbol{\xi}^{\dagger}\right\} \\
\textrm{or}\,\,\,\longrightarrow h_{k\sigma k^{\prime}\sigma^{\prime}}^{\textrm{ph}\dagger}\frac{\partial}{\partial\xi_{k^{\prime}\sigma^{\prime},k\sigma}}+\textrm{Tr}\left\{ \boldsymbol{h}^{\textrm{ph}*}\boldsymbol{\xi}^{\dagger}\left[\boldsymbol{I}_{\textrm{p}}+\boldsymbol{\xi}\boldsymbol{\xi}^{\dagger}\right]^{-1}\right\} .
\end{multline}
Next, 
\begin{multline}
h_{k\sigma k^{\prime}\sigma^{\prime}}^{\textrm{ph}}\left[\sum_{\alpha\beta}\xi_{k\sigma,\beta}\xi_{\alpha,k^{\prime}\sigma^{\prime}}\frac{\partial}{\partial\xi_{\alpha\beta}}-\xi_{k\sigma,k^{\prime}\sigma^{\prime}}\right]\\
\longrightarrow h_{k\sigma k^{\prime}\sigma^{\prime}}^{\textrm{ph}}\left[\sum_{\alpha\beta}\xi_{k\sigma,\beta}\xi_{\alpha,k^{\prime}\sigma^{\prime}}\frac{\partial}{\partial\xi_{\alpha\beta}}-\xi_{k\sigma,k^{\prime}\sigma^{\prime}}\right]+h_{k\sigma k^{\prime}\sigma^{\prime}}^{\textrm{ph}}\sum_{\alpha\beta}\xi_{k\sigma,\beta}\xi_{\alpha,k^{\prime}\sigma^{\prime}}\left\langle \beta\right|\left[\boldsymbol{I}_{\textrm{h}}+\boldsymbol{\xi}^{\dagger}\boldsymbol{\xi}\right]^{-1}\boldsymbol{\xi}^{\dagger}\left|\alpha\right\rangle \\
=h_{k\sigma k^{\prime}\sigma^{\prime}}^{\textrm{ph}}\left[\sum_{\alpha\beta}\xi_{k\sigma,\beta}\xi_{\alpha,k^{\prime}\sigma^{\prime}}\frac{\partial}{\partial\xi_{\alpha\beta}}-\xi_{k\sigma,k^{\prime}\sigma^{\prime}}\right]+h_{k\sigma k^{\prime}\sigma^{\prime}}^{\textrm{ph}}\left\langle k\sigma\right|\boldsymbol{\xi}\left[\boldsymbol{I}_{\textrm{h}}+\boldsymbol{\xi}^{\dagger}\boldsymbol{\xi}\right]^{-1}\boldsymbol{\xi}^{\dagger}\boldsymbol{\xi}\left|k^{\prime}\sigma^{\prime}\right\rangle \\
=h_{k\sigma k^{\prime}\sigma^{\prime}}^{\textrm{ph}}\left[\sum_{\alpha\beta}\xi_{k\sigma,\beta}\xi_{\alpha,k^{\prime}\sigma^{\prime}}\frac{\partial}{\partial\xi_{\alpha\beta}}-\xi_{k\sigma,k^{\prime}\sigma^{\prime}}\right]+\textrm{Tr}\left\{ \boldsymbol{h}^{\textrm{ph}T}\boldsymbol{\xi}\left[\boldsymbol{I}_{\textrm{h}}+\boldsymbol{\xi}^{\dagger}\boldsymbol{\xi}\right]^{-1}\boldsymbol{\xi}^{\dagger}\boldsymbol{\xi}\right\} \\
=h_{k\sigma k^{\prime}\sigma^{\prime}}^{\textrm{ph}}\sum_{\alpha\beta}\xi_{k\sigma,\beta}\xi_{\alpha,k^{\prime}\sigma^{\prime}}\frac{\partial}{\partial\xi_{\alpha\beta}}+\textrm{Tr}\left\{ \boldsymbol{h}^{\textrm{ph}T}\boldsymbol{\xi}\left[\boldsymbol{I}_{\textrm{h}}+\boldsymbol{\xi}^{\dagger}\boldsymbol{\xi}\right]^{-1}\boldsymbol{\xi}^{\dagger}\boldsymbol{\xi}\right\} -\textrm{Tr}\left\{ \boldsymbol{h}^{\textrm{ph}T}\boldsymbol{\xi}\right\} \\
=h_{k\sigma k^{\prime}\sigma^{\prime}}^{\textrm{ph}}\sum_{\alpha\beta}\xi_{k\sigma,\beta}\xi_{\alpha,k^{\prime}\sigma^{\prime}}\frac{\partial}{\partial\xi_{\alpha\beta}}+\textrm{Tr}\left\{ \boldsymbol{h}^{\textrm{ph}T}\boldsymbol{\xi}\left(\left[\boldsymbol{I}_{\textrm{h}}+\boldsymbol{\xi}^{\dagger}\boldsymbol{\xi}\right]^{-1}\boldsymbol{\xi}^{\dagger}\boldsymbol{\xi}-\boldsymbol{I}_{\textrm{h}}\right)\right\} \\
=h_{k\sigma k^{\prime}\sigma^{\prime}}^{\textrm{ph}}\sum_{\alpha\beta}\xi_{k\sigma,\beta}\xi_{\alpha,k^{\prime}\sigma^{\prime}}\frac{\partial}{\partial\xi_{\alpha\beta}}-\textrm{Tr}\left\{ \boldsymbol{h}^{\textrm{ph}T}\boldsymbol{\xi}\left[\boldsymbol{I}_{\textrm{h}}+\boldsymbol{\xi}^{\dagger}\boldsymbol{\xi}\right]^{-1}\right\} .
\end{multline}
Therefore, the Hermitian conjugate term corresponds to 
\begin{multline}
-h_{k\sigma k^{\prime}\sigma^{\prime}}^{\textrm{ph}\dagger}\frac{\partial}{\partial\xi_{k^{\prime}\sigma^{\prime},k\sigma}}+h_{k\sigma k^{\prime}\sigma^{\prime}}^{\textrm{ph}}\left[\sum_{\alpha\beta}\xi_{k\sigma,\beta}\xi_{\alpha,k^{\prime}\sigma^{\prime}}\frac{\partial}{\partial\xi_{\alpha\beta}}-\xi_{k\sigma,k^{\prime}\sigma^{\prime}}\right]\\
\longrightarrow-h_{k\sigma k^{\prime}\sigma^{\prime}}^{\textrm{ph}\dagger}\frac{\partial}{\partial\xi_{k^{\prime}\sigma^{\prime},k\sigma}}+h_{k\sigma k^{\prime}\sigma^{\prime}}^{\textrm{ph}}\sum_{\alpha\beta}\xi_{k\sigma,\beta}\xi_{\alpha,k^{\prime}\sigma^{\prime}}\frac{\partial}{\partial\xi_{\alpha\beta}}\\
-\textrm{Tr}\left\{ \boldsymbol{h}^{\textrm{ph}T}\boldsymbol{\xi}\left[\boldsymbol{I}_{\textrm{h}}+\boldsymbol{\xi}^{\dagger}\boldsymbol{\xi}\right]^{-1}\right\} -\textrm{Tr}\left\{ \left[\boldsymbol{I}_{\textrm{h}}+\boldsymbol{\xi}^{\dagger}\boldsymbol{\xi}\right]^{-1}\boldsymbol{\xi}^{\dagger}\boldsymbol{h}^{\textrm{ph}T\dagger}\right\} .
\end{multline}

\subsection{$D$-algebra for commutators with projections on normalized coherent
states}

We have:

\begin{multline*}
\frac{\partial}{\partial\xi_{\alpha\beta}^{*}}\textrm{det}\left(\left[\boldsymbol{I}_{\textrm{h}}+\boldsymbol{\xi}^{\dagger}\boldsymbol{\xi}\right]^{-1}\right)\\
=\frac{\partial}{\partial\xi_{\beta\alpha}^{\dagger}}\textrm{det}\left(\left[\boldsymbol{I}_{\textrm{h}}+\boldsymbol{\xi}^{\dagger}\boldsymbol{\xi}\right]^{-1}\right)=-\textrm{det}\left(\left[\boldsymbol{I}_{\textrm{h}}+\boldsymbol{\xi}^{\dagger}\boldsymbol{\xi}\right]^{-1}\right)\left\langle \alpha\right|\boldsymbol{\xi}\left[\boldsymbol{I}_{\textrm{h}}+\boldsymbol{\xi}^{\dagger}\boldsymbol{\xi}\right]^{-1}\left|\beta\right\rangle 
\end{multline*}
and
\begin{multline*}
\frac{\partial}{\partial\xi_{\alpha\beta}^{*}}\textrm{det}\left(\left[\boldsymbol{I}_{\textrm{p}}+\boldsymbol{\xi}\boldsymbol{\xi}^{\dagger}\right]^{-1}\right)\\
=\frac{\partial}{\partial\xi_{\beta\alpha}^{\dagger}}\textrm{det}\left(\left[\boldsymbol{I}_{\textrm{p}}+\boldsymbol{\xi}\boldsymbol{\xi}^{\dagger}\right]^{-1}\right)=-\textrm{det}\left(\left[\boldsymbol{I}_{\textrm{p}}+\boldsymbol{\xi}\boldsymbol{\xi}^{\dagger}\right]^{-1}\right)\left\langle \alpha\right|\left[\boldsymbol{I}_{\textrm{p}}+\boldsymbol{\xi}\boldsymbol{\xi}^{\dagger}\right]^{-1}\boldsymbol{\xi}\left|\beta\right\rangle 
\end{multline*}

The conjugate correspondences:

\begin{equation}
\widehat{c}_{k^{\prime}\sigma^{\prime}}^{\textrm{p}\dagger}\widehat{c}_{k\sigma}^{\textrm{p}}\left|\boldsymbol{\xi}\right\rangle =\sum_{\beta}\xi_{k\sigma,\beta}^{*}\frac{\partial}{\partial\xi_{k^{\prime}\sigma^{\prime}\beta}^{*}}\left|\boldsymbol{\xi}\right\rangle ,
\end{equation}
\begin{equation}
h_{k\sigma k^{\prime}\sigma^{\prime}}^{\textrm{pp}*}\widehat{c}_{k^{\prime}\sigma^{\prime}}^{\textrm{p}\dagger}\widehat{c}_{k\sigma}^{\textrm{p}}\left|\boldsymbol{\xi}\right\rangle =h_{k\sigma k^{\prime}\sigma^{\prime}}^{\textrm{pp}*}\sum_{\beta}\xi_{k\sigma,\beta}^{*}\frac{\partial}{\partial\xi_{k^{\prime}\sigma^{\prime}\beta}^{*}}\left|\boldsymbol{\xi}\right\rangle ,
\end{equation}
\begin{multline}
h_{k\sigma k^{\prime}\sigma^{\prime}}^{\textrm{pp}*}\widehat{c}_{k^{\prime}\sigma^{\prime}}^{\textrm{p}\dagger}\widehat{c}_{k\sigma}^{\textrm{p}}\mathcal{N}^{-1}\left(\boldsymbol{\xi}\right)\left|\boldsymbol{\xi}\right\rangle \\
=\left\{ h_{k\sigma k^{\prime}\sigma^{\prime}}^{\textrm{pp}*}\sum_{\beta}\xi_{k\sigma,\beta}^{*}\frac{\partial}{\partial\xi_{k^{\prime}\sigma^{\prime}\beta}^{*}}-\mathcal{N}\left(\boldsymbol{\xi}\right)h_{k\sigma k^{\prime}\sigma^{\prime}}^{\textrm{pp}*}\sum_{\beta}\xi_{k\sigma,\beta}^{*}\frac{\partial}{\partial\xi_{k^{\prime}\sigma^{\prime}\beta}^{*}}\textrm{det}\left(\left[\boldsymbol{I}_{\textrm{p}}+\boldsymbol{\xi}\boldsymbol{\xi}^{\dagger}\right]^{-1}\right)\right\} \mathcal{N}^{-1}\left(\boldsymbol{\xi}\right)\left|\boldsymbol{\xi}\right\rangle \\
=\left\{ h_{k\sigma k^{\prime}\sigma^{\prime}}^{\textrm{pp}*}\sum_{\beta}\xi_{k\sigma,\beta}^{*}\frac{\partial}{\partial\xi_{k^{\prime}\sigma^{\prime}\beta}^{*}}+h_{k\sigma k^{\prime}\sigma^{\prime}}^{\textrm{pp}*}\sum_{\beta}\xi_{k\sigma,\beta}^{*}\left\langle k^{\prime}\sigma^{\prime}\right|\left[\boldsymbol{I}_{\textrm{p}}+\boldsymbol{\xi}\boldsymbol{\xi}^{\dagger}\right]^{-1}\boldsymbol{\xi}\left|\beta\right\rangle \right\} \mathcal{N}^{-1}\left(\boldsymbol{\xi}\right)\left|\boldsymbol{\xi}\right\rangle \\
=\left\{ h_{k\sigma k^{\prime}\sigma^{\prime}}^{\textrm{pp}*}\sum_{\beta}\xi_{k\sigma,\beta}^{*}\frac{\partial}{\partial\xi_{k^{\prime}\sigma^{\prime}\beta}^{*}}+h_{k\sigma k^{\prime}\sigma^{\prime}}^{\textrm{pp}*}\left\langle k^{\prime}\sigma^{\prime}\right|\left[\boldsymbol{I}_{\textrm{p}}+\boldsymbol{\xi}\boldsymbol{\xi}^{\dagger}\right]^{-1}\boldsymbol{\xi}\boldsymbol{\xi}^{\dagger}\left|k\sigma\right\rangle \right\} \mathcal{N}^{-1}\left(\boldsymbol{\xi}\right)\left|\boldsymbol{\xi}\right\rangle \\
=\left\{ h_{k\sigma k^{\prime}\sigma^{\prime}}^{\textrm{pp}*}\sum_{\beta}\xi_{k\sigma,\beta}^{*}\frac{\partial}{\partial\xi_{k^{\prime}\sigma^{\prime}\beta}^{*}}+\textrm{Tr}\left\{ \boldsymbol{h}^{\textrm{pp}*}\left[\boldsymbol{I}_{\textrm{p}}+\boldsymbol{\xi}\boldsymbol{\xi}^{\dagger}\right]^{-1}\boldsymbol{\xi}\boldsymbol{\xi}^{\dagger}\right\} \right\} \mathcal{N}^{-1}\left(\boldsymbol{\xi}\right)\left|\boldsymbol{\xi}\right\rangle \\
=\left\{ h_{k\sigma k^{\prime}\sigma^{\prime}}^{\textrm{pp}*}\sum_{\beta}\xi_{k\sigma,\beta}^{*}\frac{\partial}{\partial\xi_{k^{\prime}\sigma^{\prime}\beta}^{*}}+\textrm{Tr}\left\{ \boldsymbol{h}^{\textrm{pp}\dagger*}\left[\boldsymbol{I}_{\textrm{p}}+\boldsymbol{\xi}\boldsymbol{\xi}^{\dagger}\right]^{-1}\boldsymbol{\xi}\boldsymbol{\xi}^{\dagger}\right\} \right\} \mathcal{N}^{-1}\left(\boldsymbol{\xi}\right)\left|\boldsymbol{\xi}\right\rangle \\
=\left\{ h_{k\sigma k^{\prime}\sigma^{\prime}}^{\textrm{pp}*}\sum_{\beta}\xi_{k\sigma,\beta}^{*}\frac{\partial}{\partial\xi_{k^{\prime}\sigma^{\prime}\beta}^{*}}+\textrm{Tr}\left\{ \boldsymbol{h}^{\textrm{pp}T}\boldsymbol{\xi}\boldsymbol{\xi}^{\dagger}\left[\boldsymbol{I}_{\textrm{p}}+\boldsymbol{\xi}\boldsymbol{\xi}^{\dagger}\right]^{-1}\right\} \right\} \mathcal{N}^{-1}\left(\boldsymbol{\xi}\right)\left|\boldsymbol{\xi}\right\rangle .
\end{multline}
Therefore, the commutator 
\begin{multline}
-i\left[h_{k\sigma k^{\prime}\sigma^{\prime}}^{\textrm{pp}}\widehat{c}_{k\sigma}^{\textrm{p}\dagger}\widehat{c}_{k^{\prime}\sigma^{\prime}}^{\textrm{p}},\frac{\left|\boldsymbol{\xi}\right\rangle \left\langle \boldsymbol{\xi}\right|}{\mathcal{N}\left(\boldsymbol{\xi}\right)}\right]\\
=\left\{ -ih_{k\sigma k^{\prime}\sigma^{\prime}}^{\textrm{pp}*}\sum_{\beta}\xi_{k\sigma,\beta}^{*}\frac{\partial}{\partial\xi_{k^{\prime}\sigma^{\prime}\beta}^{*}}-i\textrm{Tr}\left\{ \boldsymbol{h}^{\textrm{pp}T}\boldsymbol{\xi}\boldsymbol{\xi}^{\dagger}\left[\boldsymbol{I}_{\textrm{p}}+\boldsymbol{\xi}\boldsymbol{\xi}^{\dagger}\right]^{-1}\right\} \right.\\
\left.+ih_{k\sigma k^{\prime}\sigma^{\prime}}^{\textrm{pp}}\sum_{\beta}\xi_{k\sigma,\beta}\frac{\partial}{\partial\xi_{k^{\prime}\sigma^{\prime}\beta}}+i\textrm{Tr}\left\{ \boldsymbol{h}^{\textrm{pp}T}\boldsymbol{\xi}\boldsymbol{\xi}^{\dagger}\left[\boldsymbol{I}_{\textrm{p}}+\boldsymbol{\xi}\boldsymbol{\xi}^{\dagger}\right]^{-1}\right\} \right\} \frac{\left|\boldsymbol{\xi}\right\rangle \left\langle \boldsymbol{\xi}\right|}{\mathcal{N}\left(\boldsymbol{\xi}\right)}\\
=\left\{ -ih_{k\sigma k^{\prime}\sigma^{\prime}}^{\textrm{pp}*}\sum_{\beta}\xi_{k\sigma,\beta}^{*}\frac{\partial}{\partial\xi_{k^{\prime}\sigma^{\prime}\beta}^{*}}+ih_{k\sigma k^{\prime}\sigma^{\prime}}^{\textrm{pp}}\sum_{\beta}\xi_{k\sigma,\beta}\frac{\partial}{\partial\xi_{k^{\prime}\sigma^{\prime}\beta}}\right\} \frac{\left|\boldsymbol{\xi}\right\rangle \left\langle \boldsymbol{\xi}\right|}{\mathcal{N}\left(\boldsymbol{\xi}\right)}.
\end{multline}
Next, 
\begin{equation}
\widehat{c}_{k^{\prime}\sigma^{\prime}}^{\textrm{h}\dagger}\widehat{c}_{k\sigma}^{\textrm{h}}\left|\boldsymbol{\xi}\right\rangle =\sum_{\alpha}\xi_{\alpha,k\sigma}^{*}\frac{\partial}{\partial\xi_{\alpha k^{\prime}\sigma^{\prime}}^{*}}\left|\boldsymbol{\xi}\right\rangle ,
\end{equation}
\begin{equation}
h_{k\sigma k^{\prime}\sigma^{\prime}}^{\textrm{hh}*}\widehat{c}_{k^{\prime}\sigma^{\prime}}^{\textrm{h}\dagger}\widehat{c}_{k\sigma}^{\textrm{h}}\left|\boldsymbol{\xi}\right\rangle =h_{k\sigma k^{\prime}\sigma^{\prime}}^{\textrm{hh}*}\sum_{\alpha}\xi_{\alpha,k\sigma}^{*}\frac{\partial}{\partial\xi_{\alpha k^{\prime}\sigma^{\prime}}^{*}}\left|\boldsymbol{\xi}\right\rangle ,
\end{equation}
\begin{multline}
h_{k\sigma k^{\prime}\sigma^{\prime}}^{\textrm{hh}*}\widehat{c}_{k^{\prime}\sigma^{\prime}}^{\textrm{h}\dagger}\widehat{c}_{k\sigma}^{\textrm{h}}\mathcal{N}^{-1}\left(\boldsymbol{\xi}\right)\left|\boldsymbol{\xi}\right\rangle \\
=\left\{ h_{k\sigma k^{\prime}\sigma^{\prime}}^{\textrm{hh}*}\sum_{\alpha}\xi_{\alpha,k\sigma}^{*}\frac{\partial}{\partial\xi_{\alpha k^{\prime}\sigma^{\prime}}^{*}}-\mathcal{N}\left(\boldsymbol{\xi}\right)h_{k\sigma k^{\prime}\sigma^{\prime}}^{\textrm{hh}*}\sum_{\alpha}\xi_{\alpha,k\sigma}^{*}\frac{\partial}{\partial\xi_{\alpha k^{\prime}\sigma^{\prime}}^{*}}\textrm{det}\left(\left[\boldsymbol{I}_{\textrm{h}}+\boldsymbol{\xi}^{\dagger}\boldsymbol{\xi}\right]^{-1}\right)\right\} \mathcal{N}^{-1}\left(\boldsymbol{\xi}\right)\left|\boldsymbol{\xi}\right\rangle \\
=\left\{ h_{k\sigma k^{\prime}\sigma^{\prime}}^{\textrm{hh}*}\sum_{\alpha}\xi_{\alpha,k\sigma}^{*}\frac{\partial}{\partial\xi_{\alpha k^{\prime}\sigma^{\prime}}^{*}}+h_{k\sigma k^{\prime}\sigma^{\prime}}^{\textrm{hh}*}\sum_{\alpha}\xi_{\alpha,k\sigma}^{*}\left\langle \alpha\right|\boldsymbol{\xi}\left[\boldsymbol{I}_{\textrm{h}}+\boldsymbol{\xi}^{\dagger}\boldsymbol{\xi}\right]^{-1}\left|k^{\prime}\sigma^{\prime}\right\rangle \right\} \mathcal{N}^{-1}\left(\boldsymbol{\xi}\right)\left|\boldsymbol{\xi}\right\rangle \\
=\left\{ h_{k\sigma k^{\prime}\sigma^{\prime}}^{\textrm{hh}*}\sum_{\alpha}\xi_{\alpha,k\sigma}^{*}\frac{\partial}{\partial\xi_{\alpha k^{\prime}\sigma^{\prime}}^{*}}+h_{k\sigma k^{\prime}\sigma^{\prime}}^{\textrm{hh}*}\left\langle k\sigma\right|\boldsymbol{\xi}^{\dagger}\boldsymbol{\xi}\left[\boldsymbol{I}_{\textrm{h}}+\boldsymbol{\xi}^{\dagger}\boldsymbol{\xi}\right]^{-1}\left|k^{\prime}\sigma^{\prime}\right\rangle \right\} \mathcal{N}^{-1}\left(\boldsymbol{\xi}\right)\left|\boldsymbol{\xi}\right\rangle \\
=\left\{ h_{k\sigma k^{\prime}\sigma^{\prime}}^{\textrm{hh}*}\sum_{\alpha}\xi_{\alpha,k\sigma}^{*}\frac{\partial}{\partial\xi_{\alpha k^{\prime}\sigma^{\prime}}^{*}}+\textrm{Tr}\left\{ \boldsymbol{h}^{\textrm{hh}\dagger}\boldsymbol{\xi}^{\dagger}\boldsymbol{\xi}\left[\boldsymbol{I}_{\textrm{h}}+\boldsymbol{\xi}^{\dagger}\boldsymbol{\xi}\right]^{-1}\right\} \right\} \mathcal{N}^{-1}\left(\boldsymbol{\xi}\right)\left|\boldsymbol{\xi}\right\rangle \\
=\left\{ h_{k\sigma k^{\prime}\sigma^{\prime}}^{\textrm{hh}*}\sum_{\alpha}\xi_{\alpha,k\sigma}^{*}\frac{\partial}{\partial\xi_{\alpha k^{\prime}\sigma^{\prime}}^{*}}+\textrm{Tr}\left\{ \boldsymbol{h}^{\textrm{hh}}\left[\boldsymbol{I}_{\textrm{h}}+\boldsymbol{\xi}^{\dagger}\boldsymbol{\xi}\right]^{-1}\boldsymbol{\xi}^{\dagger}\boldsymbol{\xi}\right\} \right\} \mathcal{N}^{-1}\left(\boldsymbol{\xi}\right)\left|\boldsymbol{\xi}\right\rangle .
\end{multline}
Therefore, the commutator is 
\begin{multline}
-i\left[h_{k\sigma k^{\prime}\sigma^{\prime}}^{\textrm{hh}}\widehat{c}_{k\sigma}^{\textrm{h}\dagger}\widehat{c}_{k^{\prime}\sigma^{\prime}}^{\textrm{h}},\frac{\left|\boldsymbol{\xi}\right\rangle \left\langle \boldsymbol{\xi}\right|}{\mathcal{N}\left(\boldsymbol{\xi}\right)}\right]\\
=\left\{ -ih_{k\sigma k^{\prime}\sigma^{\prime}}^{\textrm{hh}*}\sum_{\alpha}\xi_{\alpha,k\sigma}^{*}\frac{\partial}{\partial\xi_{\alpha k^{\prime}\sigma^{\prime}}^{*}}-i\textrm{Tr}\left\{ \boldsymbol{h}^{\textrm{hh}}\left[\boldsymbol{I}_{\textrm{h}}+\boldsymbol{\xi}^{\dagger}\boldsymbol{\xi}\right]^{-1}\boldsymbol{\xi}^{\dagger}\boldsymbol{\xi}\right\} \right.\\
\left.+ih_{k\sigma k^{\prime}\sigma^{\prime}}^{\textrm{hh}}\sum_{\alpha}\xi_{\alpha,k\sigma}\frac{\partial}{\partial\xi_{\alpha k^{\prime}\sigma^{\prime}}}+i\textrm{Tr}\left\{ \boldsymbol{h}^{\textrm{hh}}\left[\boldsymbol{I}_{\textrm{h}}+\boldsymbol{\xi}^{\dagger}\boldsymbol{\xi}\right]^{-1}\boldsymbol{\xi}^{\dagger}\boldsymbol{\xi}\right\} \right\} \frac{\left|\boldsymbol{\xi}\right\rangle \left\langle \boldsymbol{\xi}\right|}{\mathcal{N}\left(\boldsymbol{\xi}\right)}\\
=\left\{ -ih_{k\sigma k^{\prime}\sigma^{\prime}}^{\textrm{hh}*}\sum_{\alpha}\xi_{\alpha,k\sigma}^{*}\frac{\partial}{\partial\xi_{\alpha k^{\prime}\sigma^{\prime}}^{*}}+ih_{k\sigma k^{\prime}\sigma^{\prime}}^{\textrm{hh}}\sum_{\alpha}\xi_{\alpha,k\sigma}\frac{\partial}{\partial\xi_{\alpha k^{\prime}\sigma^{\prime}}}\right\} \frac{\left|\boldsymbol{\xi}\right\rangle \left\langle \boldsymbol{\xi}\right|}{\mathcal{N}\left(\boldsymbol{\xi}\right)}.
\end{multline}
Next, 
\begin{multline}
h_{k\sigma k^{\prime}\sigma^{\prime}}^{\textrm{ph}\dagger*}\widehat{c}_{k^{\prime}\sigma^{\prime}}^{\textrm{p}\dagger}\widehat{c}_{k\sigma}^{\textrm{h}\dagger}\left|\boldsymbol{\xi}\right\rangle =-h_{k\sigma k^{\prime}\sigma^{\prime}}^{\textrm{ph}\dagger*}\frac{\partial}{\partial\xi_{k^{\prime}\sigma^{\prime},k\sigma}^{*}}\left|\boldsymbol{\xi}\right\rangle ,
\end{multline}
\begin{multline}
h_{k\sigma k^{\prime}\sigma^{\prime}}^{\textrm{ph}\dagger*}\widehat{c}_{k^{\prime}\sigma^{\prime}}^{\textrm{p}\dagger}\widehat{c}_{k\sigma}^{\textrm{h}\dagger}\mathcal{N}^{-1}\left(\boldsymbol{\xi}\right)\left|\boldsymbol{\xi}\right\rangle =\left[-h_{k\sigma k^{\prime}\sigma^{\prime}}^{\textrm{ph}\dagger*}\frac{\partial}{\partial\xi_{k^{\prime}\sigma^{\prime},k\sigma}^{*}}-\textrm{Tr}\left\{ \boldsymbol{h}^{\textrm{ph}}\left[\boldsymbol{I}_{\textrm{h}}+\boldsymbol{\xi}^{T}\boldsymbol{\xi}^{*}\right]^{-1}\boldsymbol{\xi}^{T}\right\} \right]\mathcal{N}^{-1}\left(\boldsymbol{\xi}\right)\left|\boldsymbol{\xi}\right\rangle \\
=\left[-h_{k\sigma k^{\prime}\sigma^{\prime}}^{\textrm{ph}\dagger*}\frac{\partial}{\partial\xi_{k^{\prime}\sigma^{\prime},k\sigma}^{*}}-\textrm{Tr}\left\{ \left[\boldsymbol{I}_{\textrm{h}}+\boldsymbol{\xi}^{T}\boldsymbol{\xi}^{*}\right]^{-1}\boldsymbol{\xi}^{T}\boldsymbol{h}^{\textrm{ph}}\right\} \right]\mathcal{N}^{-1}\left(\boldsymbol{\xi}\right)\left|\boldsymbol{\xi}\right\rangle \\
=\left[-h_{k\sigma k^{\prime}\sigma^{\prime}}^{\textrm{ph}\dagger*}\frac{\partial}{\partial\xi_{k^{\prime}\sigma^{\prime},k\sigma}^{*}}-\textrm{Tr}\left\{ \boldsymbol{h}^{\textrm{ph}T}\boldsymbol{\xi}\left[\boldsymbol{I}_{\textrm{h}}+\boldsymbol{\xi}^{\dagger}\boldsymbol{\xi}\right]^{-1}\right\} \right]\mathcal{N}^{-1}\left(\boldsymbol{\xi}\right)\left|\boldsymbol{\xi}\right\rangle .
\end{multline}
Next,
\begin{multline}
h_{k\sigma k^{\prime}\sigma^{\prime}}^{\textrm{ph}*}\widehat{c}_{k^{\prime}\sigma^{\prime}}^{\textrm{h}}\widehat{c}_{k\sigma}^{\textrm{p}}\mathcal{N}^{-1}\left(\boldsymbol{\xi}\right)\left|\boldsymbol{\xi}\right\rangle =\left(h_{k\sigma k^{\prime}\sigma^{\prime}}^{\textrm{ph}*}\sum_{\alpha\beta}\xi_{k\sigma,\beta}^{*}\xi_{\alpha,k^{\prime}\sigma^{\prime}}^{*}\frac{\partial}{\partial\xi_{\alpha\beta}^{*}}-\textrm{Tr}\left\{ \boldsymbol{h}^{\textrm{ph}T*}\boldsymbol{\xi}^{*}\left[\boldsymbol{I}_{\textrm{h}}+\boldsymbol{\xi}^{\dagger*}\boldsymbol{\xi}^{*}\right]^{-1}\right\} \right)\mathcal{N}^{-1}\left(\boldsymbol{\xi}\right)\left|\boldsymbol{\xi}\right\rangle \\
=\left(h_{k\sigma k^{\prime}\sigma^{\prime}}^{\textrm{ph}*}\sum_{\alpha\beta}\xi_{k\sigma,\beta}^{*}\xi_{\alpha,k^{\prime}\sigma^{\prime}}^{*}\frac{\partial}{\partial\xi_{\alpha\beta}^{*}}-\textrm{Tr}\left\{ \left[\boldsymbol{I}_{\textrm{h}}+\boldsymbol{\xi}^{\dagger}\boldsymbol{\xi}\right]^{-1}\boldsymbol{\xi}^{\dagger}\boldsymbol{h}^{\textrm{ph}T\dagger}\right\} \right)\mathcal{N}^{-1}\left(\boldsymbol{\xi}\right)\left|\boldsymbol{\xi}\right\rangle .
\end{multline}
Therefore, the commutator is:
\begin{multline}
-i\left[h_{k\sigma k^{\prime}\sigma^{\prime}}^{\textrm{ph}}\widehat{c}_{k\sigma}^{\textrm{p}\dagger}\widehat{c}_{k^{\prime}\sigma^{\prime}}^{\textrm{h}\dagger}+h_{k\sigma k^{\prime}\sigma^{\prime}}^{\textrm{ph}\dagger}\widehat{c}_{k\sigma}^{\textrm{h}}\widehat{c}_{k^{\prime}\sigma^{\prime}}^{\textrm{p}},\frac{\left|\boldsymbol{\xi}\right\rangle \left\langle \boldsymbol{\xi}\right|}{\mathcal{N}\left(\boldsymbol{\xi}\right)}\right]\\
=\left\{ -ih_{k\sigma k^{\prime}\sigma^{\prime}}^{\textrm{ph}*}\sum_{\alpha\beta}\xi_{k\sigma,\beta}^{*}\xi_{\alpha,k^{\prime}\sigma^{\prime}}^{*}\frac{\partial}{\partial\xi_{\alpha\beta}^{*}}+i\textrm{Tr}\left\{ \left[\boldsymbol{I}_{\textrm{h}}+\boldsymbol{\xi}^{\dagger}\boldsymbol{\xi}\right]^{-1}\boldsymbol{\xi}^{\dagger}\boldsymbol{h}^{\textrm{ph}T\dagger}\right\} \right.\\
+ih_{k\sigma k^{\prime}\sigma^{\prime}}^{\textrm{ph}\dagger*}\frac{\partial}{\partial\xi_{k^{\prime}\sigma^{\prime},k\sigma}^{*}}+i\textrm{Tr}\left\{ \boldsymbol{h}^{\textrm{ph}T}\boldsymbol{\xi}\left[\boldsymbol{I}_{\textrm{h}}+\boldsymbol{\xi}^{\dagger}\boldsymbol{\xi}\right]^{-1}\right\} \\
+ih_{k\sigma k^{\prime}\sigma^{\prime}}^{\textrm{ph}}\sum_{\alpha\beta}\xi_{k\sigma,\beta}\xi_{\alpha,k^{\prime}\sigma^{\prime}}\frac{\partial}{\partial\xi_{\alpha\beta}}-i\textrm{Tr}\left\{ \boldsymbol{h}^{\textrm{ph}T}\boldsymbol{\xi}\left[\boldsymbol{I}_{\textrm{h}}+\boldsymbol{\xi}^{\dagger}\boldsymbol{\xi}\right]^{-1}\right\} \\
\left.-ih_{k\sigma k^{\prime}\sigma^{\prime}}^{\textrm{ph}\dagger}\frac{\partial}{\partial\xi_{k^{\prime}\sigma^{\prime},k\sigma}}-i\textrm{Tr}\left\{ \left[\boldsymbol{I}_{\textrm{h}}+\boldsymbol{\xi}^{\dagger}\boldsymbol{\xi}\right]^{-1}\boldsymbol{\xi}^{\dagger}\boldsymbol{h}^{\textrm{ph}T\dagger}\right\} \right\} \frac{\left|\boldsymbol{\xi}\right\rangle \left\langle \boldsymbol{\xi}\right|}{\mathcal{N}\left(\boldsymbol{\xi}\right)}\\
=\left\{ -ih_{k\sigma k^{\prime}\sigma^{\prime}}^{\textrm{ph}*}\sum_{\alpha\beta}\xi_{k\sigma,\beta}^{*}\xi_{\alpha,k^{\prime}\sigma^{\prime}}^{*}\frac{\partial}{\partial\xi_{\alpha\beta}^{*}}+ih_{k\sigma k^{\prime}\sigma^{\prime}}^{\textrm{ph}\dagger*}\frac{\partial}{\partial\xi_{k^{\prime}\sigma^{\prime},k\sigma}^{*}}\right.\\
\left.+h_{k\sigma k^{\prime}\sigma^{\prime}}^{\textrm{ph}}\sum_{\alpha\beta}\xi_{k\sigma,\beta}\xi_{\alpha,k^{\prime}\sigma^{\prime}}\frac{\partial}{\partial\xi_{\alpha\beta}}-ih_{k\sigma k^{\prime}\sigma^{\prime}}^{\textrm{ph}\dagger}\frac{\partial}{\partial\xi_{k^{\prime}\sigma^{\prime},k\sigma}}\right\} \frac{\left|\boldsymbol{\xi}\right\rangle \left\langle \boldsymbol{\xi}\right|}{\mathcal{N}\left(\boldsymbol{\xi}\right)}.
\end{multline}

\subsubsection{Change the order of derivatives and coefficients}

We have for commutators with the reordered derivatives:

\begin{multline}
-i\left[h_{k\sigma k^{\prime}\sigma^{\prime}}^{\textrm{pp}}\widehat{c}_{k\sigma}^{\textrm{p}\dagger}\widehat{c}_{k^{\prime}\sigma^{\prime}}^{\textrm{p}},\frac{\left|\boldsymbol{\xi}\right\rangle \left\langle \boldsymbol{\xi}\right|}{\mathcal{N}\left(\boldsymbol{\xi}\right)}\right]\\
=\left\{ -i\frac{\partial}{\partial\xi_{k^{\prime}\sigma^{\prime}\beta}^{*}}h_{k\sigma k^{\prime}\sigma^{\prime}}^{\textrm{pp}*}\xi_{k\sigma,\beta}^{*}+i\frac{\partial}{\partial\xi_{k^{\prime}\sigma^{\prime}\beta}}h_{k\sigma k^{\prime}\sigma^{\prime}}^{\textrm{pp}}\sum_{\beta}\xi_{k\sigma,\beta}\right\} \frac{\left|\boldsymbol{\xi}\right\rangle \left\langle \boldsymbol{\xi}\right|}{\mathcal{N}\left(\boldsymbol{\xi}\right)}\\
+\left\{ ih_{k\sigma k\sigma}^{\textrm{pp}*}M_{\textrm{h}}-ih_{k\sigma k\sigma}^{\textrm{pp}}M_{\textrm{h}}\right\} \frac{\left|\boldsymbol{\xi}\right\rangle \left\langle \boldsymbol{\xi}\right|}{\mathcal{N}\left(\boldsymbol{\xi}\right)}\\
\left(\textrm{the last line is zero since diagonal elements are real for Hermitian matrix}\right)\\
=\left\{ -i\frac{\partial}{\partial\xi_{k^{\prime}\sigma^{\prime}\beta}^{*}}h_{k\sigma k^{\prime}\sigma^{\prime}}^{\textrm{pp}*}\xi_{k\sigma,\beta}^{*}+i\frac{\partial}{\partial\xi_{k^{\prime}\sigma^{\prime}\beta}}h_{k\sigma k^{\prime}\sigma^{\prime}}^{\textrm{pp}}\sum_{\beta}\xi_{k\sigma,\beta}\right\} \frac{\left|\boldsymbol{\xi}\right\rangle \left\langle \boldsymbol{\xi}\right|}{\mathcal{N}\left(\boldsymbol{\xi}\right)}.
\end{multline}
Next, 
\begin{multline}
-i\left[h_{k\sigma k^{\prime}\sigma^{\prime}}^{\textrm{hh}}\widehat{c}_{k\sigma}^{\textrm{h}\dagger}\widehat{c}_{k^{\prime}\sigma^{\prime}}^{\textrm{h}},\frac{\left|\boldsymbol{\xi}\right\rangle \left\langle \boldsymbol{\xi}\right|}{\mathcal{N}\left(\boldsymbol{\xi}\right)}\right]\\
=\left\{ -i\frac{\partial}{\partial\xi_{\alpha k^{\prime}\sigma^{\prime}}^{*}}h_{k\sigma k^{\prime}\sigma^{\prime}}^{\textrm{hh}*}\xi_{\alpha,k\sigma}^{*}+i\frac{\partial}{\partial\xi_{\alpha k^{\prime}\sigma^{\prime}}}h_{k\sigma k^{\prime}\sigma^{\prime}}^{\textrm{hh}}\xi_{\alpha,k\sigma}\right\} \frac{\left|\boldsymbol{\xi}\right\rangle \left\langle \boldsymbol{\xi}\right|}{\mathcal{N}\left(\boldsymbol{\xi}\right)}\\
+\left\{ ih_{k\sigma k\sigma}^{\textrm{hh}*}M_{\textrm{p}}-ih_{k\sigma k\sigma}^{\textrm{hh}}M_{\textrm{p}}\right\} \frac{\left|\boldsymbol{\xi}\right\rangle \left\langle \boldsymbol{\xi}\right|}{\mathcal{N}\left(\boldsymbol{\xi}\right)}\\
=\left\{ -i\frac{\partial}{\partial\xi_{\alpha k^{\prime}\sigma^{\prime}}^{*}}h_{k\sigma k^{\prime}\sigma^{\prime}}^{\textrm{hh}*}\xi_{\alpha,k\sigma}^{*}+i\frac{\partial}{\partial\xi_{\alpha k^{\prime}\sigma^{\prime}}}h_{k\sigma k^{\prime}\sigma^{\prime}}^{\textrm{hh}}\xi_{\alpha,k\sigma}\right\} \frac{\left|\boldsymbol{\xi}\right\rangle \left\langle \boldsymbol{\xi}\right|}{\mathcal{N}\left(\boldsymbol{\xi}\right)}.
\end{multline}
Next, 
\begin{multline}
-i\left[h_{k\sigma k^{\prime}\sigma^{\prime}}^{\textrm{ph}}\widehat{c}_{k\sigma}^{\textrm{p}\dagger}\widehat{c}_{k^{\prime}\sigma^{\prime}}^{\textrm{h}\dagger}+h_{k\sigma k^{\prime}\sigma^{\prime}}^{\textrm{ph}\dagger}\widehat{c}_{k\sigma}^{\textrm{h}}\widehat{c}_{k^{\prime}\sigma^{\prime}}^{\textrm{p}},\frac{\left|\boldsymbol{\xi}\right\rangle \left\langle \boldsymbol{\xi}\right|}{\mathcal{N}\left(\boldsymbol{\xi}\right)}\right]\\
=\left\{ -i\sum_{\alpha\beta}\frac{\partial}{\partial\xi_{\alpha\beta}^{*}}h_{k\sigma k^{\prime}\sigma^{\prime}}^{\textrm{ph}*}\xi_{k\sigma,\beta}^{*}\xi_{\alpha,k^{\prime}\sigma^{\prime}}^{*}+ih_{k\sigma k^{\prime}\sigma^{\prime}}^{\textrm{ph}\dagger*}\frac{\partial}{\partial\xi_{k^{\prime}\sigma^{\prime},k\sigma}^{*}}\right.\\
\left.+i\sum_{\alpha\beta}\frac{\partial}{\partial\xi_{\alpha\beta}}h_{k\sigma k^{\prime}\sigma^{\prime}}^{\textrm{ph}}\xi_{k\sigma,\beta}\xi_{\alpha,k^{\prime}\sigma^{\prime}}-ih_{k\sigma k^{\prime}\sigma^{\prime}}^{\textrm{ph}\dagger}\frac{\partial}{\partial\xi_{k^{\prime}\sigma^{\prime},k\sigma}}\right\} \frac{\left|\boldsymbol{\xi}\right\rangle \left\langle \boldsymbol{\xi}\right|}{\mathcal{N}\left(\boldsymbol{\xi}\right)}\\
+\left\{ ih_{k\sigma k^{\prime}\sigma^{\prime}}^{\textrm{ph}*}\xi_{k\sigma,k^{\prime}\sigma^{\prime}}^{*}M_{\textrm{h}}+ih_{k\sigma k^{\prime}\sigma^{\prime}}^{\textrm{ph}*}\xi_{k\sigma,k^{\prime}\sigma^{\prime}}^{*}M_{\textrm{p}}-ih_{k\sigma k^{\prime}\sigma^{\prime}}^{\textrm{ph}}\xi_{k\sigma,k^{\prime}\sigma^{\prime}}M_{\textrm{h}}-ih_{k\sigma k^{\prime}\sigma^{\prime}}^{\textrm{ph}}\xi_{k\sigma,k^{\prime}\sigma^{\prime}}M_{\textrm{p}}\right\} \frac{\left|\boldsymbol{\xi}\right\rangle \left\langle \boldsymbol{\xi}\right|}{\mathcal{N}\left(\boldsymbol{\xi}\right)}\\
=\left\{ -i\sum_{\alpha\beta}\frac{\partial}{\partial\xi_{\alpha\beta}^{*}}h_{k\sigma k^{\prime}\sigma^{\prime}}^{\textrm{ph}*}\xi_{k\sigma,\beta}^{*}\xi_{\alpha,k^{\prime}\sigma^{\prime}}^{*}+ih_{k\sigma k^{\prime}\sigma^{\prime}}^{\textrm{ph}\dagger*}\frac{\partial}{\partial\xi_{k^{\prime}\sigma^{\prime},k\sigma}^{*}}\right.\\
\left.+i\sum_{\alpha\beta}\frac{\partial}{\partial\xi_{\alpha\beta}}h_{k\sigma k^{\prime}\sigma^{\prime}}^{\textrm{ph}}\xi_{k\sigma,\beta}\xi_{\alpha,k^{\prime}\sigma^{\prime}}-ih_{k\sigma k^{\prime}\sigma^{\prime}}^{\textrm{ph}\dagger}\frac{\partial}{\partial\xi_{k^{\prime}\sigma^{\prime},k\sigma}}\right\} \frac{\left|\boldsymbol{\xi}\right\rangle \left\langle \boldsymbol{\xi}\right|}{\mathcal{N}\left(\boldsymbol{\xi}\right)}\\
+Mi\left\{ h_{k\sigma k^{\prime}\sigma^{\prime}}^{\textrm{ph}*}\xi_{k\sigma,k^{\prime}\sigma^{\prime}}^{*}-h_{k\sigma k^{\prime}\sigma^{\prime}}^{\textrm{ph}}\xi_{k\sigma,k^{\prime}\sigma^{\prime}}\right\} \frac{\left|\boldsymbol{\xi}\right\rangle \left\langle \boldsymbol{\xi}\right|}{\mathcal{N}\left(\boldsymbol{\xi}\right)}.
\end{multline}

\subsection{\label{subsec:Covariant-form-of}Covariant form of the differential
correspondences for commutators}

Now we should reorder 
\begin{equation}
\det g_{\alpha,\overline{\beta}}=\det\left\{ \left(\boldsymbol{I}_{\textrm{h}}+\boldsymbol{\xi}^{\dagger}\boldsymbol{\xi}\right)^{-1}\right\} ^{M}=\det\left\{ \left(\boldsymbol{I}_{\textrm{p}}+\boldsymbol{\xi}\boldsymbol{\xi}^{\dagger}\right)^{-1}\right\} ^{M}.
\end{equation}
and the partial derivatives. We have:
\begin{multline}
\det g_{\alpha,\overline{\beta}}\frac{\partial}{\partial\xi_{\alpha\beta}}=\frac{\partial}{\partial\xi_{\alpha\beta}}\det g_{\alpha,\overline{\beta}}-\left(\frac{\partial}{\partial\xi_{\alpha\beta}}\det g_{\alpha,\overline{\beta}}\right)\\
=\frac{\partial}{\partial\xi_{\alpha\beta}}\det g_{\alpha,\overline{\beta}}-\left(\frac{\partial}{\partial\xi_{\alpha\beta}}\det\left\{ \left(\boldsymbol{I}_{\textrm{h}}+\boldsymbol{\xi}^{\dagger}\boldsymbol{\xi}\right)^{-1}\right\} ^{M}\right)\\
=\frac{\partial}{\partial\xi_{\alpha\beta}}\det g_{\alpha,\overline{\beta}}-\left(\frac{\partial}{\partial\xi_{\alpha\beta}}\left\{ \mathcal{N}^{-1}\left(\boldsymbol{\xi}\right)\right\} ^{M}\right)\\
=\frac{\partial}{\partial\xi_{\alpha\beta}}\det g_{\alpha,\overline{\beta}}-M\left\{ \mathcal{N}^{-1}\left(\boldsymbol{\xi}\right)\right\} ^{M-1}\left(\frac{\partial}{\partial\xi_{\alpha\beta}}\mathcal{N}^{-1}\left(\boldsymbol{\xi}\right)\right)\\
=\frac{\partial}{\partial\xi_{\alpha\beta}}\det g_{\alpha,\overline{\beta}}+M\left\{ \mathcal{N}^{-1}\left(\boldsymbol{\xi}\right)\right\} ^{M-1}\mathcal{N}^{-1}\left(\boldsymbol{\xi}\right)\left\langle \beta\right|\left[\boldsymbol{I}_{\textrm{h}}+\boldsymbol{\xi}^{\dagger}\boldsymbol{\xi}\right]^{-1}\boldsymbol{\xi}^{\dagger}\left|\alpha\right\rangle \\
=\frac{\partial}{\partial\xi_{\alpha\beta}}\det g_{\alpha,\overline{\beta}}+M\det g_{\alpha,\overline{\beta}}\left\langle \beta\right|\left[\boldsymbol{I}_{\textrm{h}}+\boldsymbol{\xi}^{\dagger}\boldsymbol{\xi}\right]^{-1}\boldsymbol{\xi}^{\dagger}\left|\alpha\right\rangle .
\end{multline}
Or:
\begin{multline}
\det g_{\alpha,\overline{\beta}}\frac{\partial}{\partial\xi_{\alpha\beta}}\\
=\frac{\partial}{\partial\xi_{\alpha\beta}}\det g_{\alpha,\overline{\beta}}+2M\det g_{\alpha,\overline{\beta}}\left\langle \beta\right|\boldsymbol{\xi}^{\dagger}\left[\boldsymbol{I}_{\textrm{p}}+\boldsymbol{\xi}\boldsymbol{\xi}^{\dagger}\right]^{-1}\left|\alpha\right\rangle .
\end{multline}
The conjugated equality:
\begin{multline}
\det g_{\alpha,\overline{\beta}}\frac{\partial}{\partial\xi_{\alpha\beta}^{*}}=\frac{\partial}{\partial\xi_{\alpha\beta}^{*}}\det g_{\alpha,\overline{\beta}}-\left(\frac{\partial}{\partial\xi_{\alpha\beta}^{*}}\det g_{\alpha,\overline{\beta}}\right)\\
=\frac{\partial}{\partial\xi_{\alpha\beta}^{*}}\det g_{\alpha,\overline{\beta}}-M\left\{ \mathcal{N}^{-1}\left(\boldsymbol{\xi}\right)\right\} ^{M-1}\left(\frac{\partial}{\partial\xi_{\alpha\beta}^{*}}\mathcal{N}^{-1}\left(\boldsymbol{\xi}\right)\right)\\
=\frac{\partial}{\partial\xi_{\alpha\beta}^{*}}\det g_{\alpha,\overline{\beta}}+M\det g_{\alpha,\overline{\beta}}\left\langle \alpha\right|\boldsymbol{\xi}\left[\boldsymbol{I}_{\textrm{h}}+\boldsymbol{\xi}^{\dagger}\boldsymbol{\xi}\right]^{-1}\left|\beta\right\rangle ,
\end{multline}
or
\begin{multline}
\det g_{\alpha,\overline{\beta}}\frac{\partial}{\partial\xi_{\alpha\beta}^{*}}\\
=\frac{\partial}{\partial\xi_{\alpha\beta}^{*}}\det g_{\alpha,\overline{\beta}}+M\det g_{\alpha,\overline{\beta}}\left\langle \alpha\right|\left[\boldsymbol{I}_{\textrm{p}}+\boldsymbol{\xi}\boldsymbol{\xi}^{\dagger}\right]^{-1}\boldsymbol{\xi}\left|\beta\right\rangle .
\end{multline}
We substitute these to the equations for the commutators:
\begin{multline}
-i\left[h_{k\sigma k^{\prime}\sigma^{\prime}}^{\textrm{pp}}\widehat{c}_{k\sigma}^{\textrm{p}\dagger}\widehat{c}_{k^{\prime}\sigma^{\prime}}^{\textrm{p}},\det g_{\alpha,\overline{\beta}}\frac{\left|\boldsymbol{\xi}\right\rangle \left\langle \boldsymbol{\xi}\right|}{\mathcal{N}\left(\boldsymbol{\xi}\right)}\right]\\
=\left\{ -i\frac{\partial}{\partial\xi_{k^{\prime}\sigma^{\prime}\beta}^{*}}h_{k\sigma k^{\prime}\sigma^{\prime}}^{\textrm{pp}*}\xi_{k\sigma,\beta}^{*}+i\frac{\partial}{\partial\xi_{k^{\prime}\sigma^{\prime}\beta}}h_{k\sigma k^{\prime}\sigma^{\prime}}^{\textrm{pp}}\xi_{k\sigma,\beta}\right\} \det g_{\alpha,\overline{\beta}}\frac{\left|\boldsymbol{\xi}\right\rangle \left\langle \boldsymbol{\xi}\right|}{\mathcal{N}\left(\boldsymbol{\xi}\right)}\\
+Mi\left\{ -h_{k\sigma k^{\prime}\sigma^{\prime}}^{\textrm{pp}*}\xi_{k\sigma,\beta}^{*}\left\langle k^{\prime}\sigma^{\prime}\right|\left[\boldsymbol{I}_{\textrm{p}}+\boldsymbol{\xi}\boldsymbol{\xi}^{\dagger}\right]^{-1}\boldsymbol{\xi}\left|\beta\right\rangle +h_{k\sigma k^{\prime}\sigma^{\prime}}^{\textrm{pp}}\xi_{k\sigma,\beta}\left\langle \beta\right|\boldsymbol{\xi}^{\dagger}\left[\boldsymbol{I}_{\textrm{p}}+\boldsymbol{\xi}\boldsymbol{\xi}^{\dagger}\right]^{-1}\left|k^{\prime}\sigma^{\prime}\right\rangle \right\} \det g_{\alpha,\overline{\beta}}\frac{\left|\boldsymbol{\xi}\right\rangle \left\langle \boldsymbol{\xi}\right|}{\mathcal{N}\left(\boldsymbol{\xi}\right)}\\
=\left\{ -i\frac{\partial}{\partial\xi_{k^{\prime}\sigma^{\prime}\beta}^{*}}h_{k\sigma k^{\prime}\sigma^{\prime}}^{\textrm{pp}*}\xi_{k\sigma,\beta}^{*}+i\frac{\partial}{\partial\xi_{k^{\prime}\sigma^{\prime}\beta}}h_{k\sigma k^{\prime}\sigma^{\prime}}^{\textrm{pp}}\xi_{k\sigma,\beta}\right\} \det g_{\alpha,\overline{\beta}}\frac{\left|\boldsymbol{\xi}\right\rangle \left\langle \boldsymbol{\xi}\right|}{\mathcal{N}\left(\boldsymbol{\xi}\right)}\\
+Mi\left\{ -h_{k\sigma k^{\prime}\sigma^{\prime}}^{\textrm{pp}*}\left\langle k^{\prime}\sigma^{\prime}\right|\left[\boldsymbol{I}_{\textrm{p}}+\boldsymbol{\xi}\boldsymbol{\xi}^{\dagger}\right]^{-1}\boldsymbol{\xi}\boldsymbol{\xi}^{\dagger}\left|k\sigma\right\rangle +h_{k\sigma k^{\prime}\sigma^{\prime}}^{\textrm{pp}}\left\langle k\sigma\right|\boldsymbol{\xi}\boldsymbol{\xi}^{\dagger}\left[\boldsymbol{I}_{\textrm{p}}+\boldsymbol{\xi}\boldsymbol{\xi}^{\dagger}\right]^{-1}\left|k^{\prime}\sigma^{\prime}\right\rangle \right\} \det g_{\alpha,\overline{\beta}}\frac{\left|\boldsymbol{\xi}\right\rangle \left\langle \boldsymbol{\xi}\right|}{\mathcal{N}\left(\boldsymbol{\xi}\right)}\\
=\left\{ -i\frac{\partial}{\partial\xi_{k^{\prime}\sigma^{\prime}\beta}^{*}}h_{k\sigma k^{\prime}\sigma^{\prime}}^{\textrm{pp}*}\xi_{k\sigma,\beta}^{*}+i\frac{\partial}{\partial\xi_{k^{\prime}\sigma^{\prime}\beta}}h_{k\sigma k^{\prime}\sigma^{\prime}}^{\textrm{pp}}\xi_{k\sigma,\beta}\right\} \det g_{\alpha,\overline{\beta}}\frac{\left|\boldsymbol{\xi}\right\rangle \left\langle \boldsymbol{\xi}\right|}{\mathcal{N}\left(\boldsymbol{\xi}\right)}\\
+Mi\left\{ -\textrm{Tr}\left\{ \boldsymbol{h}^{\textrm{pp}*}\left[\boldsymbol{I}_{\textrm{p}}+\boldsymbol{\xi}\boldsymbol{\xi}^{\dagger}\right]^{-1}\boldsymbol{\xi}\boldsymbol{\xi}^{\dagger}\right\} +\textrm{Tr}\left\{ \boldsymbol{\xi}\boldsymbol{\xi}^{\dagger}\left[\boldsymbol{I}_{\textrm{p}}+\boldsymbol{\xi}\boldsymbol{\xi}^{\dagger}\right]^{-1}\boldsymbol{h}^{\textrm{pp}*}\right\} \right\} \det g_{\alpha,\overline{\beta}}\frac{\left|\boldsymbol{\xi}\right\rangle \left\langle \boldsymbol{\xi}\right|}{\mathcal{N}\left(\boldsymbol{\xi}\right)}\\
=\left\{ -i\frac{\partial}{\partial\xi_{k^{\prime}\sigma^{\prime}\beta}^{*}}h_{k\sigma k^{\prime}\sigma^{\prime}}^{\textrm{pp}*}\xi_{k\sigma,\beta}^{*}+i\frac{\partial}{\partial\xi_{k^{\prime}\sigma^{\prime}\beta}}h_{k\sigma k^{\prime}\sigma^{\prime}}^{\textrm{pp}}\xi_{k\sigma,\beta}\right\} \det g_{\alpha,\overline{\beta}}\frac{\left|\boldsymbol{\xi}\right\rangle \left\langle \boldsymbol{\xi}\right|}{\mathcal{N}\left(\boldsymbol{\xi}\right)}.
\end{multline}
Next:

\begin{multline}
-i\left[h_{k\sigma k^{\prime}\sigma^{\prime}}^{\textrm{hh}}\widehat{c}_{k\sigma}^{\textrm{h}\dagger}\widehat{c}_{k^{\prime}\sigma^{\prime}}^{\textrm{h}},\det g_{\alpha,\overline{\beta}}\frac{\left|\boldsymbol{\xi}\right\rangle \left\langle \boldsymbol{\xi}\right|}{\mathcal{N}\left(\boldsymbol{\xi}\right)}\right]\\
=\left\{ -i\frac{\partial}{\partial\xi_{\alpha k^{\prime}\sigma^{\prime}}^{*}}h_{k\sigma k^{\prime}\sigma^{\prime}}^{\textrm{hh}*}\xi_{\alpha,k\sigma}^{*}+i\frac{\partial}{\partial\xi_{\alpha k^{\prime}\sigma^{\prime}}}h_{k\sigma k^{\prime}\sigma^{\prime}}^{\textrm{hh}}\xi_{\alpha,k\sigma}\right\} \det g_{\alpha,\overline{\beta}}\frac{\left|\boldsymbol{\xi}\right\rangle \left\langle \boldsymbol{\xi}\right|}{\mathcal{N}\left(\boldsymbol{\xi}\right)}\\
+Mi\left\{ -h_{k\sigma k^{\prime}\sigma^{\prime}}^{\textrm{hh}*}\xi_{\alpha,k\sigma}^{*}\left\langle \alpha\right|\boldsymbol{\xi}\left[\boldsymbol{I}_{\textrm{h}}+\boldsymbol{\xi}^{\dagger}\boldsymbol{\xi}\right]^{-1}\left|k^{\prime}\sigma^{\prime}\right\rangle +h_{k\sigma k^{\prime}\sigma^{\prime}}^{\textrm{hh}}\xi_{\alpha,k\sigma}\left\langle k^{\prime}\sigma^{\prime}\right|\left[\boldsymbol{I}_{\textrm{h}}+\boldsymbol{\xi}^{\dagger}\boldsymbol{\xi}\right]^{-1}\boldsymbol{\xi}^{\dagger}\left|\alpha\right\rangle \right\} \det g_{\alpha,\overline{\beta}}\frac{\left|\boldsymbol{\xi}\right\rangle \left\langle \boldsymbol{\xi}\right|}{\mathcal{N}\left(\boldsymbol{\xi}\right)}\\
=\left\{ -i\frac{\partial}{\partial\xi_{\alpha k^{\prime}\sigma^{\prime}}^{*}}h_{k\sigma k^{\prime}\sigma^{\prime}}^{\textrm{hh}*}\xi_{\alpha,k\sigma}^{*}+i\frac{\partial}{\partial\xi_{\alpha k^{\prime}\sigma^{\prime}}}h_{k\sigma k^{\prime}\sigma^{\prime}}^{\textrm{hh}}\xi_{\alpha,k\sigma}\right\} \det g_{\alpha,\overline{\beta}}\frac{\left|\boldsymbol{\xi}\right\rangle \left\langle \boldsymbol{\xi}\right|}{\mathcal{N}\left(\boldsymbol{\xi}\right)}\\
+Mi\left\{ -h_{k\sigma k^{\prime}\sigma^{\prime}}^{\textrm{hh}*}\left\langle k\sigma\right|\boldsymbol{\xi}^{\dagger}\boldsymbol{\xi}\left[\boldsymbol{I}_{\textrm{h}}+\boldsymbol{\xi}^{\dagger}\boldsymbol{\xi}\right]^{-1}\left|k^{\prime}\sigma^{\prime}\right\rangle +h_{k\sigma k^{\prime}\sigma^{\prime}}^{\textrm{hh}}\left\langle k^{\prime}\sigma^{\prime}\right|\left[\boldsymbol{I}_{\textrm{h}}+\boldsymbol{\xi}^{\dagger}\boldsymbol{\xi}\right]^{-1}\boldsymbol{\xi}^{\dagger}\boldsymbol{\xi}\left|k\sigma\right\rangle \right\} \det g_{\alpha,\overline{\beta}}\frac{\left|\boldsymbol{\xi}\right\rangle \left\langle \boldsymbol{\xi}\right|}{\mathcal{N}\left(\boldsymbol{\xi}\right)}\\
=\left\{ -i\frac{\partial}{\partial\xi_{\alpha k^{\prime}\sigma^{\prime}}^{*}}h_{k\sigma k^{\prime}\sigma^{\prime}}^{\textrm{hh}*}\xi_{\alpha,k\sigma}^{*}+i\frac{\partial}{\partial\xi_{\alpha k^{\prime}\sigma^{\prime}}}h_{k\sigma k^{\prime}\sigma^{\prime}}^{\textrm{hh}}\xi_{\alpha,k\sigma}\right\} \det g_{\alpha,\overline{\beta}}\frac{\left|\boldsymbol{\xi}\right\rangle \left\langle \boldsymbol{\xi}\right|}{\mathcal{N}\left(\boldsymbol{\xi}\right)}\\
+Mi\left\{ -\textrm{Tr}\left\{ \boldsymbol{\xi}^{\dagger}\boldsymbol{\xi}\left[\boldsymbol{I}_{\textrm{h}}+\boldsymbol{\xi}^{\dagger}\boldsymbol{\xi}\right]^{-1}\boldsymbol{h}^{\textrm{hh}\dagger}\right\} +\textrm{Tr}\left\{ \boldsymbol{h}^{\textrm{hh}}\left[\boldsymbol{I}_{\textrm{h}}+\boldsymbol{\xi}^{\dagger}\boldsymbol{\xi}\right]^{-1}\boldsymbol{\xi}^{\dagger}\boldsymbol{\xi}\right\} \right\} \det g_{\alpha,\overline{\beta}}\frac{\left|\boldsymbol{\xi}\right\rangle \left\langle \boldsymbol{\xi}\right|}{\mathcal{N}\left(\boldsymbol{\xi}\right)}\\
=\left\{ -i\frac{\partial}{\partial\xi_{\alpha k^{\prime}\sigma^{\prime}}^{*}}h_{k\sigma k^{\prime}\sigma^{\prime}}^{\textrm{hh}*}\xi_{\alpha,k\sigma}^{*}+i\frac{\partial}{\partial\xi_{\alpha k^{\prime}\sigma^{\prime}}}h_{k\sigma k^{\prime}\sigma^{\prime}}^{\textrm{hh}}\xi_{\alpha,k\sigma}\right\} \det g_{\alpha,\overline{\beta}}\frac{\left|\boldsymbol{\xi}\right\rangle \left\langle \boldsymbol{\xi}\right|}{\mathcal{N}\left(\boldsymbol{\xi}\right)}.
\end{multline}
The most difficult commutator:
\begin{multline}
-i\left[h_{k\sigma k^{\prime}\sigma^{\prime}}^{\textrm{ph}}\widehat{c}_{k\sigma}^{\textrm{p}\dagger}\widehat{c}_{k^{\prime}\sigma^{\prime}}^{\textrm{h}\dagger}+h_{k\sigma k^{\prime}\sigma^{\prime}}^{\textrm{ph}\dagger}\widehat{c}_{k\sigma}^{\textrm{h}}\widehat{c}_{k^{\prime}\sigma^{\prime}}^{\textrm{p}},\det g_{\alpha,\overline{\beta}}\frac{\left|\boldsymbol{\xi}\right\rangle \left\langle \boldsymbol{\xi}\right|}{\mathcal{N}\left(\boldsymbol{\xi}\right)}\right]\\
=\left\{ -i\sum_{\alpha\beta}\frac{\partial}{\partial\xi_{\alpha\beta}^{*}}h_{k\sigma k^{\prime}\sigma^{\prime}}^{\textrm{ph}*}\xi_{k\sigma,\beta}^{*}\xi_{\alpha,k^{\prime}\sigma^{\prime}}^{*}+ih_{k\sigma k^{\prime}\sigma^{\prime}}^{\textrm{ph}\dagger*}\frac{\partial}{\partial\xi_{k^{\prime}\sigma^{\prime},k\sigma}^{*}}\right.\\
\left.+i\sum_{\alpha\beta}\frac{\partial}{\partial\xi_{\alpha\beta}}h_{k\sigma k^{\prime}\sigma^{\prime}}^{\textrm{ph}}\xi_{k\sigma,\beta}\xi_{\alpha,k^{\prime}\sigma^{\prime}}-ih_{k\sigma k^{\prime}\sigma^{\prime}}^{\textrm{ph}\dagger}\frac{\partial}{\partial\xi_{k^{\prime}\sigma^{\prime},k\sigma}}\right\} \det g_{\alpha,\overline{\beta}}\frac{\left|\boldsymbol{\xi}\right\rangle \left\langle \boldsymbol{\xi}\right|}{\mathcal{N}\left(\boldsymbol{\xi}\right)}\\
+Mi\left\{ h_{k\sigma k^{\prime}\sigma^{\prime}}^{\textrm{ph}*}\xi_{k\sigma,k^{\prime}\sigma^{\prime}}^{*}-h_{k\sigma k^{\prime}\sigma^{\prime}}^{\textrm{ph}}\xi_{k\sigma,k^{\prime}\sigma^{\prime}}\right\} \det g_{\alpha,\overline{\beta}}\frac{\left|\boldsymbol{\xi}\right\rangle \left\langle \boldsymbol{\xi}\right|}{\mathcal{N}\left(\boldsymbol{\xi}\right)}\\
+Mi\left\{ -h_{k\sigma k^{\prime}\sigma^{\prime}}^{\textrm{ph}*}\xi_{k\sigma,\beta}^{*}\xi_{\alpha,k^{\prime}\sigma^{\prime}}^{*}\left\langle \alpha\right|\boldsymbol{\xi}\left[\boldsymbol{I}_{\textrm{h}}+\boldsymbol{\xi}^{\dagger}\boldsymbol{\xi}\right]^{-1}\left|\beta\right\rangle +h_{k\sigma k^{\prime}\sigma^{\prime}}^{\textrm{ph}\dagger*}\left\langle k^{\prime}\sigma^{\prime}\right|\boldsymbol{\xi}\left[\boldsymbol{I}_{\textrm{h}}+\boldsymbol{\xi}^{\dagger}\boldsymbol{\xi}\right]^{-1}\left|k\sigma\right\rangle \right\} \det g_{\alpha,\overline{\beta}}\frac{\left|\boldsymbol{\xi}\right\rangle \left\langle \boldsymbol{\xi}\right|}{\mathcal{N}\left(\boldsymbol{\xi}\right)}\\
+Mi\left\{ h_{k\sigma k^{\prime}\sigma^{\prime}}^{\textrm{ph}}\xi_{k\sigma,\beta}\xi_{\alpha,k^{\prime}\sigma^{\prime}}\left\langle \beta\right|\left[\boldsymbol{I}_{\textrm{h}}+\boldsymbol{\xi}^{\dagger}\boldsymbol{\xi}\right]^{-1}\boldsymbol{\xi}^{\dagger}\left|\alpha\right\rangle -h_{k\sigma k^{\prime}\sigma^{\prime}}^{\textrm{ph}\dagger}\left\langle k\sigma\right|\left[\boldsymbol{I}_{\textrm{h}}+\boldsymbol{\xi}^{\dagger}\boldsymbol{\xi}\right]^{-1}\boldsymbol{\xi}^{\dagger}\left|k^{\prime}\sigma^{\prime}\right\rangle \right\} \det g_{\alpha,\overline{\beta}}\frac{\left|\boldsymbol{\xi}\right\rangle \left\langle \boldsymbol{\xi}\right|}{\mathcal{N}\left(\boldsymbol{\xi}\right)}\\
=\left\{ -i\sum_{\alpha\beta}\frac{\partial}{\partial\xi_{\alpha\beta}^{*}}h_{k\sigma k^{\prime}\sigma^{\prime}}^{\textrm{ph}*}\xi_{k\sigma,\beta}^{*}\xi_{\alpha,k^{\prime}\sigma^{\prime}}^{*}+ih_{k\sigma k^{\prime}\sigma^{\prime}}^{\textrm{ph}\dagger*}\frac{\partial}{\partial\xi_{k^{\prime}\sigma^{\prime},k\sigma}^{*}}\right.\\
\left.+i\sum_{\alpha\beta}\frac{\partial}{\partial\xi_{\alpha\beta}}h_{k\sigma k^{\prime}\sigma^{\prime}}^{\textrm{ph}}\xi_{k\sigma,\beta}\xi_{\alpha,k^{\prime}\sigma^{\prime}}-ih_{k\sigma k^{\prime}\sigma^{\prime}}^{\textrm{ph}\dagger}\frac{\partial}{\partial\xi_{k^{\prime}\sigma^{\prime},k\sigma}}\right\} \det g_{\alpha,\overline{\beta}}\frac{\left|\boldsymbol{\xi}\right\rangle \left\langle \boldsymbol{\xi}\right|}{\mathcal{N}\left(\boldsymbol{\xi}\right)}\\
+Mi\left\{ \textrm{Tr}\left\{ \boldsymbol{h}^{\textrm{ph}*}\boldsymbol{\xi}^{\dagger}\right\} -\textrm{Tr}\left\{ \boldsymbol{h}^{\textrm{ph}*\dagger}\boldsymbol{\xi}\right\} \right\} \det g_{\alpha,\overline{\beta}}\frac{\left|\boldsymbol{\xi}\right\rangle \left\langle \boldsymbol{\xi}\right|}{\mathcal{N}\left(\boldsymbol{\xi}\right)}\\
+Mi\left\{ -\textrm{Tr}\left\{ \boldsymbol{h}^{\textrm{ph}*}\boldsymbol{\xi}^{\dagger}\boldsymbol{\xi}\left[\boldsymbol{I}_{\textrm{h}}+\boldsymbol{\xi}^{\dagger}\boldsymbol{\xi}\right]^{-1}\boldsymbol{\xi}^{\dagger}\right\} +\textrm{Tr}\left\{ \boldsymbol{h}^{\textrm{ph}*\dagger}\boldsymbol{\xi}\left[\boldsymbol{I}_{\textrm{h}}+\boldsymbol{\xi}^{\dagger}\boldsymbol{\xi}\right]^{-1}\right\} \right\} \det g_{\alpha,\overline{\beta}}\frac{\left|\boldsymbol{\xi}\right\rangle \left\langle \boldsymbol{\xi}\right|}{\mathcal{N}\left(\boldsymbol{\xi}\right)}\\
+Mi\left\{ \textrm{Tr}\left\{ \boldsymbol{h}^{\textrm{ph}*\dagger}\boldsymbol{\xi}\left[\boldsymbol{I}_{\textrm{h}}+\boldsymbol{\xi}^{\dagger}\boldsymbol{\xi}\right]^{-1}\boldsymbol{\xi}^{\dagger}\boldsymbol{\xi}\right\} -\textrm{Tr}\left\{ \boldsymbol{h}^{\textrm{ph}*}\left[\boldsymbol{I}_{\textrm{h}}+\boldsymbol{\xi}^{\dagger}\boldsymbol{\xi}\right]^{-1}\boldsymbol{\xi}^{\dagger}\right\} \right\} \det g_{\alpha,\overline{\beta}}\frac{\left|\boldsymbol{\xi}\right\rangle \left\langle \boldsymbol{\xi}\right|}{\mathcal{N}\left(\boldsymbol{\xi}\right)}\\
=\left\{ -i\sum_{\alpha\beta}\frac{\partial}{\partial\xi_{\alpha\beta}^{*}}h_{k\sigma k^{\prime}\sigma^{\prime}}^{\textrm{ph}*}\xi_{k\sigma,\beta}^{*}\xi_{\alpha,k^{\prime}\sigma^{\prime}}^{*}+ih_{k\sigma k^{\prime}\sigma^{\prime}}^{\textrm{ph}\dagger*}\frac{\partial}{\partial\xi_{k^{\prime}\sigma^{\prime},k\sigma}^{*}}\right.\\
\left.+i\sum_{\alpha\beta}\frac{\partial}{\partial\xi_{\alpha\beta}}h_{k\sigma k^{\prime}\sigma^{\prime}}^{\textrm{ph}}\xi_{k\sigma,\beta}\xi_{\alpha,k^{\prime}\sigma^{\prime}}-ih_{k\sigma k^{\prime}\sigma^{\prime}}^{\textrm{ph}\dagger}\frac{\partial}{\partial\xi_{k^{\prime}\sigma^{\prime},k\sigma}}\right\} \det g_{\alpha,\overline{\beta}}\frac{\left|\boldsymbol{\xi}\right\rangle \left\langle \boldsymbol{\xi}\right|}{\mathcal{N}\left(\boldsymbol{\xi}\right)}\\
+Mi\left\{ \textrm{Tr}\left\{ \boldsymbol{h}^{\textrm{ph}*}\boldsymbol{\xi}^{\dagger}\right\} -\textrm{Tr}\left\{ \boldsymbol{h}^{\textrm{ph}*\dagger}\boldsymbol{\xi}\right\} \right\} \det g_{\alpha,\overline{\beta}}\frac{\left|\boldsymbol{\xi}\right\rangle \left\langle \boldsymbol{\xi}\right|}{\mathcal{N}\left(\boldsymbol{\xi}\right)}\\
+Mi\left\{ -\textrm{Tr}\left\{ \boldsymbol{h}^{\textrm{ph}*}\left(\boldsymbol{I}_{\textrm{h}}+\boldsymbol{\xi}^{\dagger}\boldsymbol{\xi}\right)\left[\boldsymbol{I}_{\textrm{h}}+\boldsymbol{\xi}^{\dagger}\boldsymbol{\xi}\right]^{-1}\boldsymbol{\xi}^{\dagger}\right\} +\textrm{Tr}\left\{ \boldsymbol{h}^{\textrm{ph}*\dagger}\boldsymbol{\xi}\left[\boldsymbol{I}_{\textrm{h}}+\boldsymbol{\xi}^{\dagger}\boldsymbol{\xi}\right]^{-1}\left(\boldsymbol{I}_{\textrm{h}}+\boldsymbol{\xi}^{\dagger}\boldsymbol{\xi}\right)\right\} \right\} \det g_{\alpha,\overline{\beta}}\frac{\left|\boldsymbol{\xi}\right\rangle \left\langle \boldsymbol{\xi}\right|}{\mathcal{N}\left(\boldsymbol{\xi}\right)}\\
=\left\{ -i\sum_{\alpha\beta}\frac{\partial}{\partial\xi_{\alpha\beta}^{*}}h_{k\sigma k^{\prime}\sigma^{\prime}}^{\textrm{ph}*}\xi_{k\sigma,\beta}^{*}\xi_{\alpha,k^{\prime}\sigma^{\prime}}^{*}+ih_{k\sigma k^{\prime}\sigma^{\prime}}^{\textrm{ph}\dagger*}\frac{\partial}{\partial\xi_{k^{\prime}\sigma^{\prime},k\sigma}^{*}}\right.\\
\left.+i\sum_{\alpha\beta}\frac{\partial}{\partial\xi_{\alpha\beta}}h_{k\sigma k^{\prime}\sigma^{\prime}}^{\textrm{ph}}\xi_{k\sigma,\beta}\xi_{\alpha,k^{\prime}\sigma^{\prime}}-ih_{k\sigma k^{\prime}\sigma^{\prime}}^{\textrm{ph}\dagger}\frac{\partial}{\partial\xi_{k^{\prime}\sigma^{\prime},k\sigma}}\right\} \det g_{\alpha,\overline{\beta}}\frac{\left|\boldsymbol{\xi}\right\rangle \left\langle \boldsymbol{\xi}\right|}{\mathcal{N}\left(\boldsymbol{\xi}\right)}.
\end{multline}

\subsection{Application to the Husimi master equation}

We employ these differential correspondences in the commutator of
equation (\ref{eq:Husimi_derivative}):
\begin{multline}
-i\left[\widehat{H}_{\textrm{int}}\left(t\right),\det g_{\alpha,\overline{\beta}}\mathcal{N}^{-1}\left(\boldsymbol{\xi}\right)\left|\boldsymbol{\xi}\right\rangle \left\langle \boldsymbol{\xi}\right|\otimes\widehat{1}_{s}\right]\\
=-i\left\{ \frac{\partial}{\partial\xi_{k^{\prime}\sigma^{\prime}\beta}^{*}}\widehat{h}_{k\sigma k^{\prime}\sigma^{\prime}}^{\textrm{pp}\dagger}\left(t\right)\xi_{k\sigma,\beta}^{*}-\frac{\partial}{\partial\xi_{k^{\prime}\sigma^{\prime}\beta}}\widehat{h}_{k\sigma k^{\prime}\sigma^{\prime}}^{\textrm{pp}}\left(t\right)\xi_{k\sigma,\beta}\right\} \det g_{\alpha,\overline{\beta}}\mathcal{N}^{-1}\left(\boldsymbol{\xi}\right)\left|\boldsymbol{\xi}\right\rangle \left\langle \boldsymbol{\xi}\right|\otimes\widehat{1}_{s}\\
+i\left\{ \frac{\partial}{\partial\xi_{\alpha k^{\prime}\sigma^{\prime}}^{*}}\widehat{h}_{k\sigma k^{\prime}\sigma^{\prime}}^{\textrm{hh}\dagger}\left(t\right)\xi_{\alpha,k\sigma}^{*}-\frac{\partial}{\partial\xi_{\alpha k^{\prime}\sigma^{\prime}}}\widehat{h}_{k\sigma k^{\prime}\sigma^{\prime}}^{\textrm{hh}}\left(t\right)\xi_{\alpha,k\sigma}\right\} \det g_{\alpha,\overline{\beta}}\mathcal{N}^{-1}\left(\boldsymbol{\xi}\right)\left|\boldsymbol{\xi}\right\rangle \left\langle \boldsymbol{\xi}\right|\otimes\widehat{1}_{s}\\
-i\left\{ \sum_{\alpha\beta}\frac{\partial}{\partial\xi_{\alpha\beta}^{*}}\widehat{h}_{k\sigma k^{\prime}\sigma^{\prime}}^{\textrm{ph}\dagger}\left(t\right)\xi_{k\sigma,\beta}^{*}\xi_{\alpha,k^{\prime}\sigma^{\prime}}^{*}-\widehat{h}_{k^{\prime}\sigma^{\prime}k\sigma}^{\textrm{ph}}\left(t\right)\frac{\partial}{\partial\xi_{k^{\prime}\sigma^{\prime},k\sigma}^{*}}\right.\\
\left.-\sum_{\alpha\beta}\frac{\partial}{\partial\xi_{\alpha\beta}}\widehat{h}_{k\sigma k^{\prime}\sigma^{\prime}}^{\textrm{ph}}\left(t\right)\xi_{k\sigma,\beta}\xi_{\alpha,k^{\prime}\sigma^{\prime}}+\widehat{h}_{k^{\prime}\sigma^{\prime}k\sigma}^{\textrm{ph}\dagger}\left(t\right)\frac{\partial}{\partial\xi_{k^{\prime}\sigma^{\prime},k\sigma}}\right\} \det g_{\alpha,\overline{\beta}}\mathcal{N}^{-1}\left(\boldsymbol{\xi}\right)\left|\boldsymbol{\xi}\right\rangle \left\langle \boldsymbol{\xi}\right|\otimes\widehat{1}_{s}.
\end{multline}
Therefore, we obtain:
\begin{multline}
\partial_{t}Q\left(\boldsymbol{\xi},\boldsymbol{\xi}^{*};t\right)\\
=i\frac{\partial}{\partial\xi_{k^{\prime}\sigma^{\prime}\beta}^{*}}\xi_{k\sigma,\beta}^{*}\left\langle \Psi\left(\boldsymbol{\xi};t\right)\left|\widehat{h}_{k\sigma k^{\prime}\sigma^{\prime}}^{\textrm{pp}\dagger}\left(t\right)\right|\Psi\left(\boldsymbol{\xi};t\right)\right\rangle \det g_{\alpha,\overline{\beta}}\mathcal{N}^{-1}\left(\boldsymbol{\xi}\right)\\
-i\frac{\partial}{\partial\xi_{k^{\prime}\sigma^{\prime}\beta}}\xi_{k\sigma,\beta}\left\langle \Psi\left(\boldsymbol{\xi};t\right)\left|\widehat{h}_{k\sigma k^{\prime}\sigma^{\prime}}^{\textrm{pp}}\left(t\right)\right|\Psi\left(\boldsymbol{\xi};t\right)\right\rangle \det g_{\alpha,\overline{\beta}}\mathcal{N}^{-1}\left(\boldsymbol{\xi}\right)\\
-i\frac{\partial}{\partial\xi_{\alpha k^{\prime}\sigma^{\prime}}^{*}}\xi_{\alpha,k\sigma}^{*}\left\langle \Psi\left(\boldsymbol{\xi};t\right)\left|\widehat{h}_{k\sigma k^{\prime}\sigma^{\prime}}^{\textrm{hh}\dagger}\left(t\right)\right|\Psi\left(\boldsymbol{\xi};t\right)\right\rangle \det g_{\alpha,\overline{\beta}}\mathcal{N}^{-1}\left(\boldsymbol{\xi}\right)\\
+i\frac{\partial}{\partial\xi_{\alpha k^{\prime}\sigma^{\prime}}}\xi_{\alpha,k\sigma}\left\langle \Psi\left(\boldsymbol{\xi};t\right)\left|\widehat{h}_{k\sigma k^{\prime}\sigma^{\prime}}^{\textrm{hh}}\left(t\right)\right|\Psi\left(\boldsymbol{\xi};t\right)\right\rangle \det g_{\alpha,\overline{\beta}}\mathcal{N}^{-1}\left(\boldsymbol{\xi}\right)\\
+i\frac{\partial}{\partial\xi_{\alpha\beta}^{*}}\xi_{k\sigma,\beta}^{*}\xi_{\alpha,k^{\prime}\sigma^{\prime}}^{*}\left\langle \Psi\left(\boldsymbol{\xi};t\right)\left|\widehat{h}_{k\sigma k^{\prime}\sigma^{\prime}}^{\textrm{ph}\dagger}\left(t\right)\right|\Psi\left(\boldsymbol{\xi};t\right)\right\rangle \det g_{\alpha,\overline{\beta}}\mathcal{N}^{-1}\left(\boldsymbol{\xi}\right)\\
-i\frac{\partial}{\partial\xi_{\alpha\beta}}\xi_{k\sigma,\beta}\xi_{\alpha,k^{\prime}\sigma^{\prime}}\left\langle \Psi\left(\boldsymbol{\xi};t\right)\left|\widehat{h}_{k\sigma k^{\prime}\sigma^{\prime}}^{\textrm{ph}}\left(t\right)\right|\Psi\left(\boldsymbol{\xi};t\right)\right\rangle \det g_{\alpha,\overline{\beta}}\mathcal{N}^{-1}\left(\boldsymbol{\xi}\right)\\
-i\frac{\partial}{\partial\xi_{k^{\prime}\sigma^{\prime},k\sigma}^{*}}\left\langle \Psi\left(\boldsymbol{\xi};t\right)\left|\widehat{h}_{k^{\prime}\sigma^{\prime}k\sigma}^{\textrm{ph}}\left(t\right)\right|\Psi\left(\boldsymbol{\xi};t\right)\right\rangle \det g_{\alpha,\overline{\beta}}\mathcal{N}^{-1}\left(\boldsymbol{\xi}\right)\\
+i\frac{\partial}{\partial\xi_{k^{\prime}\sigma^{\prime},k\sigma}}\left\langle \Psi\left(\boldsymbol{\xi};t\right)\left|\widehat{h}_{k^{\prime}\sigma^{\prime}k\sigma}^{\textrm{ph}\dagger}\left(t\right)\right|\Psi\left(\boldsymbol{\xi};t\right)\right\rangle \det g_{\alpha,\overline{\beta}}\mathcal{N}^{-1}\left(\boldsymbol{\xi}\right).
\end{multline}
Or, if we introduce the normalized open system averages
\begin{equation}
\left\langle \widehat{o}\right\rangle _{Q}\left(\boldsymbol{\xi},\boldsymbol{\xi}^{*};t\right)=\frac{\left\langle \Psi\left(\boldsymbol{\xi};t\right)\left|\widehat{o}\left(t\right)\right|\Psi\left(\boldsymbol{\xi};t\right)\right\rangle }{\left\Vert \Psi\left(\boldsymbol{\xi};t\right)\right\Vert ^{2}},
\end{equation}
we get the Husimi master equation:
\begin{multline*}
\partial_{t}Q\left(\boldsymbol{\xi},\boldsymbol{\xi}^{*};t\right)\\
=i\frac{\partial}{\partial\xi_{k^{\prime}\sigma^{\prime}\beta}^{*}}\xi_{k\sigma,\beta}^{*}\left\langle \widehat{h}_{k\sigma k^{\prime}\sigma^{\prime}}^{\textrm{pp}}\right\rangle _{Q}^{*}\left(\boldsymbol{\xi},\boldsymbol{\xi}^{*};t\right)Q\left(\boldsymbol{\xi},\boldsymbol{\xi}^{*};t\right)+\textrm{c.c.}\\
-i\frac{\partial}{\partial\xi_{\alpha k^{\prime}\sigma^{\prime}}^{*}}\xi_{\alpha,k\sigma}^{*}\left\langle \widehat{h}_{k\sigma k^{\prime}\sigma^{\prime}}^{\textrm{hh}}\right\rangle _{Q}\left(\boldsymbol{\xi},\boldsymbol{\xi}^{*};t\right)Q\left(\boldsymbol{\xi},\boldsymbol{\xi}^{*};t\right)+\textrm{c.c.}\\
+i\frac{\partial}{\partial\xi_{\alpha\beta}^{*}}\xi_{k\sigma,\beta}^{*}\xi_{\alpha,k^{\prime}\sigma^{\prime}}^{*}\left\langle \widehat{h}_{k\sigma k^{\prime}\sigma^{\prime}}^{\textrm{ph}}\right\rangle _{Q}^{*}\left(\boldsymbol{\xi},\boldsymbol{\xi}^{*};t\right)Q\left(\boldsymbol{\xi},\boldsymbol{\xi}^{*};t\right)+\textrm{c.c.}\\
-i\frac{\partial}{\partial\xi_{k^{\prime}\sigma^{\prime},k\sigma}^{*}}\left\langle h_{k^{\prime}\sigma^{\prime}k\sigma}^{\textrm{ph}}\right\rangle _{Q}\left(\boldsymbol{\xi},\boldsymbol{\xi}^{*};t\right)Q\left(\boldsymbol{\xi},\boldsymbol{\xi}^{*};t\right)+\textrm{c.c.}
\end{multline*}
The resulting drift vector is
\begin{multline}
\mathcal{A}_{\alpha\beta}\left(\boldsymbol{\xi},\boldsymbol{\xi}^{*};t\right)=-\left\langle h_{\alpha\beta}^{\textrm{ph}}\right\rangle _{Q}\left(\boldsymbol{\xi},\boldsymbol{\xi}^{*};t\right)+\xi_{\gamma\beta}\left\langle h_{\gamma\alpha}^{\textrm{pp}}\right\rangle _{Q}\left(\boldsymbol{\xi},\boldsymbol{\xi}^{*};t\right)-\xi_{\alpha\gamma}\left\langle h_{\gamma\beta}^{\textrm{hh}}\right\rangle _{Q}\left(\boldsymbol{\xi},\boldsymbol{\xi}^{*};t\right)+\xi_{\gamma\beta}\xi_{\alpha\delta}\left\langle h_{\gamma\delta}^{\textrm{ph}}\right\rangle _{Q}\left(\boldsymbol{\xi},\boldsymbol{\xi}^{*};t\right).
\end{multline}

\section{\label{sec:CALCULATION-OF-THE}CALCULATION OF THE DRESSED HAMILTONIAN}

We need to compute commutators with
\begin{equation}
\exp\left(\widehat{\boldsymbol{c}}^{\textrm{p}}\boldsymbol{\xi}\widehat{\boldsymbol{c}}^{\textrm{h}}\right).
\end{equation}
For this purpose, it is enough to compute 
\begin{multline}
\widehat{d}_{k\sigma}^{\textrm{p}}=\exp\left(+\widehat{\boldsymbol{c}}^{\textrm{p}}\boldsymbol{\xi}\widehat{\boldsymbol{c}}^{\textrm{h}}\right)\widehat{c}_{k\sigma}^{\textrm{p}}\exp\left(-\widehat{\boldsymbol{c}}^{\textrm{p}}\boldsymbol{\xi}\widehat{\boldsymbol{c}}^{\textrm{h}}\right)\\
=\widehat{c}_{k\sigma}^{\textrm{p}},
\end{multline}
\begin{multline}
\widehat{d}_{k\sigma}^{\textrm{h}}=\exp\left(+\widehat{\boldsymbol{c}}^{\textrm{p}}\boldsymbol{\xi}\widehat{\boldsymbol{c}}^{\textrm{h}}\right)\widehat{c}_{k\sigma}^{\textrm{h}}\exp\left(-\widehat{\boldsymbol{c}}^{\textrm{p}}\boldsymbol{\xi}\widehat{\boldsymbol{c}}^{\textrm{h}}\right)\\
=\widehat{c}_{k\sigma}^{\textrm{h}},
\end{multline}
and
\begin{equation}
\widehat{d}_{k\sigma}^{\textrm{p}+}=\exp\left(+\widehat{\boldsymbol{c}}^{\textrm{p}}\boldsymbol{\xi}\widehat{\boldsymbol{c}}^{\textrm{h}}\right)\widehat{c}_{k\sigma}^{\textrm{p}\dagger}\exp\left(-\widehat{\boldsymbol{c}}^{\textrm{p}}\boldsymbol{\xi}\widehat{\boldsymbol{c}}^{\textrm{h}}\right),
\end{equation}
\begin{equation}
\widehat{d}_{k\sigma}^{\textrm{h}+}=\exp\left(+\widehat{\boldsymbol{c}}^{\textrm{p}}\boldsymbol{\xi}\widehat{\boldsymbol{c}}^{\textrm{h}}\right)\widehat{c}_{k\sigma}^{\textrm{h}\dagger}\exp\left(-\widehat{\boldsymbol{c}}^{\textrm{p}}\boldsymbol{\xi}\widehat{\boldsymbol{c}}^{\textrm{h}}\right).
\end{equation}
In order to compute the latter two operators, we introduce the parametrized
operators
\begin{equation}
\widehat{d}_{k\sigma}^{\textrm{p}+}\left(\theta\right)=\exp\left(+\theta\widehat{\boldsymbol{c}}^{\textrm{p}}\boldsymbol{\xi}\widehat{\boldsymbol{c}}^{\textrm{h}}\right)\widehat{c}_{k\sigma}^{\textrm{p}\dagger}\exp\left(-\theta\widehat{\boldsymbol{c}}^{\textrm{p}}\boldsymbol{\xi}\widehat{\boldsymbol{c}}^{\textrm{h}}\right),
\end{equation}
\begin{equation}
\widehat{d}_{k\sigma}^{\textrm{h}+}\left(\theta\right)=\exp\left(+\theta\widehat{\boldsymbol{c}}^{\textrm{p}}\boldsymbol{\xi}\widehat{\boldsymbol{c}}^{\textrm{h}}\right)\widehat{c}_{k\sigma}^{\textrm{h}\dagger}\exp\left(-\theta\widehat{\boldsymbol{c}}^{\textrm{p}}\boldsymbol{\xi}\widehat{\boldsymbol{c}}^{\textrm{h}}\right),
\end{equation}
which obey to the differential equations
\begin{multline}
\partial_{\theta}\widehat{d}_{k\sigma}^{\textrm{p}+}\left(\theta\right)=E\left(\theta\right)\widehat{\boldsymbol{c}}^{\textrm{p}}\boldsymbol{\xi}\widehat{\boldsymbol{c}}^{\textrm{h}}\widehat{c}_{k\sigma}^{\textrm{p}\dagger}E\left(-\theta\right)-E\left(\theta\right)\widehat{c}_{k\sigma}^{\textrm{p}\dagger}\widehat{\boldsymbol{c}}^{\textrm{p}}\boldsymbol{\xi}\widehat{\boldsymbol{c}}^{\textrm{h}}E\left(-\theta\right)\\
=E\left(\theta\right)\widehat{\boldsymbol{c}}^{\textrm{p}}\boldsymbol{\xi}\widehat{\boldsymbol{c}}^{\textrm{h}}E\left(-\theta\right)E\left(\theta\right)\widehat{c}_{k\sigma}^{\textrm{p}\dagger}E\left(-\theta\right)-E\left(\theta\right)\widehat{c}_{k\sigma}^{\textrm{p}\dagger}E\left(-\theta\right)E\left(\theta\right)\widehat{\boldsymbol{c}}^{\textrm{p}}\boldsymbol{\xi}\widehat{\boldsymbol{c}}^{\textrm{h}}E\left(-\theta\right)\\
=+\left[\widehat{\boldsymbol{c}}^{\textrm{p}}\left(\theta\right)\boldsymbol{\xi}\widehat{\boldsymbol{c}}^{\textrm{h}}\left(\theta\right),\widehat{c}_{k\sigma}^{\textrm{p}\dagger}\left(\theta\right)\right]=\\
=\sum_{\alpha,\beta}\left\{ \widehat{c}_{\alpha}^{\textrm{p}}\left(\theta\right)\xi_{\alpha\beta}\widehat{c}_{\beta}^{\textrm{h}}\left(\theta\right)\widehat{c}_{k\sigma}^{\textrm{p}\dagger}\left(\theta\right)-\widehat{c}_{k\sigma}^{\textrm{p}\dagger}\left(\theta\right)\widehat{c}_{\alpha}^{\textrm{p}}\left(\theta\right)\xi_{\alpha\beta}\widehat{c}_{\beta}^{\textrm{h}}\left(\theta\right)\right\} \\
=-\sum_{\alpha,\beta}\left\{ \widehat{c}_{\alpha}^{\textrm{p}}\left(\theta\right)\xi_{\alpha\beta}\widehat{c}_{k\sigma}^{\textrm{p}\dagger}\left(\theta\right)\widehat{c}_{\beta}^{\textrm{h}}\left(\theta\right)+\widehat{c}_{k\sigma}^{\textrm{p}\dagger}\left(\theta\right)\widehat{c}_{\alpha}^{\textrm{p}}\left(\theta\right)\xi_{\alpha\beta}\widehat{c}_{\beta}^{\textrm{h}}\left(\theta\right)\right\} \\
=-\sum_{\alpha,\beta}\left\{ \widehat{c}_{\alpha}^{\textrm{p}}\left(\theta\right),\widehat{c}_{k\sigma}^{\textrm{p}\dagger}\left(\theta\right)\right\} \xi_{\alpha\beta}\widehat{c}_{\beta}^{\textrm{h}}\left(\theta\right)=-\sum_{\beta}\xi_{k\sigma,\beta}\widehat{c}_{\beta}^{\textrm{h}}\left(\theta\right)=-\sum_{\beta}\xi_{k\sigma,\beta}\widehat{c}_{\beta}^{\textrm{h}},
\end{multline}
were $E\left(\theta\right)=\exp\left(\theta\widehat{\boldsymbol{c}}^{\textrm{p}}\boldsymbol{\xi}\widehat{\boldsymbol{c}}^{\textrm{h}}\right)$
and $\beta=\left(l\varsigma\right)$ is a multiindex which runs over
modes and spin directions, and
\begin{multline}
\partial_{\theta}\widehat{d}_{k\sigma}^{\textrm{h}+}\left(\theta\right)=E\left(\theta\right)\widehat{\boldsymbol{c}}^{\textrm{p}}\boldsymbol{\xi}\widehat{\boldsymbol{c}}^{\textrm{h}}\widehat{c}_{k\sigma}^{\textrm{h}\dagger}E\left(-\theta\right)-E\left(\theta\right)\widehat{c}_{k\sigma}^{\textrm{h}\dagger}\widehat{\boldsymbol{c}}^{\textrm{p}}\boldsymbol{\xi}\widehat{\boldsymbol{c}}^{\textrm{h}}E\left(-\theta\right)\\
=\sum_{\alpha,\beta}\left\{ \widehat{c}_{\alpha}^{\textrm{p}}\left(\theta\right)\xi_{\alpha\beta}\widehat{c}_{\beta}^{\textrm{h}}\left(\theta\right)\widehat{c}_{k\sigma}^{\textrm{h}\dagger}\left(\theta\right)-\widehat{c}_{k\sigma}^{\textrm{h}\dagger}\left(\theta\right)\widehat{c}_{\alpha}^{\textrm{p}}\left(\theta\right)\xi_{\alpha\beta}\widehat{c}_{\beta}^{\textrm{h}}\left(\theta\right)\right\} \\
=\sum_{\alpha,\beta}\left\{ \widehat{c}_{\alpha}^{\textrm{p}}\left(\theta\right)\xi_{\alpha\beta}\widehat{c}_{\beta}^{\textrm{h}}\left(\theta\right)\widehat{c}_{k\sigma}^{\textrm{h}\dagger}\left(\theta\right)+\widehat{c}_{\alpha}^{\textrm{p}}\left(\theta\right)\xi_{\alpha\beta}\widehat{c}_{k\sigma}^{\textrm{h}\dagger}\left(\theta\right)\widehat{c}_{\beta}^{\textrm{h}}\left(\theta\right)\right\} \\
=\sum_{\alpha,\beta}\widehat{c}_{\alpha}^{\textrm{p}}\left(\theta\right)\xi_{\alpha\beta}\left\{ \widehat{c}_{\beta}^{\textrm{h}}\left(\theta\right),\widehat{c}_{k\sigma}^{\textrm{h}\dagger}\left(\theta\right)\right\} =\sum_{\alpha}\widehat{c}_{\alpha}^{\textrm{p}}\left(\theta\right)\xi_{\alpha,k\sigma}=\sum_{\alpha}\widehat{c}_{\alpha}^{\textrm{p}}\xi_{\alpha,k\sigma}.
\end{multline}
Therefore, integrating over $\theta$ from $0$ to $1$, we get:
\begin{equation}
\widehat{d}_{k\sigma}^{\textrm{p}+}=\widehat{c}_{k\sigma}^{\textrm{p}\dagger}-\sum_{\beta}\xi_{k\sigma,\beta}\widehat{c}_{\beta}^{\textrm{h}},
\end{equation}
\begin{equation}
\widehat{d}_{k\sigma}^{\textrm{h}+}=\widehat{c}_{k\sigma}^{\textrm{h}\dagger}+\sum_{\alpha}\widehat{c}_{\alpha}^{\textrm{p}}\xi_{\alpha,k\sigma}.
\end{equation}

\subsection{The Dressed Hamiltonian}

\begin{multline}
\widehat{H}_{\textrm{dress}}\left(t\right)=\exp\left(\widehat{\boldsymbol{c}}^{\textrm{p}}\boldsymbol{\xi}\widehat{\boldsymbol{c}}^{\textrm{h}}\right)\widehat{H}\left(t\right)\exp\left(-\widehat{\boldsymbol{c}}^{\textrm{p}}\boldsymbol{\xi}\widehat{\boldsymbol{c}}^{\textrm{h}}\right)\\
=\widehat{h}_{k\sigma k^{\prime}\sigma^{\prime}}^{\textrm{pp}}\left(t\right)\widehat{c}_{k\sigma}^{\textrm{p}\dagger}\widehat{c}_{k^{\prime}\sigma^{\prime}}^{\textrm{p}}-\widehat{h}_{k\sigma k^{\prime}\sigma^{\prime}}^{\textrm{hh}}\left(t\right)\widehat{c}_{k\sigma}^{\textrm{h}\dagger}\widehat{c}_{k^{\prime}\sigma^{\prime}}^{\textrm{h}}\\
+\widehat{h}_{k\sigma k^{\prime}\sigma^{\prime}}^{\textrm{ph}}\left(t\right)\widehat{c}_{k\sigma}^{\textrm{p}\dagger}\widehat{c}_{k^{\prime}\sigma^{\prime}}^{\textrm{h}\dagger}+\widehat{h}_{k\sigma k^{\prime}\sigma^{\prime}}^{\textrm{ph}\dagger}\left(t\right)\widehat{c}_{k\sigma}^{\textrm{h}}\widehat{c}_{k^{\prime}\sigma^{\prime}}^{\textrm{p}}\\
=\widehat{h}_{k\sigma k^{\prime}\sigma^{\prime}}^{\textrm{pp}}\left(t\right)\left\{ \widehat{c}_{k\sigma}^{\textrm{p}\dagger}-\xi_{k\sigma,\beta}\widehat{c}_{\beta}^{\textrm{h}}\right\} \widehat{c}_{k^{\prime}\sigma^{\prime}}^{\textrm{p}}-\widehat{h}_{k\sigma k^{\prime}\sigma^{\prime}}^{\textrm{hh}}\left(t\right)\left\{ \widehat{c}_{k\sigma}^{\textrm{h}\dagger}+\widehat{c}_{\alpha}^{\textrm{p}}\xi_{\alpha,k\sigma}\right\} \widehat{c}_{k^{\prime}\sigma^{\prime}}^{\textrm{h}}\\
+\widehat{h}_{k\sigma k^{\prime}\sigma^{\prime}}^{\textrm{ph}}\left(t\right)\left\{ \widehat{c}_{k\sigma}^{\textrm{p}\dagger}-\xi_{k\sigma,\beta}\widehat{c}_{\beta}^{\textrm{h}}\right\} \left\{ \widehat{c}_{k^{\prime}\sigma^{\prime}}^{\textrm{h}\dagger}+\widehat{c}_{\alpha}^{\textrm{p}}\xi_{\alpha,k^{\prime}\sigma^{\prime}}\right\} +\widehat{h}_{k\sigma k^{\prime}\sigma^{\prime}}^{\textrm{ph}\dagger}\left(t\right)\widehat{c}_{k\sigma}^{\textrm{h}}\widehat{c}_{k^{\prime}\sigma^{\prime}}^{\textrm{p}}\\
=\left\{ \widehat{h}_{k\sigma k^{\prime}\sigma^{\prime}}^{\textrm{pp}}\left(t\right)+\widehat{h}_{k\sigma\gamma}^{\textrm{ph}}\left(t\right)\xi_{k^{\prime}\sigma^{\prime},\gamma}\right\} \widehat{c}_{k\sigma}^{\textrm{p}\dagger}\widehat{c}_{k^{\prime}\sigma^{\prime}}^{\textrm{p}}\\
+\left\{ \widehat{h}_{\gamma k\sigma}^{\textrm{ph}}\left(t\right)\xi_{\gamma,k^{\prime}\sigma^{\prime}}-\widehat{h}_{k\sigma k^{\prime}\sigma^{\prime}}^{\textrm{hh}}\left(t\right)\right\} \widehat{c}_{k\sigma}^{\textrm{h}\dagger}\widehat{c}_{k^{\prime}\sigma^{\prime}}^{\textrm{h}}\\
+\left\{ \widehat{h}_{k\sigma k^{\prime}\sigma^{\prime}}^{\textrm{ph}\dagger}\left(t\right)-\widehat{h}_{\gamma k^{\prime}\sigma^{\prime}}^{\textrm{pp}}\left(t\right)\xi_{\gamma,k\sigma}+\widehat{h}_{\gamma k\sigma}^{\textrm{hh}}\left(t\right)\xi_{k^{\prime}\sigma^{\prime},\gamma}+\widehat{h}_{\gamma\delta}^{\textrm{ph}}\left(t\right)\xi_{\gamma,k\sigma}\xi_{k^{\prime}\sigma^{\prime},\delta}\right\} \widehat{c}_{k\sigma}^{\textrm{h}}\widehat{c}_{k^{\prime}\sigma^{\prime}}^{\textrm{p}}\\
+\widehat{h}_{k\sigma k^{\prime}\sigma^{\prime}}^{\textrm{ph}}\left(t\right)\left\{ \widehat{c}_{k\sigma}^{\textrm{p}\dagger}\widehat{c}_{k^{\prime}\sigma^{\prime}}^{\textrm{h}\dagger}-\xi_{k\sigma,k^{\prime}\sigma^{\prime}}\right\} .
\end{multline}


\begin{thebibliography}{18}%
\makeatletter
\providecommand \@ifxundefined [1]{%
 \@ifx{#1\undefined}
}%
\providecommand \@ifnum [1]{%
 \ifnum #1\expandafter \@firstoftwo
 \else \expandafter \@secondoftwo
 \fi
}%
\providecommand \@ifx [1]{%
 \ifx #1\expandafter \@firstoftwo
 \else \expandafter \@secondoftwo
 \fi
}%
\providecommand \natexlab [1]{#1}%
\providecommand \enquote  [1]{``#1''}%
\providecommand \bibnamefont  [1]{#1}%
\providecommand \bibfnamefont [1]{#1}%
\providecommand \citenamefont [1]{#1}%
\providecommand \href@noop [0]{\@secondoftwo}%
\providecommand \href [0]{\begingroup \@sanitize@url \@href}%
\providecommand \@href[1]{\@@startlink{#1}\@@href}%
\providecommand \@@href[1]{\endgroup#1\@@endlink}%
\providecommand \@sanitize@url [0]{\catcode `\\12\catcode `\$12\catcode
  `\&12\catcode `\#12\catcode `\^12\catcode `\_12\catcode `\%12\relax}%
\providecommand \@@startlink[1]{}%
\providecommand \@@endlink[0]{}%
\providecommand \url  [0]{\begingroup\@sanitize@url \@url }%
\providecommand \@url [1]{\endgroup\@href {#1}{\urlprefix }}%
\providecommand \urlprefix  [0]{URL }%
\providecommand \Eprint [0]{\href }%
\providecommand \doibase [0]{http://dx.doi.org/}%
\providecommand \selectlanguage [0]{\@gobble}%
\providecommand \bibinfo  [0]{\@secondoftwo}%
\providecommand \bibfield  [0]{\@secondoftwo}%
\providecommand \translation [1]{[#1]}%
\providecommand \BibitemOpen [0]{}%
\providecommand \bibitemStop [0]{}%
\providecommand \bibitemNoStop [0]{.\EOS\space}%
\providecommand \EOS [0]{\spacefactor3000\relax}%
\providecommand \BibitemShut  [1]{\csname bibitem#1\endcsname}%
\let\auto@bib@innerbib\@empty
\bibitem [{\citenamefont {Diosi}\ and\ \citenamefont
  {Strunz}(1997)}]{Diosi1997}%
  \BibitemOpen
  \bibfield  {author} {\bibinfo {author} {\bibfnamefont {L.}~\bibnamefont
  {Diosi}}\ and\ \bibinfo {author} {\bibfnamefont {W.~T.}\ \bibnamefont
  {Strunz}},\ }\href@noop {} {\bibfield  {journal} {\bibinfo  {journal} {Phys.
  Lett. A}\ }\textbf {\bibinfo {volume} {235}},\ \bibinfo {pages} {569}
  (\bibinfo {year} {1997})}\BibitemShut {NoStop}%
\bibitem [{\citenamefont {Wang}\ \emph {et~al.}(2019)\citenamefont {Wang},
  \citenamefont {Ke},\ and\ \citenamefont {Zhao}}]{Wang2019}%
  \BibitemOpen
  \bibfield  {author} {\bibinfo {author} {\bibfnamefont {Y.-C.}\ \bibnamefont
  {Wang}}, \bibinfo {author} {\bibfnamefont {Y.}~\bibnamefont {Ke}}, \ and\
  \bibinfo {author} {\bibfnamefont {Y.}~\bibnamefont {Zhao}},\ }\href@noop {}
  {\bibfield  {journal} {\bibinfo  {journal} {WIRES Comput. Mol. Sci.}\
  }\textbf {\bibinfo {volume} {9}},\ \bibinfo {pages} {e1375} (\bibinfo {year}
  {2019})}\BibitemShut {NoStop}%
\bibitem [{\citenamefont {Shao}(2004)}]{Shao2004}%
  \BibitemOpen
  \bibfield  {author} {\bibinfo {author} {\bibfnamefont {J.}~\bibnamefont
  {Shao}},\ }\href@noop {} {\bibfield  {journal} {\bibinfo  {journal} {J. Chem.
  Phys.}\ }\textbf {\bibinfo {volume} {120}},\ \bibinfo {pages} {5053}
  (\bibinfo {year} {2004})}\BibitemShut {NoStop}%
\bibitem [{\citenamefont {Yan}\ \emph {et~al.}(2004)\citenamefont {Yan},
  \citenamefont {Yang}, \citenamefont {Liu},\ and\ \citenamefont
  {Shao}}]{Yan2004}%
  \BibitemOpen
  \bibfield  {author} {\bibinfo {author} {\bibfnamefont {Y.-a.}\ \bibnamefont
  {Yan}}, \bibinfo {author} {\bibfnamefont {F.}~\bibnamefont {Yang}}, \bibinfo
  {author} {\bibfnamefont {Y.}~\bibnamefont {Liu}}, \ and\ \bibinfo {author}
  {\bibfnamefont {J.}~\bibnamefont {Shao}},\ }\href@noop {} {\bibfield
  {journal} {\bibinfo  {journal} {Chem. Phys. Lett.}\ }\textbf {\bibinfo
  {volume} {395}},\ \bibinfo {pages} {216} (\bibinfo {year}
  {2004})}\BibitemShut {NoStop}%
\bibitem [{\citenamefont {Zhou}\ and\ \citenamefont {Shao}(2008)}]{Shao2008}%
  \BibitemOpen
  \bibfield  {author} {\bibinfo {author} {\bibfnamefont {Y.}~\bibnamefont
  {Zhou}}\ and\ \bibinfo {author} {\bibfnamefont {J.}~\bibnamefont {Shao}},\
  }\href@noop {} {\bibfield  {journal} {\bibinfo  {journal} {J. Chem. Phys.}\
  }\textbf {\bibinfo {volume} {128}},\ \bibinfo {pages} {034106} (\bibinfo
  {year} {2008})}\BibitemShut {NoStop}%
\bibitem [{\citenamefont {Yan}\ and\ \citenamefont {Shao}(2016)}]{Yan2016}%
  \BibitemOpen
  \bibfield  {author} {\bibinfo {author} {\bibfnamefont {Y.-A.}\ \bibnamefont
  {Yan}}\ and\ \bibinfo {author} {\bibfnamefont {J.}~\bibnamefont {Shao}},\
  }\href@noop {} {\bibfield  {journal} {\bibinfo  {journal} {Front. Phys.}\
  }\textbf {\bibinfo {volume} {11}},\ \bibinfo {pages} {110309} (\bibinfo
  {year} {2016})}\BibitemShut {NoStop}%
\bibitem [{\citenamefont {Suess}\ \emph {et~al.}(2014)\citenamefont {Suess},
  \citenamefont {Eisfeld},\ and\ \citenamefont {Strunz}}]{Suess2014}%
  \BibitemOpen
  \bibfield  {author} {\bibinfo {author} {\bibfnamefont {D.}~\bibnamefont
  {Suess}}, \bibinfo {author} {\bibfnamefont {A.}~\bibnamefont {Eisfeld}}, \
  and\ \bibinfo {author} {\bibfnamefont {W.~T.}\ \bibnamefont {Strunz}},\
  }\href@noop {} {\bibfield  {journal} {\bibinfo  {journal} {Phys. Rev. Lett.}\
  }\textbf {\bibinfo {volume} {113}},\ \bibinfo {pages} {150403} (\bibinfo
  {year} {2014})}\BibitemShut {NoStop}%
\bibitem [{\citenamefont {Polyakov}\ and\ \citenamefont
  {Rubtsov}(2019)}]{Polyakov2019}%
  \BibitemOpen
  \bibfield  {author} {\bibinfo {author} {\bibfnamefont {E.~A.}\ \bibnamefont
  {Polyakov}}\ and\ \bibinfo {author} {\bibfnamefont {A.~N.}\ \bibnamefont
  {Rubtsov}},\ }\href@noop {} {\bibfield  {journal} {\bibinfo  {journal} {New.
  J. Phys.}\ }\textbf {\bibinfo {volume} {21}},\ \bibinfo {pages} {063004}
  (\bibinfo {year} {2019})}\BibitemShut {NoStop}%
\bibitem [{\citenamefont {Hartmann}\ and\ \citenamefont
  {Strunz}(2017)}]{Hartmann2017}%
  \BibitemOpen
  \bibfield  {author} {\bibinfo {author} {\bibfnamefont {R.}~\bibnamefont
  {Hartmann}}\ and\ \bibinfo {author} {\bibfnamefont {W.~T.}\ \bibnamefont
  {Strunz}},\ }\href@noop {} {\bibfield  {journal} {\bibinfo  {journal} {J.
  Chem. Theor. Comput.}\ }\textbf {\bibinfo {volume} {13}},\ \bibinfo {pages}
  {5834} (\bibinfo {year} {2017})}\BibitemShut {NoStop}%
\bibitem [{\citenamefont {Shabani}\ \emph {et~al.}(2017)\citenamefont
  {Shabani}, \citenamefont {Roden},\ and\ \citenamefont
  {Whaley}}]{Shabani2014}%
  \BibitemOpen
  \bibfield  {author} {\bibinfo {author} {\bibfnamefont {A.}~\bibnamefont
  {Shabani}}, \bibinfo {author} {\bibfnamefont {J.}~\bibnamefont {Roden}}, \
  and\ \bibinfo {author} {\bibfnamefont {K.~B.}\ \bibnamefont {Whaley}},\
  }\href@noop {} {\bibfield  {journal} {\bibinfo  {journal} {Phys. Rev. Lett.}\
  }\textbf {\bibinfo {volume} {112}},\ \bibinfo {pages} {113601} (\bibinfo
  {year} {2017})}\BibitemShut {NoStop}%
\bibitem [{\citenamefont {Gambetta}\ and\ \citenamefont
  {Wiseman}(2002)}]{Gambetta2002}%
  \BibitemOpen
  \bibfield  {author} {\bibinfo {author} {\bibfnamefont {J.}~\bibnamefont
  {Gambetta}}\ and\ \bibinfo {author} {\bibfnamefont {H.~M.}\ \bibnamefont
  {Wiseman}},\ }\href@noop {} {\bibfield  {journal} {\bibinfo  {journal} {Phys.
  Rev. A}\ }\textbf {\bibinfo {volume} {66}},\ \bibinfo {pages} {012108}
  (\bibinfo {year} {2002})}\BibitemShut {NoStop}%
\bibitem [{\citenamefont {Gambetta}\ and\ \citenamefont
  {Wiseman}(2003)}]{Gambetta2003}%
  \BibitemOpen
  \bibfield  {author} {\bibinfo {author} {\bibfnamefont {J.}~\bibnamefont
  {Gambetta}}\ and\ \bibinfo {author} {\bibfnamefont {H.~M.}\ \bibnamefont
  {Wiseman}},\ }\href@noop {} {\bibfield  {journal} {\bibinfo  {journal} {Phys.
  Rev. A}\ }\textbf {\bibinfo {volume} {68}},\ \bibinfo {pages} {062104}
  (\bibinfo {year} {2003})}\BibitemShut {NoStop}%
\bibitem [{\citenamefont {Han}\ \emph {et~al.}(2019)\citenamefont {Han},
  \citenamefont {Chernyak}, \citenamefont {Yan}, \citenamefont {Zheng},\ and\
  \citenamefont {Yan}}]{Han2019}%
  \BibitemOpen
  \bibfield  {author} {\bibinfo {author} {\bibfnamefont {L.}~\bibnamefont
  {Han}}, \bibinfo {author} {\bibfnamefont {V.}~\bibnamefont {Chernyak}},
  \bibinfo {author} {\bibfnamefont {Y.-A.}\ \bibnamefont {Yan}}, \bibinfo
  {author} {\bibfnamefont {X.}~\bibnamefont {Zheng}}, \ and\ \bibinfo {author}
  {\bibfnamefont {Y.}~\bibnamefont {Yan}},\ }\href@noop {} {\bibfield
  {journal} {\bibinfo  {journal} {Phys. Rev. Lett.}\ }\textbf {\bibinfo
  {volume} {123}},\ \bibinfo {pages} {0501601} (\bibinfo {year}
  {2019})}\BibitemShut {NoStop}%
\bibitem [{\citenamefont {Breuer}\ and\ \citenamefont
  {Petruccione}(2007)}]{Breuer2011}%
  \BibitemOpen
  \bibfield  {author} {\bibinfo {author} {\bibfnamefont {H.-P.}\ \bibnamefont
  {Breuer}}\ and\ \bibinfo {author} {\bibfnamefont {F.}~\bibnamefont
  {Petruccione}},\ }\href@noop {} {\emph {\bibinfo {title} {The Theory of Open
  Quantum Systems}}}\ (\bibinfo  {publisher} {Oxford University Press},\
  \bibinfo {address} {Oxford},\ \bibinfo {year} {2007})\BibitemShut {NoStop}%
\bibitem [{\citenamefont {Zhang}\ \emph {et~al.}(1990)\citenamefont {Zhang},
  \citenamefont {Feng},\ and\ \citenamefont {Gilmore}}]{Zhang1990}%
  \BibitemOpen
  \bibfield  {author} {\bibinfo {author} {\bibfnamefont {W.-M.}\ \bibnamefont
  {Zhang}}, \bibinfo {author} {\bibfnamefont {D.~H.}\ \bibnamefont {Feng}}, \
  and\ \bibinfo {author} {\bibfnamefont {R.}~\bibnamefont {Gilmore}},\
  }\href@noop {} {\bibfield  {journal} {\bibinfo  {journal} {Rev. Mod. Phys.}\
  }\textbf {\bibinfo {volume} {1990}},\ \bibinfo {pages} {867} (\bibinfo {year}
  {1990})}\BibitemShut {NoStop}%
\bibitem [{\citenamefont {Rosales-Zarate}\ and\ \citenamefont
  {Drummond}(2013)}]{RosalesZarate2013}%
  \BibitemOpen
  \bibfield  {author} {\bibinfo {author} {\bibfnamefont {L.}~\bibnamefont
  {Rosales-Zarate}}\ and\ \bibinfo {author} {\bibfnamefont {P.~D.}\
  \bibnamefont {Drummond}},\ }\href@noop {} {\bibfield  {journal} {\bibinfo
  {journal} {J. Phys. A: Math. Theor.}\ }\textbf {\bibinfo {volume} {46}},\
  \bibinfo {pages} {275203} (\bibinfo {year} {2013})}\BibitemShut {NoStop}%
\bibitem [{\citenamefont {Rosales-Zarate}\ and\ \citenamefont
  {Drummond}(2015)}]{RosalesZarate2015}%
  \BibitemOpen
  \bibfield  {author} {\bibinfo {author} {\bibfnamefont {L.}~\bibnamefont
  {Rosales-Zarate}}\ and\ \bibinfo {author} {\bibfnamefont {P.~D.}\
  \bibnamefont {Drummond}},\ }\href@noop {} {\bibfield  {journal} {\bibinfo
  {journal} {New J. Phys.}\ }\textbf {\bibinfo {volume} {17}},\ \bibinfo
  {pages} {032002} (\bibinfo {year} {2015})}\BibitemShut {NoStop}%
\bibitem [{\citenamefont {Drummond}\ and\ \citenamefont
  {Reid}()}]{Drummond2019}%
  \BibitemOpen
  \bibfield  {author} {\bibinfo {author} {\bibfnamefont {P.~D.}\ \bibnamefont
  {Drummond}}\ and\ \bibinfo {author} {\bibfnamefont {M.~D.}\ \bibnamefont
  {Reid}},\ }\href@noop {} {\enquote {\bibinfo {title} {Q-functions as models
  of physical reality},}\ }\Eprint {http://arxiv.org/abs/1909.001798}
  {arXiv:1909.001798 [quant-ph]} \BibitemShut {NoStop}%
\end{thebibliography}
\end{document}